\documentclass[final,5p,times,twocolumn]{elsarticle}

\usepackage{amsmath}
\usepackage{amssymb}
\usepackage{amsthm}
\usepackage{array}
\usepackage{amsfonts}
\usepackage{subfigure}
\usepackage{algorithmic}
\usepackage{algorithm}
\usepackage{bm}
\usepackage{graphicx}
\usepackage{multirow}
\usepackage{color}

\usepackage{colortbl}
\usepackage{booktabs}
\usepackage[american]{babel}
\usepackage{microtype}
\usepackage{gensymb}

\newtheorem{definition}{Definition}
\newtheorem{problem}{Problem}

\usepackage{enumerate}
\usepackage{enumitem}

\begin{document}


\begin{frontmatter}



\title{OTHR multitarget tracking with a GMRF model of ionospheric parameters}

\author[1,2]{Zhen Guo}
\author[1,2,3]{Zengfu Wang\corref{cor1}}
\ead{wangzengfu@nwpu.edu.cn}
\author[1,2]{Hua Lan}
\author[1,2]{Quan Pan}
\author[4]{Kun Lu}

\cortext[cor1]{Corresponding author}
\address[1]{School of Automation, Northwestern Polytechnical University, Xi'an, Shaanxi, 710072, China}
\address[2]{Key Laboratory of Information Fusion Technology, Ministry of Education, Xi'an, Shaanxi, 710072, China}
\address[3]{Faculty of Electrical Engineering, Mathematics and Computer Science, Delft University of Technology, Delft 2826 CD, the Netherlands}
\address[4]{Nanjing Research Institute of Electronics Technology, Nanjing, Jiangsu, 210039, China}

\begin{abstract}
The ionosphere is the propagation medium for radio waves transmitted by an over-the-horizon radar (OTHR).
Ionospheric parameters, typically, virtual ionospheric heights (VIHs), are required to perform coordinate registration for OTHR multitarget tracking and localization. The inaccuracy of ionospheric parameters has a significant deleterious effect on the target localization of OTHR. Therefore, to improve the localization accuracy of OTHR, it is important to develop accurate models and estimation methods of ionospheric parameters and the corresponding target tracking algorithms. In this paper, we consider the variation of the ionosphere with location and the spatial correlation of the ionosphere in OTHR target tracking.
We use a Gaussian Markov random field (GMRF) to model the VIHs, providing a more accurate representation of the VIHs for OTHR target tracking. Based on expectation-conditional maximization and GMRF modeling of the VIHs, we propose a novel joint optimization solution, called ECM-GMRF, to perform target state estimation, multipath data association and VIHs estimation simultaneously.
In ECM-GMRF, the measurements from both ionosondes and OTHR are exploited to estimate the VIHs, leading to a better estimation of the VIHs which improves the accuracy of data association and target state estimation, and vice versa. The simulation indicates the effectiveness of the proposed algorithm.
\end{abstract}




\end{frontmatter}


\section{Introduction}\label{sec:introduction}
On account of many advantages such as the ability to detect and track targets at long ranges~(typically $800$~km to $3,000$~km)
beyond the earth's horizon with comparatively low operation cost,
skywave over-the-horizon radar~(OTHR) has become increasingly important for both civil and military applications, such as maritime reconnaissance and drug enforcement~\cite{Ferraro1997, Fabrizio2013}.
As a medium and acting as a reflector for reflecting electromagnetic waves,
the earth's ionosphere makes OTHR break the limitation of radar horizon.
However, the propagation of electromagnetic waves through the inherent complicated ionosphere brings about a coordinate registration~(CR)~\cite{Wheadon1994} process for OTHR target tracking, since target locations in ground coordinate system~(i.e., latitude and longitude) are needed but OTHR receives measurements in slant coordinate system~(i.e., slant azimuth and slant range).
To carry out CR, ionospheric parameters that describe the geometric transformation between ground coordinate system and slant coordinate system, such as virtual ionospheric heights~(VIHs), are required~\cite{Pulford1998, Pulford2004}.
Location error analysis of OTHR shows that inaccuracy of ionospheric parameters is a main source of the localization error of OTHR~\cite{Zhou2008}.
Accordingly, to improve the localization accuracy of OTHR,
it is important to develop accurate models and estimation methods for ionospheric parameters~(e.g., VIHs)
and the corresponding target tracking algorithms.
In this context, the existing OTHR target tracking approaches can be classified into the following four groups.

The approaches in the first group assumed that each layer of the ionosphere is an ideal mirror,
and the VIH of each layer is a known constant.
Focusing on the multipath data association problem caused by the multilayer structure of the ionosphere,
these approaches include the multipath probabilistic data association~(MPDA)~\cite{Pulford1998},
the multiple detection multiple hypothesis tracker~(MD-MHT)~\cite{Sathyan2013},
the expectation-maximization~(EM)-based joint multipath data association and state estimation~(JMAE)~\cite{Lan2014},
the multipath Bernoulli filter~(MPBF)~\cite{Chen2014},
the multidetection probability hypothesis density~(MD-PHD) filter~\cite{Tang2015, Qin2015},
and the multipath linear multitarget integrated probabilistic data association (MP-LM-IPDA)~\cite{Huang2019}.

The second group, including \cite{Feng2014} and \cite{Lan2018},
assumed that the VIH of each layer is an unknown constant.
This being the case, it is necessary to exploit a joint optimization scheme that solves target state estimation,
multipath data association, and VIHs identification.
Specifically, the former, \cite{Feng2014}, resorted to sensor fusion of OTHR and a set of forward-based receivers;
the latter, \cite{Lan2018}, adopted a distributed expectation-conditional maximization~(ECM) framework.

The approaches in the third group made the assumption that
the VIH(s) of each layer is~(are) a random variable or a random process in time.
The improvement of reliability on the approaches is achieved by incorporating the uncertainty of VIHs into the tracking algorithms.
Examples of this group include the MPDA for uncertain coordinate registration (MPCR) where the VIH of each layer
follows a Gaussian distribution with known mean and variance~\cite{Pulford2004},
the multi-hypothesis multipath track fusion~(MPTF) where VIHs evolve with a linear dynamics model
and Gaussian distribution~\cite{Rutten2001},
and the joint estimation of target state and the bias of VIHs where VIHs are summation of known, nominal VIHs provided by ionosondes and unknown, time-varying bias~\cite{Geng2016,Geng2018}.
In~\cite{Geng2016}, by introducing the intermittently evolving dynamic model for the bias of VIHs,
the joint estimation problem was reformulated as a multi-rate state estimation with random coefficient matrices.
In~\cite{Geng2018}, modeling the bias of VIHs via a Markovian jump model,
the joint estimation problem was converted to pure state estimate with stochastic parameters by embedding the data association in the resultant measurement model with random coefficients.
Both of the two joint estimation problems were solved by the deduced linear minimum mean square error estimator with causality constraints.

Few work, belongs to the fourth group, considered full ionospheric models.
The parameters of the full ionospheric models, for example, critical frequency, height at the base of the layer and at the top,
were assumed to be known.
In~\cite{Bourgeois2005, Bourgeois2006}, the multi-quasi-parabolic model was used in CR of OTHR Nostradamus target tracking.
In~\cite{Romeo2017}, a maximum-likelihood probabilistic multihypothesis tracker for OTHR was presented,
where a 3-D ionospheric regional model from the International Reference Ionosphere~(IRI) was used
and the signal refraction in the ionosphere was modeled by ray-tracing.

The assumption of location independence of VIHs made by the first three groups significantly simplified OTHR target tracking.
However, it is far from being accurate since the ionosphere is a spatiotemporal process.
The VIHs of each layer in a large region are apparently different in different locations.
The work of the fourth group considered the location dependency of the ionosphere through the full ionospheric models.
However, the properties of the ionosphere as a spatiotemporal process, such as spatial correlation,
were not fully considered.

Much work, especially on total electron contents~(TECs) and electron density, has shown that the ionosphere is spatially correlated.
In the early stage, it was discovered that the ionosphere is horizontally spatially correlated.
In~\cite{Soicher1978}, it was found that TECs are highly correlatable in two locations separated by $13\degree$ in latitude and $5\degree$ in longitude.
In~\cite{Gail1993}, the spatial correlation coefficients of TECs was calculated,
based on which it is possible to predict mid-latitude TECs behavior over separations of up to $1200$~km given a single TECs measurement.
The correlation for mid-latitude ionosphere over the Western U.S. was showed in~\cite{Bust2001}.
Later, the spatial correlation of the ionosphere including both the horizontal and the vertical correlation was investigated~\cite{Yue2007}.
Krankowski~et al.~\cite{Krankowski2011} summarized that the correlation distance of ionosphere depends on direction and it is anisotropic.
Arikan~\cite{Arikan2007} applied random field theory to ionospheric electron density reconstruction problem and
discussed the choice of correlation functions.
Minkwitz~et al.~\cite{Minkwitz2015} proposed an approach that estimates the electron density's spatial covariance model which
reveals the different correlation lengths in latitude and longitude direction.
Liu~et al.~\cite{Liu2018} investigated the horizontal spatial correlation of globally ionospheric TECs.
The correlation scale was compared from the aspects of direction, latitude and season.
Norberg~et al.~\cite{Norberg2015, Norberg2016, Norberg2018} proposed a new ionospheric tomography method in Bayesian framework which enables an interpretable scheme to build the prior distribution based on physical and empirical information on the structure of the
ionosphere for $2$D case~\cite{Norberg2015,Norberg2016} and $3$D multi-instrument case~\cite{Norberg2018}.
In ionospheric weather forecast, an accurate correlation model of the ionosphere is used to construct background field error covariance matrix for
data assimilation.

In this paper, we consider the variation of the ionosphere with location and the spatial correlation of the ionosphere
in OTHR target tracking.
Like the approaches of the first three groups, we use VIHs as the key parameters in CR.
It is not unnatural to assume that VIHs are location dependent.
The assumption of the spatial correlation of VIHs is based upon the spatial correlation of the above-mentioned electron density and TECs,
and the dependence of VIHs on them.
As an example of this dependence, given an operating frequency of OTHR,
VIHs are linear with the height where electron density is maximum in parabolic layer model~\cite{Booker1940}.
Our purpose to consider the variation of the ionosphere with location and the spatial correlation of the ionosphere
is to create a model that more accurately represents the ionosphere in OTHR target tracking
and therefore improve the localization error of OTHR.
In practice, assuming a constant VIH for the ionosphere will unavoidably bring about a significant localization error for OTHR.
To estimate VIHs online, a network of available ionosondes, including, for example,
vertical ionosondes, quasi-vertical ionosondes, oblique ionosondes, are deployed~\cite{Wheadon1994}.
However, since the deployment of ionosondes is restricted to available areas and costs operators capital expenditures and operating expenses,
measurements of VIHs on very limited locations are obtained.
By considering the spatial correlation of VIHs,
we are able to infer or predict the VIHs of the ionosphere acting as a reflection area more accurately.

We use a discrete Gaussian Markov random field~(GMRF) to model the variation of VIHs with location and the spatial correlation of VIHs.
Based on this model, we then propose a joint tracking algorithm using ECM-based technique
for joint data association, target state estimation, and VIHs identification.
In this algorithm, we use not only the measurements of VIHs from ionosondes but also the measurements of targets from OTHR to estimate VIHs
since the latter also include the information on VIHs.
By modeling VIHs as a GMRF, we are able to infer the VIHs at the locations where the signal reflection occurs for a given target.
A better estimation of VIHs can improve the accuracy of data association and target state estimation, and vice versa.
As pointed out in~\cite{Lan2018}, joint data association and target state estimation are effective for dealing
with the coupling issue of identification risks and estimation errors~\cite{Lan2018}.
Note that~\cite{Lan2018} considered the single target tracking.
In summary, the contributions of this paper beyond~\cite{Lan2018} are as follows:
\begin{enumerate}
	\item  For the first time, we consider the variation of the ionosphere with location and the spatial correlation of ionosphere
	in OTHR target tracking. We use GMRF to model the VIHs of the ionosphere, providing a more accurate representation of VIHs for OTHR target tracking.
	\item Using ECM, we propose a joint multitarget tracking framework for data association,
	target state estimation and VIHs estimation.
	Leveraging GMRF modeling of the VIHs of the ionosphere, we use both the ionosonde measurements and the radar measurements to
	estimate the target state and the VIHs, improving target localization accuracy of OTHR.
\end{enumerate}

Our preliminary work was presented in~\cite{Guo2018} and has been substantially extended in this study.
In~\cite{Guo2018}, we took into account of spatial correlation of the VIHs and inferred the posterior distribution of the VIHs based on GMRF.
Then, the estimated VIHs was passed to the existing MPCR algorithm for single target tracking.
In this paper, we integrate the proposed VIHs model with the ECM framework,
where radar measurements are also utilized to estimate VIHs.
Additionally, the multitarget multi-detection pattern is presented for multitarget data association.
The detailed implementation for inferring the posterior distribution of the VIHs over large scale regions is also presented.

The remainder of this paper is organized as follows.
Section~\ref{Sec:problem_formulation} consists of the assumptions and models for OTHR target tracking.
The detailed derivation of the proposed algorithm, ECM-GMRF, is depicted in Section~\ref{sec:OTHRMultipleTargetTracking}.
Simulation results are presented in Section~\ref{sec:simulation} followed by concluding remarks in Section~\ref{sec:conclusion}.

Throughout this paper, the superscripts $``-1" $ and $``T" $ represent the inverse and transpose operations of a matrix, respectively; ${I\{\cdot\}}$ denotes the indicator function, which equals one if the event $\{\cdot\}$ is true, or zero otherwise;
The italic \textit{w.~r.~t.} is the abbreviation of ``with respect to".

\section{Models and problem formulation}\label{Sec:problem_formulation}
In this section, we describe the models we introduce for VIHs, ionosonde measurements,
target dynamics, and radar measurement.
Based on the models, we provide a formulation of our OTHR multitarget multipath tracking problem.

\subsection{Model of VIHs}\label{sec:Ionospheric_model}
Without loss of generality, a two-layer~(E and F) spherical mirror ionosphere is assumed.
The ionosphere region to be used for signal reflection depends on the wide surveillance area of OTHR, and is correspondingly large.
The VIHs at different locations are different due to the different characteristics of the earth's magnetic field~\cite{Shim2008}.
Based upon the existing work on the ionospheric spatial correlation~\cite{Soicher1978,KLOBUCHAR1995, Minkwitz2015}
(See more details in Section~\ref{sec:introduction}),
we assume that there exists horizontal correlation of VIHs but do not consider the vertical correlation of VIHs for simplicity.
Since, in short time periods, the correlation strength of the ionosphere at two different sites largely depends on the distance between them, which corresponds to the local Markov property~\cite{Li2009}, VIHs can be modeled as a Markov random field.
As described in~\cite{Norberg2015,Norberg2016,Norberg2018}, which assumed that electron density and TECs are Gaussian,
we assume that the VIHs are Gaussian as well.
Accordingly, we use a discrete Gaussian Markov random field~(GMRF) to describe the VIHs of both E layer and F layer.

Specifically, we take E layer as an example to illustrate our GMRF description of VIHs.
We associate E layer with an undirected graph $G = (\mathcal{V}, \mathcal{E})$,
where $\mathcal{V} = \{1, 2, \dots, N\}$ is the set of nodes in the graph,
and $\mathcal{E}$
is the set of edges $(i,j)$, $i, j \in \mathcal{V}$ and $i \neq j$.
Denote $h_{i} \in \mathbb{R}_{+}, i \in \mathcal{V}$ as the VIH of E layer at site $i$.
Henceforth, $N$ is the number of sites of E layer.
Let $\bm{h} = \{h_{i}\}_{i=1}^N$.
We assume that $\bm{h} \sim \mathcal{N} (\bm{\mu}, \Sigma)$, where $\bm{\mu}$ is the mean vector and $\Sigma$ is the covariance matrix.
Let $Q = \Sigma^{-1}$ and $\bm{\eta} = Q \bm{\mu}$.
$Q$, $\bm{\eta}$ are called precision matrix and potential vector, respectively.
The random vector of VIHs $\bm{h}$ is defined as a discrete GMRF \emph{w.~r.~t.} the labelled graph $G$ with
mean $\bm{\mu}$ and precision matrix $Q > 0$, if and only if its density has the form~\cite{Rue2005}
\begin{equation}
p(\bm{h}) = (2\pi)^{-\frac{N}{2}}|Q|^{\frac{1}{2}} \exp \left(-\frac{1}{2} (\bm{h}-\bm{\mu})^T Q(\bm{h}-\bm{\mu})\right) \,,
\end{equation}
and	$Q_{ij} \neq 0$ is equivalent to $(i,j) \in \mathcal{E}$ for all $i \neq j$.
Therefore, two VIHs at two different sites $i$ and $j$ are spatially correlated if there
exists an edge $e = (i,j) \in \mathcal{E}$ and vice versa.
For a GMRF, it is often more convenient to work with the canonical~(information) form, which is defined as \cite{Rue2005}
\begin{equation}
p(\bm{h}) \propto \exp \left(\bm{\eta}^T \bm{h} - \frac{1}{2} \bm{h}^T Q \bm{h} \right).
\end{equation}

In principle, the priors of the GMRF model, including the mean vector $\bm{\mu}$, the covariance matrix $\Sigma$,
the potential vector $\bm{\eta}$, and the precision matrix $Q$, can be learned from historical measurements of VIHs by the
standard maximum likelihood estimator.
However, for a large-scale sparse precision matrix $Q$, one may need more effective learning method~\cite{Treister2014}.
We leave the learning of the GMRF model as our future work and assume that the priors are given in this paper.

\subsection{Model of ionosonde measurement}\label{sec:subIonoModels}
The vertical incidence ionospheric sounding and the oblique incidence ionospheric sounding are two typical types of ionosondes.
We assume that OTHR measurement and ionosonde measurement are synchronized.
By assuming stationary electron density for the given time interval~\cite{Norberg2018}, the ionosonde measurements of each site of each layer can be written as,
\begin{equation}\label{equ_ZMeasurementModel}
{z}^{s}_{i, k} = g({h}^{s}_{i, k}) + {v}^{s}_{k}\,,~~~ s = \mbox{E,~F}~ \mbox{and}~ i = 1, \ldots, N,
\end{equation}
where ${z}^{s}_{i,k}$ is the time delay.
It is assumed that ${v}^{s}_{k}$ is a zero mean Gaussian noise, i.e., ${v}^{s}_{k}\sim \mathcal{N}({0}, {A_{k}^{s}})$.
The measurement function $g(\cdot)$ depends on the type of ionosondes.
For a vertical incidence ionosonde, the measurement is linearly related to the VIHs~\cite{Pignalberi2019},
\begin{equation}
g(h) = \frac{2h}{c},
\end{equation}
where $c$ is the light speed.
For an oblique incidence ionosonde with a simplified model of a flat ionosphere over a flat Earth,
the measurement function is~\cite{Zolesi2014},
\begin{equation}
g(h) = \frac{2 \sqrt{h^2 + (\bar{d}/2)^2} }{c},
\end{equation}
where $\bar{d}$ is the distance from the transmitter of the ionosonde to the receiver of the ionosonde.

We here emphasize that because of the wide-area surveillance of OTHR, the geographical limitation of the placement of ionosondes, and the construction and operation cost of ionosondes, only some sites of the ionosphere are measured by ionosondes.
The required VIHs may not be measured directly by the ionosondes.
We will discuss this later in Section~\ref{subsec:VIH_inference}.

\subsection{Model of target dynamics}
Assume that there are $L$ targets in the surveillance area of OTHR.
The state of target $l$ $(l = 1,\ldots, L)$  at scan $k$ is written as $x^{l}_{k} = [\rho^{l}_{k}, \dot{\rho}^{l}_{k}, b^{l}_{k}, \dot{b}^{l}_{k}]^T$, corresponding to ground range, ground range rate, bearing and bearing rate of the target.
The dynamics of each target is assumed to follow a discrete-time state equation written as
\begin{equation}
x^{l}_{k + 1} = f^{l}(x^{l}_{k}) + {\zeta}^{l}_{k},
\end{equation}
where $f^{l}$ is a known transition function and ${\zeta}^{l}_{k}$ is a zero-mean, white Gaussian sequence with covariance $B^{l}_{k}$.
The symbol $\bm{x}_{k} = \{ {x^{1}_{k}}, \ldots, x^{L}_{k} \}$ denotes the set of all target state at scan $k$.

\subsection{Model of OTHR measurement}
\begin{figure}[!htp]
	\centering
	\includegraphics[width=0.45\textwidth]{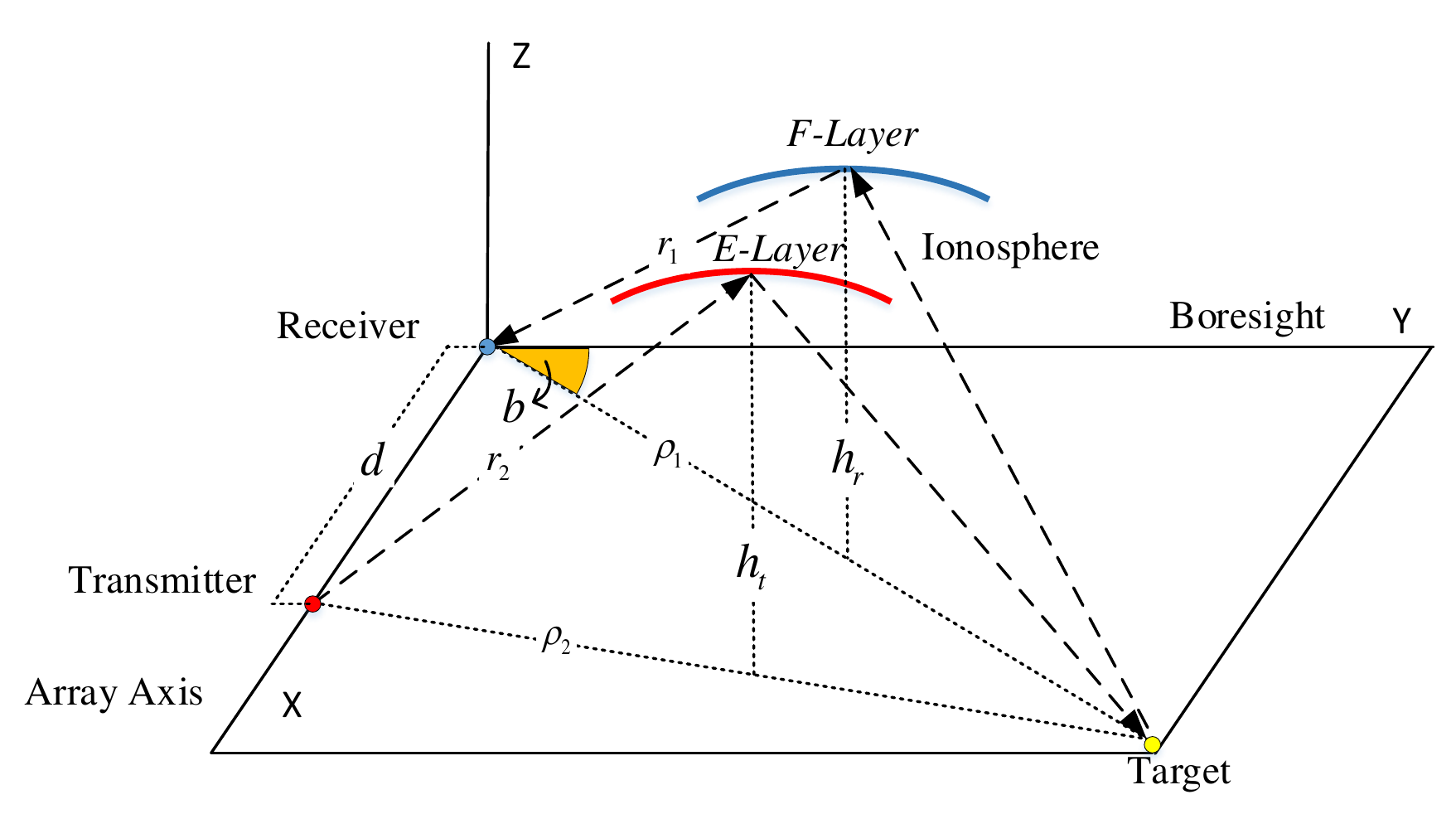}
	\caption{Illustration of OTHR propagation with EF mode~\cite{Lan2014}.}
	\label{fig_OTHR_propagation}
\end{figure}

With two ionospheric layers E and F, there are four one-hop propagation modes, EE, EF, FE and FF.
For each mode, we assume that the measurement of a target is obtained independently with known detection probability $p_d^{\gamma}$, $\gamma = 1,\dots, 4$.
For a bistatic OTHR, the multipath measurement model for OTHR is~\cite{Pulford1998, Pulford2004},
\begin{equation}
y_k =
\left\{
\begin{array}{ll}
{{u}}^1(x_k, h_k) + w^{1}_{ k} & \mbox{mode EE with   $p_d^{1}$} \\
{{u}}^2(x_k, h_k) + w^{2}_{ k} & \mbox{mode EF with   $p_d^{2}$} \\
{{u}}^3(x_k, h_k) + w^{3}_{ k} & \mbox{mode FE with   $p_d^{3}$} \\
{{u}}^4(x_k, h_k) + w^{4}_{ k} & \mbox{mode FF with   $p_d^{4}$} \\
\mbox{clutter} &\mbox{otherwise} \,,
\end{array}
\right.
\label{eq_transformation}
\end{equation}
where $y_k$ consists of slant range $r_g$, slant range rate $r_r$ and azimuth $a_z$.
The measurement noise $w^{\gamma}_k$ is zero-mean Gaussian noise with known covariance $R^{\gamma}_k$.
Fig.~\ref{fig_OTHR_propagation} illustrates OTHR propagation with EF mode.
The measurement function is specified as~\cite{Pulford1998}:
\begin{equation}
\begin{array}{l}
{r_{g}=r_{1}+r_{2}}, \\
{r_{r}=\frac{\dot{\rho}}{4}\left\{\frac{\rho}{r_{1}}+
	\frac
	{\rho-d \sin (b)	}
	{r_{2}}   \right\}}, \\
{a_{z}=\sin ^{-1}\left\{\rho \sin (b) /\left(2 r_{1}\right)\right\}},
\end{array}
\label{equ_OTHRMeasurementFunction}
\end{equation}
with
\begin{equation}
\begin{aligned}
r_{1} =r_{1}\left(\rho, h_{r}\right) &\triangleq \sqrt{(\rho / 2)^{2}+h_{r}^{2}}, \\
r_{2} =r_{2}\left(\rho, b, h_{t}\right) &\triangleq \sqrt{(\rho / 2)^{2}-d \rho \sin (b) / 2+(d / 2)^{2}+h_{t}^{2}},
\end{aligned}
\end{equation}
where $d$ is the distance from the transmitter of OTHR to the receiver of OTHR.
As shown in Table~\ref{tab:ropagation mode},
for a given propagation mode ${\gamma}$, $h_t$ and $h_r$ in Eq.~(\ref{equ_OTHRMeasurementFunction}) are replaced with the VIHs at the sites where OTHR beam reflects from the transmitter to the target and  the receiving beam reflects from the target to the receiver, respectively.

As~\cite{Pulford1998}, we assume that clutter distributes uniformly in the region of interest
and the number of clutter follows a Poisson distribution with density $\lambda$.
For a given number of false clutter measurements $N_c$, the Poisson model can be written as
\begin{equation}
\mbox{Pois}
(N_c) = \frac{(\lambda V)^{N_c}
\exp(-\lambda V)
}{N_c!}  \,,
\end{equation}
where  $V$ is the volume of the measurement space.

\subsection{Problem statement}\label{subsec:problemStatement}
\begin{figure}[!t]
	\centering
	\includegraphics[width=0.55\textwidth]{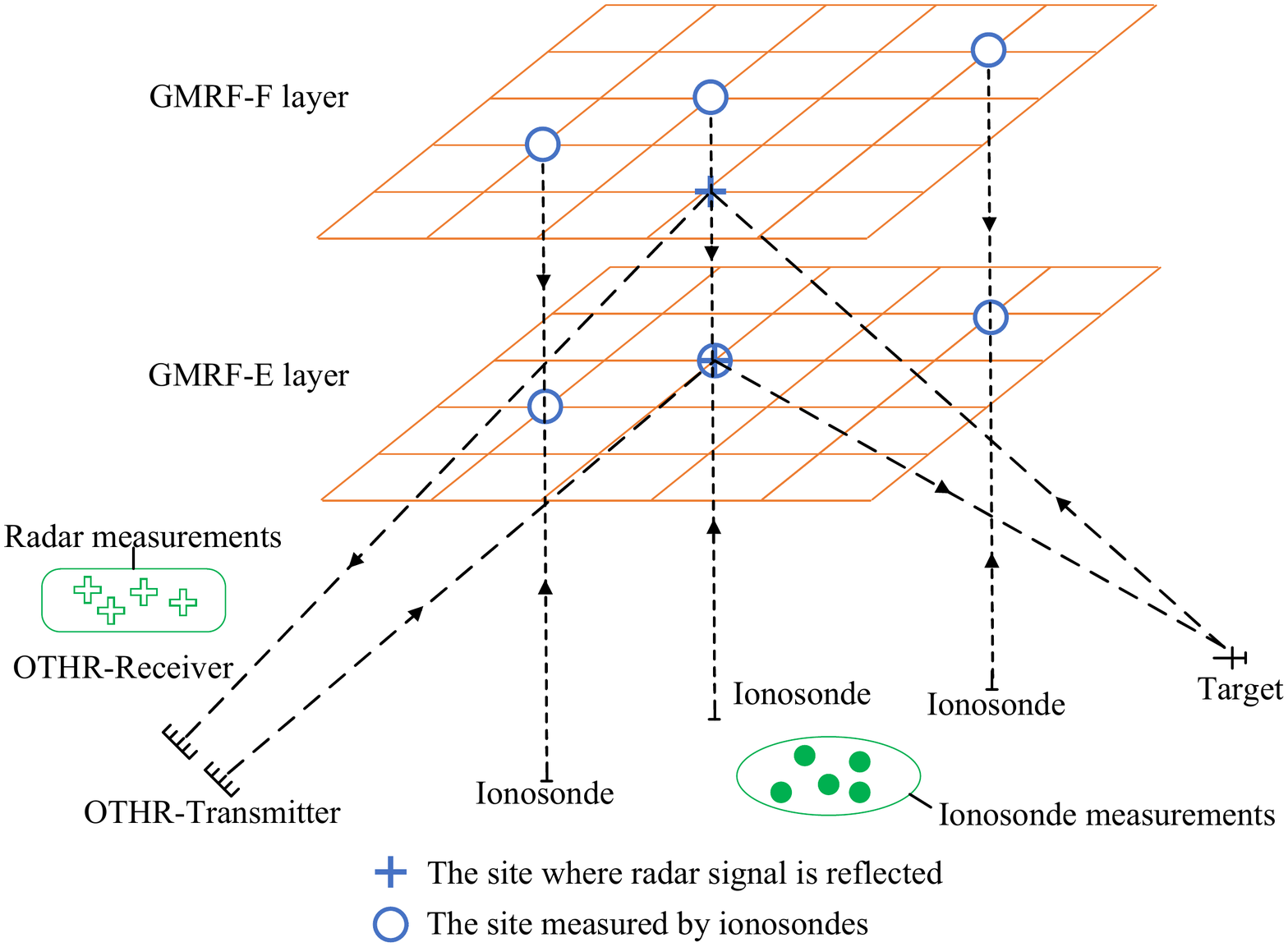}
	\caption{Illustration of the synthesis of our models on VIHs, ionosonde measurement, target and OTHR measurement~(EF propagation mode).}
	\label{fig_VIHDataStream}
\end{figure}

Taking EF propagation mode as an example, Fig.~\ref{fig_VIHDataStream} depicts the synthesis of our models on VIHs, ionosonde measurement, target and OTHR measurement.
Unlike the existing OTHR tracking algorithms mentioned in Section~\ref{sec:introduction},
we refine the description of VIHs by dividing each layer of the ionosphere into smaller subregions and modeling it by a GMRF.
As we mentioned in Section~\ref{sec:subIonoModels},
a very limited number of subregions are observed by ionosondes.
The subregions where the OTHR signal is reflected may not be observed by any ionosondes.
By modeling the VIHs of each layer as a GMRF, we are able to infer the VIHs of unobserved subregions and improve the estimation of the VIHs of
observed subregions by jointly using both OTHR measurements and ionosonde measurements,
leading to the reduction of the VIHs error and the improvement of target state estimation in return.

Let $\beta_k = [\bm{h}^{\mathrm{E}}(i_t), \bm{h}^{\mathrm{E}}(i_r), \bm{h}^{\mathrm{F}}(i_t), \bm{h}^{\mathrm{F}}(i_r) ]^T$
represent the used VIHs by a target at scan $k$,
where $\bm{h}^{\mathrm{E}}$ and  $\bm{h}^{\mathrm{F}}$  are the VIHs of E layer and F layer, respectively
.
The index $i_t$  denotes the subregion where the OTHR beam reflects from the transmitter to the target,
and $i_r$ denotes the subregion where the OTHR beam reflects from the target to the receiver.
As shown in Fig.~\ref{fig_OTHR_propagation}, $i_t$ and $i_r$ are determined by geometric transforms
among radar stations, target location and VIHs.
The indexing of the subregions corresponding to each propagation mode is shown in Table~\ref{tab:ropagation mode}.
\begin{table}[htbp]
	\caption{\upshape{Indexing propagation modes (E and F)}}
	\label{tab:ropagation mode}
	\centering
	\renewcommand\arraystretch{1.5}
	\begin{tabular}{|c|c|c|c|}
		\hline
		\textbf{Index} & \textbf{Mode} & $h_{t }$ & $h_{r }$ \\
		\hline
		$\gamma = 1$ & EE & $\bm{h}^{\mathrm{E}}(i_t)$ & $\bm{h}^{\mathrm{E}}(i_r)$ \\  \hline
		$\gamma = 2$ & EF & $\bm{h}^{\mathrm{E}}(i_t)$ & $\bm{h}^{\mathrm{F}}(i_r)$ \\  \hline
		$\gamma= 3$ & FE & $\bm{h}^{\mathrm{F}}(i_t)$ & $\bm{h}^{\mathrm{E}}(i_r)$ \\  \hline
		$\gamma= 4$ & FF & $\bm{h}^{\mathrm{F}}(i_t)$ & $\bm{h}^{\mathrm{F}}(i_r)$ \\ \hline
	\end{tabular}
\end{table}

Given the models of VIHs, ionosonde measurement, target dynamics, and radar measurement,
the aim of a tracker is to estimate target states sequences $\bm{x}_{k-\kappa:k}$ by exploiting the available data that includes OTHR measurements sequences  $Y_{k-\kappa:k}$, ionosonde measurements sequences $Z_{k-\kappa:k}$ at certain locations obtained by ionosondes.

Since the estimation of target states $\bm{x}_k$ and the estimation of the used VIHs $\bm{{{\beta}}}_k$ are coupled,
where $\bm{{{\beta}}}_k = \{ {{\beta}}_k^1, \ldots, {{\beta}}_k^{L}  \}$  denotes the set of the used VIHs of all targets,
we seek to estimate both of them jointly.
The joint probability distribution of the sequences of the target state $\bm{x}_{k-\kappa:k}$ and the used VIHs $\bm{{{\beta}}}_{k-\kappa:k}$ conditional on all the measurements is
\begin{equation}
 p \left(\bm{x}_{k-\kappa:k}, \bm{{{\beta}}}_{k-\kappa:k}  | Y_{k-\kappa:k}, Z_{k-\kappa:k}  \right).
\label{eq_joint_PDF}
\end{equation}
Accordingly, our OTHR target tracking problem can be formulated as the following joint \textit{maximum a posterior}~(MAP) estimation of the target states and the used VIHs.
\begin{problem}
Determine the sequences of the target states $\bm{x}_{k-\kappa:k}$ and the used VIHs $\bm{{{\beta}}}_{k-\kappa:k}$, i.e.,	
\begin{equation}
\{ \hat{\bm{x}}_{k-\kappa:k},~\hat{\bm{{{\beta}}}}_{k-\kappa:k}   \} ^ {\mathrm{MAP} } = \mathop { \arg	\max}_{ \bm{x}_{k-\kappa:k},~\bm{{{\beta}}}_{k-\kappa:k}}
 p \left(\bm{x}_{k-\kappa:k}, \bm{{{\beta}}}_{k-\kappa:k}  | Y_{k-\kappa:k}, Z_{k-\kappa:k}  \right)  .
\end{equation}
\label{problem_1}
\end{problem}

In Problem~\ref{problem_1}, there are two difficulties that prevent it from being solved directly.
The first one is the unknown data association~(called \textit{missing data}) $\Theta_{k-\kappa:k}$, that is, the correspondence
among a target,
a measurement and a propagation path is not known.
This means that the radar measurement is \textit{incomplete data}.
The second one is the above-mentioned coupling of the target state and the used VIHs.
EM, which works iteratively,
is an effective way to deal with the MAP estimation problem with missing data~\cite{Dempster1977}.
Each iteration of EM involves an expectation step~(E-step), which creates a function for the expectation of the log-likelihood evaluated using current estimation for the parameters, and a maximization step~(M-step), which computes parameters
by maximizing the function formulated in the E-step.
EM has provided performance improvement for OTHR target tracking~\cite{Pulford1997,Lan2014,Lan2018}
although the GMRF model of VIHs was not considered.
In this paper, we leverage EM framework as well and develop the OTHR multitarget tracking algorithm.
We next will describe the details of the proposed algorithm.

\section{OTHR multitarget tracking}\label{sec:OTHRMultipleTargetTracking}
As we mentioned in Section~\ref{subsec:problemStatement}, in the EM framework,
OTHR measurements are incomplete data due to the unknown data association.
Both OTHR measurements and ionosonde measurements contain the information on the used VIHs.
For a clear representation, below we rename each variables in the EM framework:
\begin{itemize}
\item \textit{Missing data}: data association $\Theta_{k-\kappa:k}$;
\item \textit{Incomplete data}: OTHR measurements and ionosonde measurements, i.e., $ \mathcal{Y} \stackrel{\vartriangle}{=} ( Y_{k-\kappa:k}, Z_{k-\kappa:k} )$ ;
\item \textit{Unknown parameters}: target states and the used VIHs, i.e.,
	$ \varPhi \stackrel{\vartriangle}{=}  (\bm{x}_{k-\kappa:k}, \bm{{{\beta}}}_{k-\kappa:k})$;
\item \textit{Complete data}: $\mathcal{X} \stackrel{\vartriangle}{=}   (Y_{k-\kappa:k}, Z_{k-\kappa:k},  \Theta_{k-\kappa:k} )$.
\end{itemize}

In the vein of~\cite{Dempster1977}, omitting time subscript for the sake of simplicity, we define the conditional expectation $\mathcal{Q}$-function associated with Problem \ref{problem_1} as,
\begin{equation}
\mathcal{Q} \left( \varPhi^{\prime}| \varPhi \right)
\stackrel{\vartriangle}{=}
\mathbb{E}\left( \log p \left( \mathcal{X} | \varPhi^{\prime}  \right) | \mathcal{Y}, \varPhi \right).
\label{eq_QfunctionDefinition}
\end{equation}
Denote the logarithm of the a prior density of $\varPhi$ as $ \mathcal{G}(\varPhi)$.
Then the EM iteration $\varPhi^{(r)} \rightarrow \varPhi^{(r + 1)} $ is defined as follows.
\begin{align}
&\mbox{\text{E-step}}: \mbox{Compute}~  \mathcal{Q}  \left(     \varPhi| \varPhi^{(r)} \right) \,.  \notag \\
&\mbox{\text{M-step}}: \mbox{Choose} ~ \varPhi^{(r+1)} \mbox{to maximizes} \notag \\
&~~~~~~~~~ \mathcal{Q}  \left( \varPhi| \varPhi^{(r)}  \right) + \mathcal{G}(\varPhi)  \,.
\label{eq_M_step}
\end{align}

One well-known appealing property of the EM algorithm is that the posterior density of $\varPhi$ increases monotonically, i.e.,
$p \left( \varPhi^{(r + 1)}  | \mathcal{Y}  \right)  \geq p \left( \varPhi^{(r)}  | \mathcal{Y}  \right) $
with equality holding at the stationary points (local minima, maxima and saddle points) of the posterior distribution~\cite{Wu1983}.
Based on Eqs.~(\ref{eq_QfunctionDefinition}),~(\ref{eq_M_step}),
we present the detailed process of the EM iteration for our OTHR target tracking as follows.
\begin{itemize}
	\item \textbf{Step 1:} Select the initial guess $ \{\bm{{x}}_{k-\kappa:k}^{(r = 1)}, \bm{{{\beta}}} ^ {(r = 1)} _ {k-\kappa:k} \}$;
	\item \textbf{Step 2:} Using the current guess $\{\bm{{x}}_{k-\kappa:k}^{(r)}, \bm{{{\beta}}} ^ {(r)} _ {k-\kappa:k}\}$ to
	calculate the posterior distribution of $\Theta_{k-\kappa:k}$;
	\item \textbf{Step 3:} Throw away the current guess $\{\bm{{x}}_{k-\kappa:k}^{(r)}, \bm{{{\beta}}} ^ {(r)} _ {k-\kappa:k}\}$ while keep the distribution of $\Theta_{k-\kappa:k}$ ;
	\item \textbf{Step 4:} Evaluate the conditional expectation $\mathcal{Q}(\varPhi | \varPhi^{(r)})$ with the conditional distribution of $\Theta_{k-\kappa:k}$;
	\item \textbf{Step 5:} Make a new guess $\{\bm{{x}}_{k-\kappa:k}^{(r + 1)}, \bm{{{\beta}}} ^ {(r +1)} _ {k-\kappa:k} \} $ that maximizes   $\mathcal{Q}(\varPhi | \varPhi^{(r)}) + \mathcal{G}(\varPhi)$;
	\item \textbf{Step 6:} Check for the convergence of either the parameters or the log likelihood. If the convergence criterion is not fulfilled, let $r = r + 1$ and go back to Step 2.
\end{itemize}

The goal of the M-step~(i.e., \textbf{Step 5}) is to maximize Eq.~(\ref{eq_M_step}) over parameters $\bm{x}_{k-\kappa:k}$ and $\bm{{{\beta}}}_{k-\kappa:k}$.
However, due to the high dimensional parameter space and the coupling of the target states and the used VIHs,
it is too complicated to maximize Eq.~(\ref{eq_M_step}) over parameters  $\bm{x}_{k-\kappa:k}$ and $\bm{{{\beta}}}_{k-\kappa:k}$ directly at the same time.
ECM~\cite{Meng1993}, which shares all the appealing convergence properties of EM, is adopted here to reduced the computational complexity.
Accordingly, the M-step defined in Eq.~(\ref{eq_M_step}) is replaced by the following two CM-steps:
\begin{align}
\mathcal{Q}  \left(  \bm{x}_{k-\kappa:k}^{(r + 1)}, \bm{{{\beta}}}^ {(r)} _ {k-\kappa:k}| \varPhi^{(r)}  \right) &    +
\mathcal{G} \left(      \bm{x}_{k-\kappa:k}^{(r + 1)}, \bm{{{\beta}}}^ {(r)} _ {k-\kappa:k}  \right)  \notag \\
& \geq \mathcal{Q} \left(
{\varPhi} ^{} | {\varPhi}^{(r)}    \right) + \mathcal{G}(\varPhi)
\label{eq_x_r_inECM}
\\
\mathcal{Q}    \left( {\varPhi} ^{(r+1)} | {\varPhi}^{(r)}    \right) +   \mathcal{G}  \left(\varPhi   ^{(r+1)}   \right)
& \geq   \notag  \\
\mathcal{Q}  \left(  \bm{x}_{k-\kappa:k}^{(r + 1)}, \bm{{{\beta}}} _ {k-\kappa:k}^{(r)}| \varPhi^{(r)}  \right) &    +
\mathcal{G} \left(      \bm{x}_{k-\kappa:k}^{(r + 1)}, \bm{{{\beta}}} _ {k-\kappa:k}^{(r)}  \right),
\label{eq_h_r_inECM}
\end{align}
where $r$ corresponds to the $r$-th iteration.

\begin{figure}[!htp]
	\centering
	\includegraphics[width=0.5\textwidth]{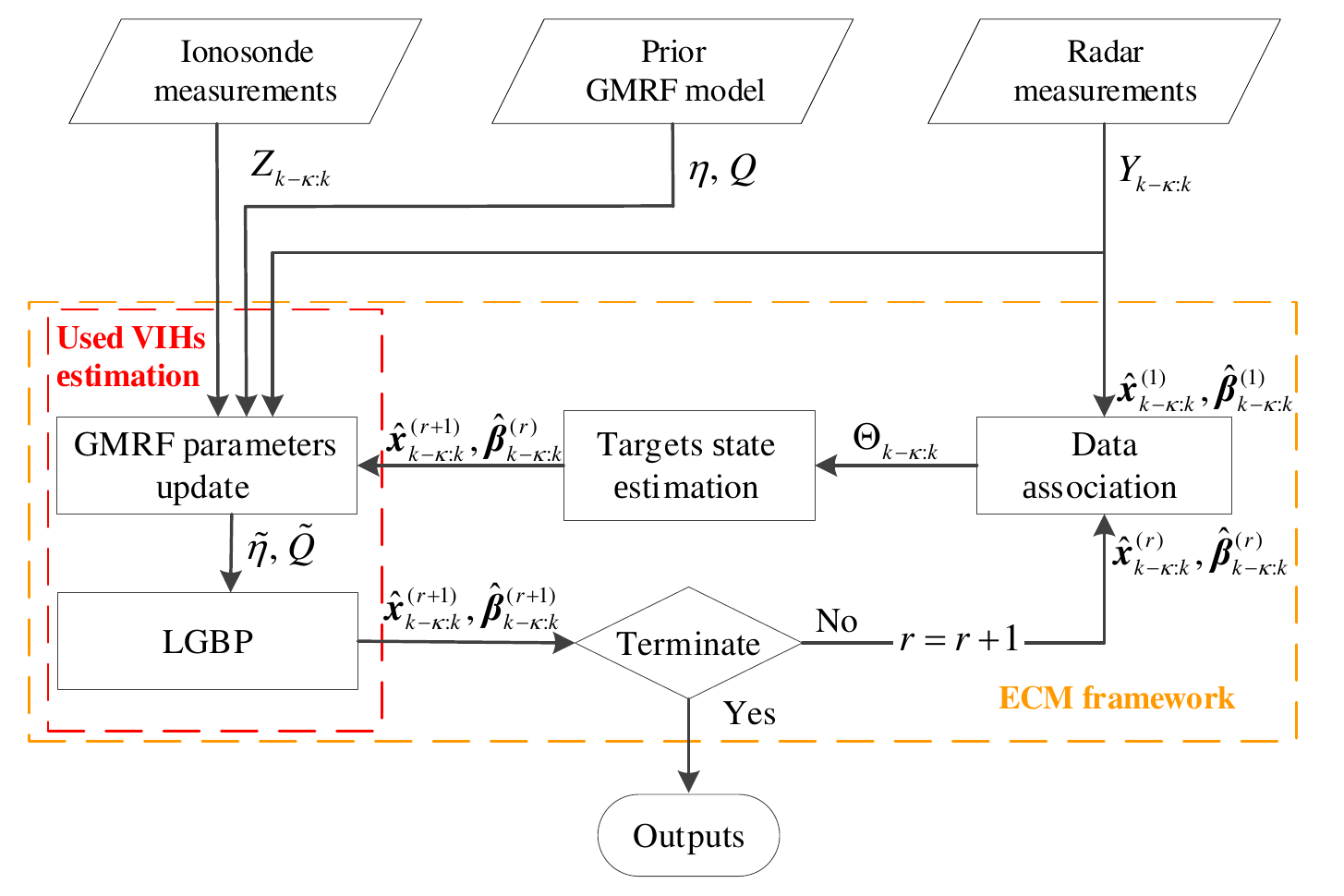}
	\caption{The diagram of ECM-GMRF.}
	\label{fig_ECM-CVIHAlgorithmFlow}
\end{figure}

To this end, the diagram of our proposed algorithm, ECM-GMRF, is shown in Fig.~\ref{fig_ECM-CVIHAlgorithmFlow}.
In Fig.~\ref{fig_ECM-CVIHAlgorithmFlow}, the modules in orange dotted box represent the entire ECM process.
The E-step, including the using of multitarget multidetection pattern and the calculation of the posterior distribution of each association event, is represented by \textit{data association} block.
The \textit{target state estimation} block corresponds to Eq.~(\ref{eq_x_r_inECM}), i.e., the first step of the CM-step.
The red dotted box represents the second step of the CM-step, i.e., Eq.~(\ref{eq_h_r_inECM}).

In the rest of this section, we will present the multitarget multidetection pattern and define data association events in Section~\ref{subsec_multiTargetMultiDetectionPattern}.
Next we will specify the derivation of E-step in Section~\ref{subsec:EStep}.
Then we will elaborate the above two CM-steps, i.e., target state estimation and inference of VIHs, in Section~\ref{subsec_targetsStateEstimation} and Section~\ref{subsec:VIH_inference}, respectively.

\subsection{Data association}\label{subsec_multiTargetMultiDetectionPattern}
In OTHR, the data association uncertainties include the unknown number of target-originated measurements for each target,
the unknown measurement source as well as the unknown measurement mode.
We here extend the multidetection pattern in~\cite{Habtemariam2013} to multitarget case.
The multitarget multidetection pattern is formulated with the following assumptions:
\begin{itemize}
\item
One measurement can only be associated with at most one target through a propagation mode.
\item For each target, at most one measurement can be generated through a propagation mode.
\end{itemize}

Assume that there are $N_{m,k}$
 OTHR measurements at scan $k$.
For the sake of simplicity, we omit the time subscript in the following description within this subsection.
We have following definitions.
\begin{definition}
For each target $l ~(l = 1,\dots, L)$,  define the number of target-originated measurements from target $l$ as $ \varphi^{l}~ (\varphi^{l} = 0,1,\dots, \varphi_{\text{max}} ^{l} )$:
\begin{itemize}
\item $ ``\varphi^{l} = 0" $: the target exists, but no OTHR measurement is from target $l$;
\item $ ``\varphi^{l} > 0" $: the target exists, and $\varphi^{l}$ out of
$N_{m,k}$ OTHR measurements are originated from target $l$,
\end{itemize}
where $\varphi_{\text{max}}^{l}$ equals the number of propagation modes that can be associated with OTHR measurements.
\end{definition}

To reduce the computational cost of the data association, the validation gate can be adopted for each propagation mode.
For a certain propagation mode ${\gamma}$, $\gamma=  1, \dots, 4$, the commonly used elliptical gate is~\cite{Pulford2004},
\begin{equation}\label{eq_validation_gate}
Y ^ {\gamma, l}  \stackrel{\vartriangle}{=} \left \{ y \in Y \mid ~
(y - \hat{y}^{\gamma, l})^T(S^{\gamma, l})^{-1} (y - \hat{y}^{\gamma, l})
\leq  V ^ {\gamma, l} \right \},
\end{equation}
where $\hat{y}^{\gamma, l}$ and $ S^{\gamma, l}$ represent the measurement prediction and innovation covariance of target $l$ through mode ${\gamma}$, respectively, and are given by Eq.~(\ref{eq_measure_predict}) and Eq.~(\ref{eq_mea_cov_predic}) later.
The scalar constants $V^{\gamma, l}$ are chosen to make the gate probability, namely, the probability that $y$ lies in the gate equal to $p_{g}^{\gamma, l}$~\cite{Pulford1998}.
With the validation gate, $\varphi_{max} ^{l} $ is calculated as
\begin{equation}
\varphi_{max} ^{l} =  \sum_{\gamma = 1}^{4} I  \{ |Y^ {\gamma, l}| \geq 1 \}.
\end{equation}
\begin{definition}[Association instance for each target]\label{def_association_instance_onetarget}
	For target $l$, $l = 1,\dots, L$, the possible association instance is
		\begin{equation}\label{eq_association_each_target}
		\begin{aligned}
	    &\Psi
		_{\varphi^{l}, n  } =
		\left \{
		\begin{array}{cl}
		\emptyset, & \text{if} \, \varphi^{l} = 0, \\
		\Big\{
		 \big (  {\epsilon_{1, l}},~ {\varrho_{1, l}} \big ) ,
		\ldots,
		\big (  {\epsilon_{\varphi^{l}, l}}, ~ {\varrho_{\varphi^{l}, l}} \big )
		\Big\},
		&
		\text{if} \, \varphi^{l} > 0 \,,
		\end{array}
		\right.
		\end{aligned}
		\end{equation}
	with the constraint that there is no repeated element in vector $\bm{\epsilon_{l}}$, where $\bm{\epsilon_{{l}}} = (\epsilon_{1, l},\dots, \epsilon_{\varphi^{l}, l} )$, and
	\begin{itemize}
		\item $\epsilon_{t, l}$~($t = 1, \dots, \varphi^{l}$) denotes the propagation mode that fulfills $|Y^ {\epsilon_{t, l}, l}| \geq 1$,
		\item $\varrho_{t, l}$~($t = 1, \dots, \varphi^{l}$) denotes the index of the measurement which is chosen from the measurement set $Y^{\epsilon_{t, l}, l}$,
		\item $\big (   {\epsilon_{t, l}}, ~{\varrho_{t, l}} \big )   $
		denotes a possible association among the measurement, the propagation mode and the target,
		\item $n$ denotes the index that represents the event under the chosen $\varphi^{l}$,
		\item ($\varphi^{l},n$) denotes a unique association instance for target $l$.
	\end{itemize}
\end{definition}

As we can see, the association instance for target $l$ is actually a set of possible association
$ \big (   {\epsilon_{t, l}}, ~{\varrho_{t, l}} \big )   $
~($t = 1, \dots, \varphi^{l}$).
Based on the definition of association instance for each single target, the association event for all targets is given as follows.
\begin{definition}[Association event for all targets]\label{def_association_event_alltarget}
	One of the feasible \textit{association event} for all targets is defined as
	\begin{equation}
	\theta_{\bm{\varphi},\bm{n}} = \bigcup_{l = 1}^{L}
	\Psi_{\varphi^{l},n} \,,
	\label{eq_association_event_for_all}
	\end{equation}
	with the constraint that there is no repeated element in vector $\bm{\varrho_{}}$,
	where the vector $\bm{\varrho} = ( \bm{\varrho_{1}},\dots, \bm{ \varrho_{{T}}} )$ and  $\bm{\varrho_{{l}}} = (\varrho_{1, l},\dots, \varrho_{\varphi^{l}, l} )$.
\end{definition}
The constraints in Definition~\ref{def_association_instance_onetarget} and Definition~\ref{def_association_event_alltarget} ensure that the assumptions at the beginning of this subsection are fulfilled.
Accordingly, the event space of data association
$\Theta_{k}$
 is $\{\theta_{\chi}\}_{\chi = 1}^{
 	{ N_{a,k}} }$,
where
$ N_{a,k}$
is the total number of association events for all targets at scan
{$k$}
.

\subsection{Derivation of E-step}\label{subsec:EStep}
Here we present the derivation of the E-step, i.e., the calculation of the $\mathcal{Q}$-function.
The complete data log-likelihood in Eq.~(\ref{eq_QfunctionDefinition}) is calculated as,
\begin{equation}
\label{eq:logXPhi}
\begin{aligned}
&\log p(\mathcal{X}|\varPhi^{\prime})\\
&= \log~ p(Y_{k-\kappa:k} ,Z_{k-\kappa:k},\Theta_{k- \kappa:k} |\bm{x}^{\prime}_{k-\kappa:k}, \bm{{{\beta}}}^{\prime}_{k-\kappa:k} ) \\
& = \log  p(Y_{k-\kappa:k} ,Z_{k-\kappa:k} |\Theta_{k- \kappa:k}, \bm{x}^{\prime}_{k-\kappa:k}, \bm{{{\beta}}}^{\prime}_{k-\kappa:k} ) \\
&~~~~ + \log  p(\Theta_{k- \kappa:k}| \bm{x}^{\prime}_{k-\kappa:k}, \bm{{{\beta}}}^{\prime}_{k-\kappa:k} ) \\
& = \sum_{\tau = k-\kappa}^k \log p(Y_\tau | \bm{x}^{\prime}_{\tau}, \bm{{{\beta}}}^{\prime}_{\tau} ,    {\Theta}_{\tau})
+  \sum_{\tau = k-\kappa}^{k} \log p( {\Theta}_{\tau}|\bm{x}^{\prime}_{\tau}, \bm{{{\beta}}}^{\prime}_{\tau}) \\
&~~~~   +  \sum_{\tau = k-\kappa}^k \log p(Z_\tau|\bm{{{\beta}}}^{\prime}_{\tau}).
\end{aligned}
\end{equation}
Substituting Eq.~(\ref{eq:logXPhi}) into Eq.~(\ref{eq_QfunctionDefinition}), the conditional expectation of the $\mathcal{Q}$-function can be expressed as,
\begin{equation}
\begin{aligned}
\mathcal{Q}  \left(     \varPhi^{\prime}| \varPhi    \right)
& = \sum_{\Theta_{k-\kappa:k}} \log p( \mathcal{X}|  \varPhi^{\prime}  )   p(\Theta_{k-\kappa:k}|\mathcal{Y}, \varPhi) \\
& =  \sum_{\tau = k-\kappa}^k   \sum_{\chi = 1} ^
{N_{a,\tau}}
  \omega_{\tau}(\chi)  \log p(Y_\tau   | \bm{x}^{\prime}_{\tau}, \bm{{{\beta}}}^{\prime}_{\tau} , {\theta}_{\chi}   ) \\
&~~~~  +    \sum_{\tau = k-\kappa}^{k}  \sum_{\chi = 1} ^
{N_{a,\tau}}
\omega_{\tau}(\chi)   \log p({\theta}_{\chi} |\bm{x}^{\prime}_{\tau}, \bm{{{\beta}}}^{\prime}_{\tau}) \\
& ~~~~   +  \sum_{\tau = k-\kappa}^k \log p(Z_\tau|\bm{
{{\beta}}
}^{\prime}_{\tau}),
\end{aligned}
\label{equ_QInProbabilityDensity}
\end{equation}
where $\omega_{\tau}(\chi)$ is the posterior probability of the association event ${\theta}_{\chi}$,
and it can be calculated by the Bayes rule
\begin{equation}
\begin{aligned}
\omega_{\tau}(\chi) & = p (\Theta_{\tau} = \theta_{\chi} |\mathcal{Y}, \varPhi ) \\
& =  p (\theta_{\chi} |\bm{{x}}_{\tau}^{}, \bm{{{\beta}}} _ {\tau}, Y_{\tau}) \\
& = \frac{\pi_{\tau}(\chi) p(Y_{\tau}|\bm{{x}}_{\tau}, \bm{{{\beta}}} _ {\tau}, \theta_{\chi} )  }
{ \sum_{i = 1}^
{N_{a,\tau}}
 \pi_{\tau}(i) p(Y_{\tau}|\bm{{x}}_{\tau}, \bm{{{\beta}}} _ {\tau}, \theta_{i}) }.
\end{aligned}
\label{equ_calculationOfOmega}
\end{equation}

In Eq.~(\ref{equ_calculationOfOmega}), $\pi_{\tau}(\chi)$ is the a prior probability of ${\theta}_{\chi}$, which is calculated as,
\begin{equation}
\begin{aligned}
\pi_{\tau}(\chi)
&=
p({\theta}_{\chi} |\bm{{x}}_{\tau}, \bm{{{\beta}}} _ {\tau}) \\
&=
\frac{1}{ \delta }   	\prod_{l}
\bigg( 	\prod_{t = 1}^{\varphi^{l}}
p_d^ {\epsilon_{t, l}}
p_g^ {\epsilon_{t, l}, l }
\mbox{Pois}
(
|Y_{\tau}^{\epsilon_{t, l}, l} |
-1)   \\
&~~~~~~~~~~~~~~~ \times
\prod_{\epsilon' \in \bm{\zeta}_{l}}
(1 - p_d^{\epsilon'}
p_g^{\epsilon', l}
)
\mbox{Pois}
(
|Y_{\tau}^{\epsilon', l} |
)
\bigg),
\end{aligned}
\end{equation}
where $\delta$ is a normalization constant, and $\bm{\zeta}_{l} =  \{1, \ldots, 4\} / \bm{\epsilon_{l}}$
denotes the modes which are not associated with any OTHR measurements in ${\theta}_{\chi}$.

By the OTHR measurement Eq.~(\ref{equ_ZMeasurementModel}) and the ionosonde measurement Eq.~(\ref{eq_transformation}),
the rest probability distributions in Eq.~(\ref{equ_QInProbabilityDensity}) are given as,
\begin{align}
p({\theta}_{\chi} |\bm{x}^{\prime}_{\tau}, \bm{{{\beta}}}^{\prime}_{\tau}) &=  \pi_{\tau}(\chi),\\
p(Z_\tau|\bm{{{\beta}}}^{\prime}_{\tau})  &=
\prod_{ h\in \bm{{{\beta}}}^{\prime}  }
\mathcal{N} \left(  g(h)  ,    A_{\tau}^s   \right), \\
p(Y_\tau   | \bm{x}^{\prime}_{\tau}, \bm{{{\beta}}}^{\prime}_{\tau} , {\theta}_{\chi}   )  &=
\prod_{ l:\varphi^{l}>0  }
\prod_{t = 1}^{\varphi^{l}}  \mathcal{N}  \left(
y_{\tau}(\varrho_{t, l})
;
{{u}}^{\epsilon_{t, l}}({x}_{\tau} ^ {\prime l}, {{{\beta}}}_{\tau}^{\prime l}), R_{\tau} ^ {\epsilon_{t, l}}  \right),
\label{equ_likelihoodOfRadarMeasurement}
\end{align}
where $y_{\tau}(\varrho_{t, l})$  denotes the $\varrho_{t, l}$th measurement in $Y_{\tau}$.

\subsection{Targets state estimation} \label{subsec_targetsStateEstimation}
In this section, our goal is to achieve the first step of CM, i.e., Eq.~(\ref{eq_x_r_inECM}).
The log of the a \textit{prior} density $\mathcal{G}(\varPhi)$ in Eq.~(\ref{eq_x_r_inECM}) is calculated as,
\begin{equation}
\begin{aligned}
&\mathcal{G}(\varPhi) \\
& =   \log p(\bm{x}_{k-\kappa:k}) + \log p(\bm{{{\beta}}}_{k-\kappa:k}) \\
& = \sum_{\tau=k-\kappa}^{k} \log p\left(\bm{x}_{\tau} | \bm{x}_{\tau-1}\right) + \log  p\left( \bm{x}_{k -\kappa -1}\right) + \sum_{\tau=k-\kappa}^{k} \log p\left(\bm{{{\beta}}}_{\tau} \right) \\
& = \sum_{\tau=k-\kappa}^{k} \sum_{l  = 1}^{L} \log \mathcal{N} \left(  x_{\tau} ^{l};   f^{l}( x_{\tau - 1} ^{l}), B_k^{l}           \right) \\
& ~~~~  + \sum_{\tau=k-\kappa}^{k} \sum_{{{\beta}}_i \in \bm{{{\beta}}_{\tau}} }^{  } \log \mathcal{N} \left( {{\beta}}_i; \mu_i, \Sigma_{ii}         \right) \,,
\end{aligned}
\label{equ_priorDensity}
\end{equation}
where  $\mu_i$ and $\Sigma_{ii}$ are the parameters of the marginal distribution of VIH $i$.
Separating the irrelevant terms with target state $\bm{x}_{\tau}$ in the right side of Eq.~(\ref{eq_x_r_inECM}) and using
Eqs.~(\ref{equ_QInProbabilityDensity})-(\ref{equ_priorDensity})
and the rule of sum of squared forms of Gaussians~\cite{Petersen2012}, the right side of Eq.~(\ref{eq_x_r_inECM}) can be rewritten as
\begin{equation}
\label{eq_Q_function_unfold}
\begin{aligned}
& \mathcal{Q}(\varPhi | \varPhi^{(r)}) + \mathcal{G}(\varPhi) \\
& =  \sum_{\tau = k-\kappa}^k    \sum_{l  = 1}^{L}     \log p( {x}^{l} _{\tau}|  {x}^{l} _{\tau -1})
+  \sum_{l  = 1}^{L}     \log p( {x}^{l} _{k - \kappa - 1})
\\
& ~~- \frac{1}{2}   \sum_{\tau = k-\kappa}^k
\sum_{l = 1} ^{L}  \sum_{\gamma = 1}^{4}
%
 \Big[  \tilde{y}_{\tau}^{\gamma,l} -{{u}}^{\gamma}(x_{\tau}^{l}, {{{\beta}}}_{\tau}^{l })  \Big]
 \big( \tilde{R}^{\gamma, l}_{\tau} \big) ^ {-1}
 \Big[ \tilde{y}_{\tau}^{\gamma,l} -{{u}}^{\gamma}(x_{\tau}^{l}, {{{\beta}}}_{\tau}^{l })  \Big] ^T
\\
& ~~+  \big[ \mbox{terms independent of}  ~{\bm{x}}_{k-\kappa:k} \big]
\end{aligned}
\end{equation}
with
\begin{equation} \label{eq_eequivalent_othr_measurement}
\tilde{y}_{\tau}^{\gamma,l} =  \frac{\sum\nolimits_{\chi \in
{{M}}
_{\tau}^{\gamma,l}   }  \omega^{(r)}_{\tau}(\chi) y^{\gamma,l} (\chi) }{\sum\nolimits_{\chi \in {{M}}_{\tau}^{\gamma,l}  } \omega^{(r)}_{\tau}(\chi)  },
\tilde{R}^{\gamma, l}_{\tau} = \frac{ R^{\gamma}_{\tau} }{\sum\nolimits_{\chi \in
{{M}}
_{\tau}^{\gamma,l}   } \omega^{(r)}_{\tau}(\chi)  },
\end{equation}
where $
{{M}}
_{\tau}^{\gamma,l}$ is a subset of $ \{1, \dots, N_{a, \tau} \}$ and $\chi \in
{{M}}
_{\tau}^{\gamma,l}$
indicates that  there is an OTHR measurement $y^{\gamma,l} (\chi)$  which  associates with target $l$ via propagation mode ${\gamma}$ in the
$\chi$th association event.

Henceforth, the realization of Eq.~(\ref{eq_x_r_inECM}) is a matter of maximizing Eq.~(\ref{eq_Q_function_unfold}) regarding to $\bm{x}_{k-\kappa:k}$ with the assumption that $\bm{{{\beta}}}_{k-\kappa:k}$ is known.
This procedure can be accomplished by applying an appropriate smoother for each target~\cite{Logothetis2002,Lan2018}, i.e.,
\begin{align}
\label{eq_x_hat}
\hat{{x}}^{l}_{\tau|k-\kappa:k} &= \mathbb{E}  \big[{x}^{l}_{\tau} | y_{k-\kappa:k}, {{{\beta}}}^{l}_{k-\kappa:k}    \big  ], \\
\hat{P}^{l}_{\tau|k-\kappa:k} &=  \mathbb{E} \big[({x}^{l}_{\tau} - \hat{{x}}^{l}_{\tau|k-\kappa:k}) (\bullet)^{T}  | y_{k-\kappa:k}, {{{\beta}}}^{l}_{k-\kappa:k} \big].
\label{eq_P_hat}
\end{align}
In Algorithm~\ref{alg_target_state_smoother}, we use the nonlinear smoother unscented Rauch-Tang-Strieble algorithm ~\cite{Sarkka2008, Lan2018} to estimate the target state $\hat{{x}}^{l}_{k-\kappa:k}$.
See the following Section \ref{sec:estimator} for the detailed calculation of Line 3 and Line 8 in Algorithm~\ref{alg_target_state_smoother}.
The calculation of Line 12 in Algorithm~\ref{alg_target_state_smoother} is elaborated in Section \ref{sec:smoother}.

\begin{algorithm}[t]
	\caption{Data association and target state estimation}
	\label{alg_target_state_smoother}
	\begin{algorithmic}[1]
		\REQUIRE $ \{Y_{k-\kappa:k}, R_{k-\kappa:k}\}$ and $ \{ \bm{\hat{x}}_{k - \kappa - 1} , \bm{\hat{{{\beta}}}}_{k-\kappa:k} \} $\\
		\ENSURE state estimation:  $\{\hat{\bm{x}}_{k-\kappa: k}, \hat{P}_{k-\kappa: k}\}$		
		\FOR {each time $\tau = k- \kappa: k$}
		\FOR {each target $l = 1, \dots, L$ }
		\STATE Calculate the state and measurement prediction by Eqs.~(\ref{eq_target_state_predict}-\ref{eq_mea_cov_predic}) ;
		\STATE Select measurement subset $Y_{\tau} ^{\gamma,l} $ by Eq.~(\ref{eq_validation_gate});
		\ENDFOR
		\STATE Generate all the association events through Eqs.~(\ref{eq_validation_gate}-\ref{eq_association_event_for_all});
		\STATE Calculate \textit{posterior} association probability $\omega_{\tau}(\chi)$ by Eq.~(\ref{equ_calculationOfOmega}) for $\chi = 1, \dots, N_{a, \tau}$;
		\STATE Calculate the state estimation by Eq.~(\ref{eq_filter_target});
		\ENDFOR
		\STATE Set smoothed estimation $\hat{x}^{l} _ {k|k-\kappa:k} = \hat{x}^{l} _ {k}$ and $\hat{P}^{l} _ {k|k-\kappa:k} = \hat{P}^{l} _ {k}$  for $l = 1, \dots, L$  ;
		\FOR {each time $\tau = k-1: k- \kappa$}
		\STATE Calculate the smoothed estimation by Eqs.~(\ref{eq_sigma_points}-\ref{eq_smoother_result}).
		\ENDFOR
	\end{algorithmic}
\end{algorithm}

\subsubsection{Estimator}\label{sec:estimator}
The state prediction for target $l$ is performed by
\begin{align}
\label{eq_target_state_predict}
\hat{x}_{\tau|\tau - 1} ^ {l} &= f^{l}(\hat{x}_{\tau - 1|\tau-1}^{l}), \\
\hat{P}_{\tau|\tau - 1} ^ {l-} &= J_{f} ^{l} \hat{P}_{\tau - 1|\tau - 1} ^ {l} (J_{f} ^{l} )^{T} + B^{l}_{\tau} ,
\end{align}
where $J_{f} ^{l} $ is the Jacobian matrix of transition function
{$f^{l}(\cdot)$}
with respect to $\hat{x}_{\tau|\tau-1}^{l}$.
Then the measurement prediction is
\begin{equation}
\hat{y}^{\gamma, l}_{\tau|\tau-1} = {{u}}^{\gamma}(\hat{x}_{\tau|\tau-1} ^ {l}, \hat{{{\beta}}}^{l}_{\tau} ),  l = 1, \dots, 4,
\label{eq_measure_predict}
\end{equation}
where $\hat{{{\beta}}}^{l}_{\tau}$ is the VIHs used by target $l$ at time $\tau$.
The measurement prediction covariance $S_{\tau}^{\gamma,l} $ for target $l$ through propagation mode ${\gamma}$ is
\begin{equation}
\begin{aligned}
S_{\tau} ^ {\gamma,l} = &  J_{u} ^{\gamma,l}   \hat{P}_{\tau|\tau-1} ^ {l} \big(J_{u} ^{\gamma,l}\big) ^{T}
+  {R}^{\gamma},
\label{eq_mea_cov_predic}
\end{aligned}
\end{equation}
where $J_{u} ^{\gamma,l}$ is the Jacobian matrix of measurement function
${{u}}^{\gamma}(\cdot)$
\emph{w.~r.~t.} $\hat{x}_{\tau|\tau-1}^{l}$.

Then, the state update for target $l$ can be expressed as follows according to the operation rules of block matrix \cite{Petersen2012},
\begin{equation}\label{eq_filter_target}
\begin{aligned}
\hat{x}^{l}_{\tau|\tau}    & = \hat{x}_{\tau|\tau - 1} ^{l}   +  K^{l}_{\tau|\tau}   \cdot \nu ^{l}_{\tau|\tau},  \\
\hat{P}^{l}_{\tau|\tau}    & = \mathbb{E} \Big\{ \big( {x}^{l}_{\tau} - \hat{x}^{l}_{\tau|\tau}    \big)   \big( \bullet \big) ^{T}  \Big\},
\end{aligned}
\end{equation}
where the corresponding innovation is
\begin{equation}
\nu ^{l}_{\tau|\tau}   =
\left[
\begin{aligned}
\tilde{y}^{1,l}_{\tau} &- \hat{y}^{1, l}_{\tau|\tau-1}  \\
& ~\vdots \\
\tilde{y}^{4,l}_{\tau} &- \hat{y}^{4, l}_{\tau|\tau-1}  \\
\end{aligned}
\right].
\end{equation}
The Kalman gain $K^{l}_{\tau|\tau} $ is given as
\begin{equation}
K^{l}_{\tau|\tau}   = \hat {P}_{\tau|\tau-1} ^{l}  \big(J_{uc}^{l} \big) ^{T} / S_{\tau}^{l},
\end{equation}
where
\begin{equation}
\begin{aligned}
S_{\tau}^{l}  & = J_{uc}^{l}\hat {P}_{\tau|\tau-1} ^{l}  \big(J_{uc}^{l}\big) ^{T} + R^{l}_c \,,\\
J_{uc}^{l}  & = \big[J_{u} ^{1, l}, \dots, J_{u} ^{4, l} \big]^{T} \,, \\
R^{l}_c & =
\left[
\begin{array}{cccc}
\tilde{R}^{1,l} & 0        & \dots  & 0 \\
0 		 & \tilde{R}^{2,l} & \dots  & 0 \\
\vdots   & \vdots   & \ddots & \vdots \\
0 		 & 0 		& \dots  & \tilde{R}^{4,l}
\end{array}
\right] \,.
\end{aligned}
\end{equation}

\subsubsection{Smoother}\label{sec:smoother}
At first, the sigma points $(i = 1,\dots, \sigma)$ are sampled.
\begin{equation}
\label{eq_sigma_points}
\begin{aligned}
&\phi_{\tau} ^{l}(0) = \hat{x}^{l}_{\tau}, \\
&\phi_{\tau} ^{l}(i) =  \hat{x}^{l}_{\tau} + \sqrt{\sigma + \varsigma} \left[ \sqrt{\hat{P}^{l}_{\tau}}    \right] _{i}, \\
&\phi_{\tau} ^{l}(i + \sigma) = \hat{x}^{l}_{\tau} - \sqrt{\sigma + \varsigma} \left[ \sqrt{\hat{P}^{l}_{\tau}}    \right] _{i},
\end{aligned}
\end{equation}
where $\varsigma$ is a scaling factor and $\sigma$ is the dimension of $\hat{x}^{l}_{\tau}$.

Then, the sigma points are propagated:
\begin{equation}
\phi^{l}_{\tau + 1}(i) = f^{l}(\phi^{l}_{\tau}(i)) , i = 1,\dots, \sigma.
\end{equation}

Next, the predicted state, the predicted covariance  and the cross-covariance  are computed as:
\begin{equation}
\begin{aligned}
&x_{\tau + 1} ^{ l-} = \sum_{i = 0}^{2\sigma} W_{0} \phi_{\tau + 1}^{l}(i),\\
&P_{\tau + 1} ^{ l-} = \sum_{i = 0}^{2\sigma} W_{i}
\Big [   \phi_{\tau + 1}^{l}(i) -  x_{\tau + 1} ^{ l-}  \Big ]
\Big [ \bullet    \Big ]  ^{T}
+ B_{\tau} ^{l},\\
&O_{\tau + 1} ^{l } = \sum_{i = 0}^{2\sigma} W_{i}  \left[ \phi_{\tau}^{l}(i) -  \hat{x}_{\tau} ^{ l} \right] \left[ \phi_{\tau +1}^{l}(i) -  x_{\tau + 1} ^{ l-} \right]^{T},
\end{aligned}
\end{equation}
where the weights $W_0 = \varsigma / (\varsigma + \sigma)$, $W_{i} = 1 /(2(\varsigma + \sigma)) $.

Finally, the smoothed estimation is calculated as:
\begin{equation}
\label{eq_smoother_result}
\begin{aligned}
D_{\tau} ^{l} &= O_{\tau + 1} ^{l }  \left[P_{\tau + 1} ^{ l-}\right]^{-1},\\
\hat{x}^{l} _ {\tau|k-\kappa:k} &= \hat{x}^{l}_{\tau} + D_{\tau} ^{l}  \left[ \hat{x}^{l} _ {\tau + 1|k-\kappa:k}
-x_{\tau + 1} ^{ l-}   \right], \\
\hat{P}^{l} _ {\tau|k-\kappa:k} &=\hat{P}^{l}_{\tau} + D_{\tau} ^{l}   \left[ \hat{P}^{l} _ {\tau + 1|k-\kappa:k} - P_{\tau + 1} ^{ l-}              \right]  \big( D_{\tau} ^{l}  \big) ^{T}.
\end{aligned}
\end{equation}

\subsection{Inference of VIHs } \label{subsec:VIH_inference}
In this section, we focus on the implementation of the second step of CM, i.e., Eq.~(\ref{eq_h_r_inECM}).
Separating the irrelevant terms with the used VIHs $\bm{{{\beta}}}_{\tau}$ in the right side of Eq.~(\ref{eq_h_r_inECM})
and using Eqs.~(\ref{equ_QInProbabilityDensity})-(\ref{equ_priorDensity})
and the rule of sum of squared forms of Gaussians~\cite{Petersen2012},
the right side of  Eq.~(\ref{eq_h_r_inECM}) can be rewritten as
\begin{equation}
\label{eq_Q_function_unfold_wrtVIH}
\begin{aligned}
&\mathcal{Q}  \left(  \bm{x}_{k-\kappa:k}^{(r + 1)}, \bm{{{\beta}}} _ {k-\kappa:k}^{(r)}| \varPhi^{(r)}  \right)  +
\mathcal{G} \left(      \bm{x}_{k-\kappa:k}^{(r + 1)}, \bm{{{\beta}}} _ {k-\kappa:k}^{(r)}  \right) \\
& = \sum_{\tau = k-\kappa}^k \log p(\tilde{Y}_\tau | \bm{{{\beta}}}^{}_{\tau} )
+  \sum_{\tau = k-\kappa}^k \log p(Z_\tau|\bm{{{\beta}}}^{}_{\tau}) \\
& ~~ + \sum_{\tau = k-\kappa}^k \log p(\bm{{{\beta}}}_{\tau})
+  \big[ \mbox{terms independent of}  ~{\bm{{{\beta}}}}_{k-\kappa:k} \big] \\
& = \log p(\bm{{{\beta}}}_{k-\kappa:k}|\tilde{Y}_{k-\kappa:k},  Z_{k-\kappa:k}) \\
& ~~~~~~~+  \big[ \mbox{terms independent of}  ~{\bm{{{\beta}}}}_{k-\kappa:k} \big],
\end{aligned}
\end{equation}
where  $\tilde{Y}_\tau  \stackrel{\vartriangle}{=}  \{\tilde{y}_{\tau}^{\gamma,l}  \}_{\gamma = 1,\ldots, 4, ~l = 1,\ldots, L } $ represents the equivalent OTHR measurement set and
\begin{equation} \label{eq_equivalent_mea_likelihood_VIH}
p(\tilde{Y}_\tau | \bm{{{\beta}}}^{}_{\tau} ) =
\prod_{l = 1} ^{L}  \prod_{\gamma = 1}^4 \mathcal{N} \big(\tilde{y}_{\tau}^{\gamma,l}; {{u}}^{\gamma}(x_{\tau}^{l}, {{{\beta}}}_{\tau}^{l }), \tilde{R}^{\gamma, l}_{\tau} \big).
\end{equation}

Therefore, by Eq.~(\ref{eq_Q_function_unfold_wrtVIH}) and given the target state, Eq.~(\ref{eq_h_r_inECM})
can be expressed as,
\begin{equation}\label{equ_maximizationRegardToHbar}
\hat{\bm{{{\beta}}}}_{k-\kappa:k}  ^ {\mathrm{MAP} } = \mathop { \arg	\max} _{\bm{{{\beta}}}_{k-\kappa:k}} p(\bm{{{\beta}}}_{k-\kappa:k}| \tilde{Y}_{k-\kappa:k},  Z_{k-\kappa:k}).
\end{equation}
To maximize Eq.~(\ref{equ_maximizationRegardToHbar}) \emph{w.~r.~t.} $\bm{{{\beta}}}^{}_{k-\kappa:k}$,
we need to infer the posterior marginal distribution of the each used VIH; this is a typical inference problem on probabilistic graphical models.
For Gaussian graphical models of moderate size,
exact inference can be solved by algorithms such as direct matrix inversion, Cholesky factorization, and nested dissection.
However, these algorithms cannot be used for large-scale problems due to the computational complexity~\cite{Liu2012}.
Exploiting the structure of the GMRF, the \textit{message passing} approach can significantly reduce the computational cost and
is adopted to infer the VIHs.

\begin{figure}[!htb]
	\centering
	\includegraphics[width=0.45\textwidth]{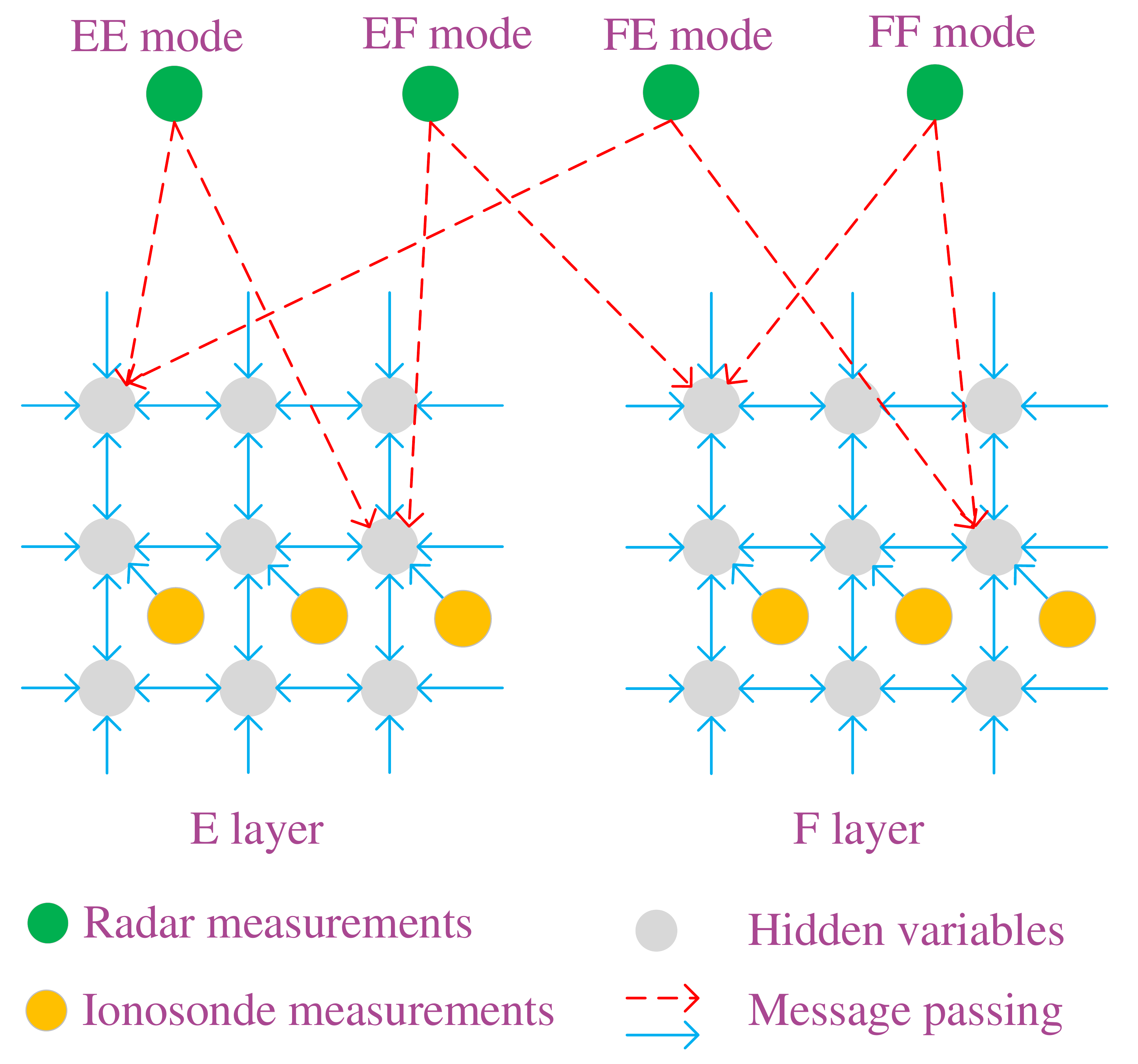}
	\caption{Illustration of the message passing flow for the inference of the VIHs.}
	\label{fig_messagepassingflow}
\end{figure}

Fig.~\ref{fig_messagepassingflow} shows the message passing flow for the used VIHs for one target.
The VIH of E layer and the VIH of F layer are linked through the OTHR measurements generated by propagation mode EF and FE.
We combine two GMRFs into one, i.e., $\mathcal{E} = \{ \mathcal{E}^{\mathrm{E}} \cup \mathcal{E}^{\mathrm{F}} \}$,  $\mathcal{V} = \{ \mathcal{V}^{\mathrm{E}} \cup \mathcal{V}^{\mathrm{F}} \}$. That is,
\begin{equation}
\bm{h} =
\left[
\begin{array}{c}
{\bm{h}}^\mathrm{E}  \\
{\bm{h}}^\mathrm{F}
\end{array}
\right] \,,
Q =
\left[
\begin{array}{cc}
{Q}^\mathrm{E} , & \bm{0}\\
\bm{0},& {{Q}}^\mathrm{F}
\end{array}
\right] \,,
\bm{\eta} =
\left[
\begin{array}{c}
{\bm{\eta}}^\mathrm{E}  \\
{\bm{\eta}}^\mathrm{F}
\end{array}
\right].
\end{equation}

Omitting time subscript for the sake of simplicity, given the equivalent OTHR measurements set $\tilde{Y}$ and ionosonde measurements set $Z$,
the joint posterior distribution of all VIHs $\bm{h}$ at one scan can be written in a factored form as
\begin{equation}
\begin{aligned}
p(\bm{h}| \tilde{Y}, Z)  \propto p(\bm{h}, \tilde{Y}, Z) = p(\bm{h})  p(\tilde{Y}| \bm{h}  ) p(Z|  \bm{h}  ),
\end{aligned}
\label{eq_factorationPosteriorDistribution}
\end{equation}
where $p(\bm{h})$ is modeled by the GMRF and
\begin{equation}
p(\bm{h}) \propto \exp (- \frac{1}{2} \bm{h}^{T}Q\bm{h} + \bm{\eta}^{T}\bm{h}    ).
\label{equ_GMRFDistribution}
\end{equation}

By the ionosonde measurement function Eq.~(\ref{equ_ZMeasurementModel}), the likelihood function $p(Z|\bm{h})$ can be written as
\begin{equation}
\begin{aligned}
p(Z|  \bm{h} )  \propto  \prod_{s = \mathrm{E},\mathrm{F} } \prod_{i \in \Omega_{z} } \exp \Big(  g(h_i)  (A^s)^{-1} z^{s}_i - \frac{1}{2}    g(h_i)  (A^s)^{-1} (g(h_i)    \Big),
\end{aligned}
\label{equ_likelihoodFunctionIonosonde}
\end{equation}
where $\Omega_{z} $ denotes the measured VIHs by ionosondes.
By Eq.~(\ref{eq_equivalent_mea_likelihood_VIH}),
\begin{equation}
\begin{aligned}
p(\tilde{Y}|  \bm{h} ) &=  p(\tilde{Y}|  \bm{{{\beta}}} ) \\
&\propto
\prod_{l = 1}^{L} \prod_{\gamma = 1}^{4} \exp
\Big(  \big( {{u}}^{\gamma}(x^{l}, {{\beta}}^{l})  \big)^{T} {(\tilde{R}^{\gamma, l})}^{-1} \tilde{y}^{\gamma,l}   \\
&~~~~ ~~~~  - \frac{1}{2}     \big( {{u}}^{\gamma}(x^{l}, {{\beta}}^{l})  \big)^{T} {(\tilde{R}^{\gamma, l})}^{-1}    {{u}}^{\gamma}(x^{l}, {{\beta}}^{l}) \Big).
\end{aligned}
\label{equ_likelihoodfunctionOTHR}
\end{equation}

Since the OTHR measurement function $ {{u}}^{\gamma}(x^{l}, {{\beta}}^{l}) $ is nonlinear and $g(h_i)$ is nonlinear for oblique incidence ionosondes,
we use the first order Taylor expansion for linear approximation,
\begin{align}
\label{equ_tarlorExpansionIonosondeMeasureFunc}
g(h_i) &= g(h\degree_i) + \frac{\partial g}{\partial h_i}  {(h\degree_i)}(h_i - h\degree_i)  \,,\\ \notag
{{u}}^{\gamma}(x^{l}, {{\beta}}^{l}) & = {{u}}^{\gamma}(\hat{x}^{l}, h\degree_t,h\degree_r) +
(h_t - h\degree_t)\frac{\partial {{u}}^{\gamma}}{\partial h_t}  (\hat{x}^{l},  h\degree_t,h\degree_r) \\
&~~~~ +  (h_r - h\degree_r) \frac{\partial {{u}}^{\gamma}}{\partial h_r}  (\hat{x}^{l},h\degree_t,h\degree_r),
{~~{\gamma}=1,\dots, 4 \,,}
\label{equ_tarlorExpansionOTHRMeasureFunc}
\end{align}
where $h\degree_i, h\degree_t ~\mbox{and}~ h\degree_r$ are chosen as the corresponding VIHs means in the GRMF model.
In order to facilitate the expression, we simplify the symbol as,
\begin{equation*}
\begin{aligned}
&U^{\gamma,l}
\stackrel{\vartriangle}{=}   {{u}}^{\gamma}(\hat{x}^{l}, h\degree_t,h\degree_r) \,,
U^{\gamma,l}_t \stackrel{\vartriangle}{=}  \frac{\partial {{u}}^{\gamma}}{\partial h_t}  (\hat{x}^{l},h\degree_t,h\degree_r)  \,,~ \\
&U^{\gamma,l}_r \stackrel{\vartriangle}{=}   \frac{\partial {{u}}^{\gamma}}{\partial h_r}  (\hat{x}^{l},h\degree_t,h\degree_r) \,.
\end{aligned}
\end{equation*}	

Then the likelihood function Eq.~(\ref{equ_likelihoodFunctionIonosonde}) can be rewritten as,
\begin{equation}
\begin{aligned}
&p(Z|  \bm{h} )   \propto    \prod_{s = \mathrm{E},\mathrm{F} } \prod_{i \in \Omega_{z} } \exp
\Big( - \frac{1}{2}  {h_i}^2   \Delta_{Q z}   + {h_i}  \Delta_{\eta z}   \Big),
\end{aligned}
\label{equ_ZGaussianLinear}
\end{equation}
where
\begin{align}
\Delta_{Q z} &=  \Big[ \frac{\partial g}{\partial h_i}  {(h\degree_i)}  \Big]^2 / (A^s), \notag  \\
\Delta_{\eta z} &= \Big[ \frac{\partial g}{\partial h_i}  {(h\degree_i)} \Big] \Big[ \frac{\partial g}{\partial h_i}  {(h\degree_i)} h\degree_i -  g(h\degree_i) + z^s_i \Big] / (A^s)   \,. \notag
\end{align}
The likelihood function Eq.~(\ref{equ_likelihoodfunctionOTHR}) can be rewritten as,
\begin{equation}
\begin{aligned}
p(\tilde{Y}| \bm{h})  \propto
\prod_{l = 1}^{L} \prod_{\gamma = 1}^{4} \exp
\Big( &
-\frac{1}{2} h_t^2 \Delta_{Q t} + h_t \Delta_{\eta t}  \\
& -\frac{1}{2} h_r^2 \Delta_{Q r} + h_r \Delta_{\eta r}
- h_t h_r \Delta_{Qtr}
\Big),
\end{aligned}
\label{equ_YConditionhLinearExpansion}
\end{equation}
where
\begin{equation*}
\begin{aligned}
	\Delta_{Q t} &= (U^{\gamma,l}_t)^T {(\tilde{R}^{\gamma, l})}^{-1} U^{\gamma,l}_t  \,, \\
	\Delta_{Q r} &=  (U^{\gamma,l}_r)^T {(\tilde{R}^{\gamma, l})}^{-1} U^{\gamma,l}_r \,,  \\
	\Delta_{Qtr} &= (U^{\gamma,l}_t)^T {(\tilde{R}^{\gamma, l})}^{-1} U^{\gamma,l}_r \,,  \\
	\Delta_{\eta t} &=  (U^{\gamma,l}_t)^T {(\tilde{R}^{\gamma, l})}^{-1}
	\Big(
	h\degree_t U^{\gamma,l}_t + h\degree_r   U^{\gamma,l}_r  +     \tilde{y}^{\gamma,l} -    U^{\gamma,l}
	\Big)   \,,  \\
	\Delta_{\eta r} &=
	(U^{\gamma,l}_r)^T {(\tilde{R}^{\gamma, l})}^{-1}
	\Big(
	h\degree_r  U^{\gamma,l}_r + h\degree_t   U^{\gamma,l}_t  +     \tilde{y}^{\gamma,l} -    U^{\gamma,l}
	\Big)\,.
\end{aligned}
\end{equation*}

From~Eqs.~(\ref{equ_GMRFDistribution})-(\ref{equ_YConditionhLinearExpansion}), it is seen that the posterior distribution of $\bm{h}$ is approximately Gaussian and Eq.~(\ref{eq_factorationPosteriorDistribution})  can be written as,
\begin{equation}
p(\bm{h}|\tilde{Y},Z) \propto \exp( - \frac{1}{2} \bm{h}^{T} \tilde{Q} \bm{h} +  \tilde{\bm{\eta}}^{T}\bm{h}   ) \,,
\label{equ_postierDistributionGaussian}
\end{equation}
which means that OTHR measurements and ionosonde measurements will only update the potential vector ${\eta}$ and the information matrix ${Q}$. That is, for the subregion $i$ which is involved in any measurement procedure,
\begin{align}
\tilde{Q}_{ii} &= {Q}_{ii} +
\sum_{l = 1} ^{L}  \sum_{\gamma=1}^{4}
\Big[
I  \big( {{u}}^{{\gamma},l, t}(h_i)  \big) \Delta_{Q t}   \notag  \\
&~ ~~~~~~~~~~~ + I  \big( {{u}}^{{\gamma},l, r}(h_i)  \big) \Delta_{Q r}
\Big]  +   I \big( g(h_i) \big) \Delta_{Q z}   \,,  \\
\tilde{\eta}_{i} &= {\eta}_{i} +
\sum_{l = 1} ^{L}  \sum_{\gamma=1}^{4}
\Big[
I  \big( {{u}}^{{\gamma},l, t}(h_i)  \big) \Delta_{\eta t}   \notag \\
&~ ~~~~~~~~~~~ + I  \big( {{u}}^{{\gamma},l, r}(h_i)  \big) \Delta_{\eta r}
\Big]  + I \big( g(h_i) \big) \Delta_{\eta z} \,,
\end{align}
and for the subregion pair~$(i,j)$ which is involved in any OTHR measurement procedure,
\begin{align}
\tilde{Q}_{ij} & =\tilde{Q} _{ji} = {Q}_{ij}
+  \sum_{l = 1} ^{L}  \sum_{\gamma=1}^{4}
\Big[
I  \big( {{u}}^{{\gamma},l, t}(h_i)  \big)
\times I  \big( {{u}}^{{\gamma},l, r}(h_j)  \big) \Delta_{Q tr}
\Big]  \,,
\end{align}
where $I  \big( {{u}}^{{\gamma},l,t}(h_i)  \big)$ equals one if
subregion $i$ is the subregion where OTHR beam reflects from the transmitter to the target
$l$ through mode ${\gamma}$, $I  \big( {{u}}^{{\gamma},l,r}(h_i)  \big) $
equals one if subregion $i$ is the subregion where the receiving beam reflects from the target
$l$ through mode ${\gamma}$ to the receiver, and $I \big( g(h_i) \big)$ equals one if $h_i$ is measured by a certain ionosonde.

For the graph with loops as shown in Fig.~\ref{fig_messagepassingflow}, loopy Gaussian belief propagation~\cite{Frey1998,Papachristoudis2015}~(LGBP)
is used to achieve approximate inference in GMRF.
The \textit{flooding schedule} is adopted here to organize the message passing schedule,
which simultaneously passes a message across every edge in both directions at each iteration.

The posterior distribution~(\ref{equ_postierDistributionGaussian}) can be factored into pairwise GMRF form as~\cite{Malioutov2006},
\begin{equation}
\begin{aligned}
p(\bm{h}| \tilde{Y}, Z) \propto \prod_{i, j \in \mathcal{E}} \psi_{i j}\left(h_{i}, h_{j}\right) \prod_{i \in \mathcal{V}} \psi_{i}\left(h_{i}, \tilde{Y}, Z\right)
\end{aligned}
\label{eq_factorationPosteriorDistributionpairwiseGMRF}
\end{equation}
in terms of node and edge \textit{potential function},
\begin{equation}
\psi_{i}\left( h_{i},\tilde{Y},Z   \right)=\exp \left( -\frac{1}{2} \tilde{Q}_{ii} h_{i}^{2} + \tilde{ \bm{\eta} }_{i} h_{i} \right),
\end{equation}
and
\begin{equation}
\psi_{i j}\left(h_{i}, h_{j}  \right)  =  \exp \left(  - h_i \tilde{Q}_{ij} h_j    \right)  .
\end{equation}
At each iteration of the LGBP algorithm, for each node $i \in V$, the message $m_{ij}(h_j)$ sent to each neighboring node $j \in N_e(i)$ is:
\begin{equation}
\begin{aligned}
m_{ij}\left(h_{j}\right) \propto \int \psi_{i j}\left(h_{i}, h_{j}\right) \psi_{i}\left(h_{i}, \tilde{Y}, Z \right)
{\prod_{n \in N_e(i) \backslash j} m_{n i}}
\left(h_{i}\right) d h_{i}.
\end{aligned}
\end{equation}
At any iteration, each node can produce an approximation $q_{i}(h_{i})$ to the
 marginal distribution  $p(h_i| \tilde{Y}, Z  ) $   by combining incoming
messages with the local evidence potential,
\begin{equation}
q_{i}\left(h_{i}\right) \propto \psi_{i}\left(h_{i}, \tilde{Y}, Z\right) \prod_{j \in N_e(i)} m_{j i}\left(h_{i}\right).
\end{equation}
In the GMRF $G = (\mathcal{V}, \mathcal{E} )$, the set of messages $m_{ij}(h_j)$  can be represented by $\{
\Delta Q_{i\rightarrow j}\cup \Delta \eta_{i\rightarrow j} \}_{(i,j) \in \mathcal{E}}$~\cite{Papachristoudis2015}.
The messages are initialized as $\Delta Q_{i\rightarrow j}^{0} = 0$ and $\Delta \eta_{i\rightarrow j}^{0} = 0$ for all $(i,j) \in \mathcal{E}$.
The LGBP consists of two steps:
\begin{enumerate}
	\item Message passing:\\	
	For $(i,j) \in \mathcal{E}$:	
	\begin{align}
	\Delta Q_{i\rightarrow j}^{t} &= -\tilde{Q}_{ji}(\hat{Q}^{(t-1)}_{i\backslash j})^{-1}\tilde{Q}_{ij},\\
	\Delta \eta_{i\rightarrow j}^{t} &= -\tilde{Q}_{ji}(\hat{Q}^{(t-1)}_{i\backslash j})^{-1}\hat{\eta}^{t-1}_{i\backslash j},
	\end{align}
	where
	\begin{align}
		\hat{Q}^{(t-1)}_{i\backslash j} &= \tilde{Q}_{ii} +
			\sum\nolimits_{
				n\in   N_e(i)  \backslash j
			} \Delta Q^{(t-1)}_{n\rightarrow i}	,\\
		\hat{\eta}^{(t-1)}_{i\backslash j} &= \tilde{\eta}_{i} +
			\sum\nolimits_{n\in  N_e(i)  \backslash j} \Delta \eta^{(t-1)}_{n\rightarrow i}.
	\end{align}

	The messages are updated based on previous messages at each iteration $t$.
	The fixed-point messages are denoted as $\Delta Q^{\ast}_{i \rightarrow j}$ and  $\Delta \eta^{\ast}_{i \rightarrow j}$ if the messages converge.\\
	
	\item Computation of means and variances: \\	
	For the used VIHs  $h_i \in {{\beta}}^{l} $ and $l = 1,\ldots, L$ :	
	\begin{align}
	\label{eq_Q_infer}
	\hat{Q}_{i} &= \tilde{Q}_{ii} +
	\sum\nolimits_{j\in N_e(i)} \Delta Q^{\ast}_{j\rightarrow i}
	,\\
	\label{eq_h_infer}
	\hat{\eta}_{i} &= \tilde{\eta}_{i} +
	\sum\nolimits_{j\in N_e(i)}
	\Delta \eta^{\ast}_{j\rightarrow i}.
	\end{align}
	The variances and means are computed based on the fixed-point messages and can be obtained by $\hat{\Sigma}_{ii} = 	\hat{Q}_{i}^{-1} $ and $\hat{\mu}_i = \hat{Q}_{i}^{-1}\hat{\eta}_{i}$.
\end{enumerate}

The output of LGBP is $h_i \sim \mathcal{N} (\hat{\mu}_i, \hat{\Sigma}_{ii})$.
Note that LGBP has no convergence guarantees, but when convergence is reached the estimated means equals the true ones~\cite{Weiss2001}.

\section{Simulation and analysis}\label{sec:simulation}
In this section, numerical simulations are implemented to verify the performance of ECM-GMRF.
We compare ECM-GMRF with MD-JPDAF~\cite{Habtemariam2013}.
The statistical results of the two algorithms are based on 400 Monte Carlo runs.

\subsection{Scenario settings}
\begin{figure}[!htb]
\centering
\includegraphics[width=0.35\textwidth]{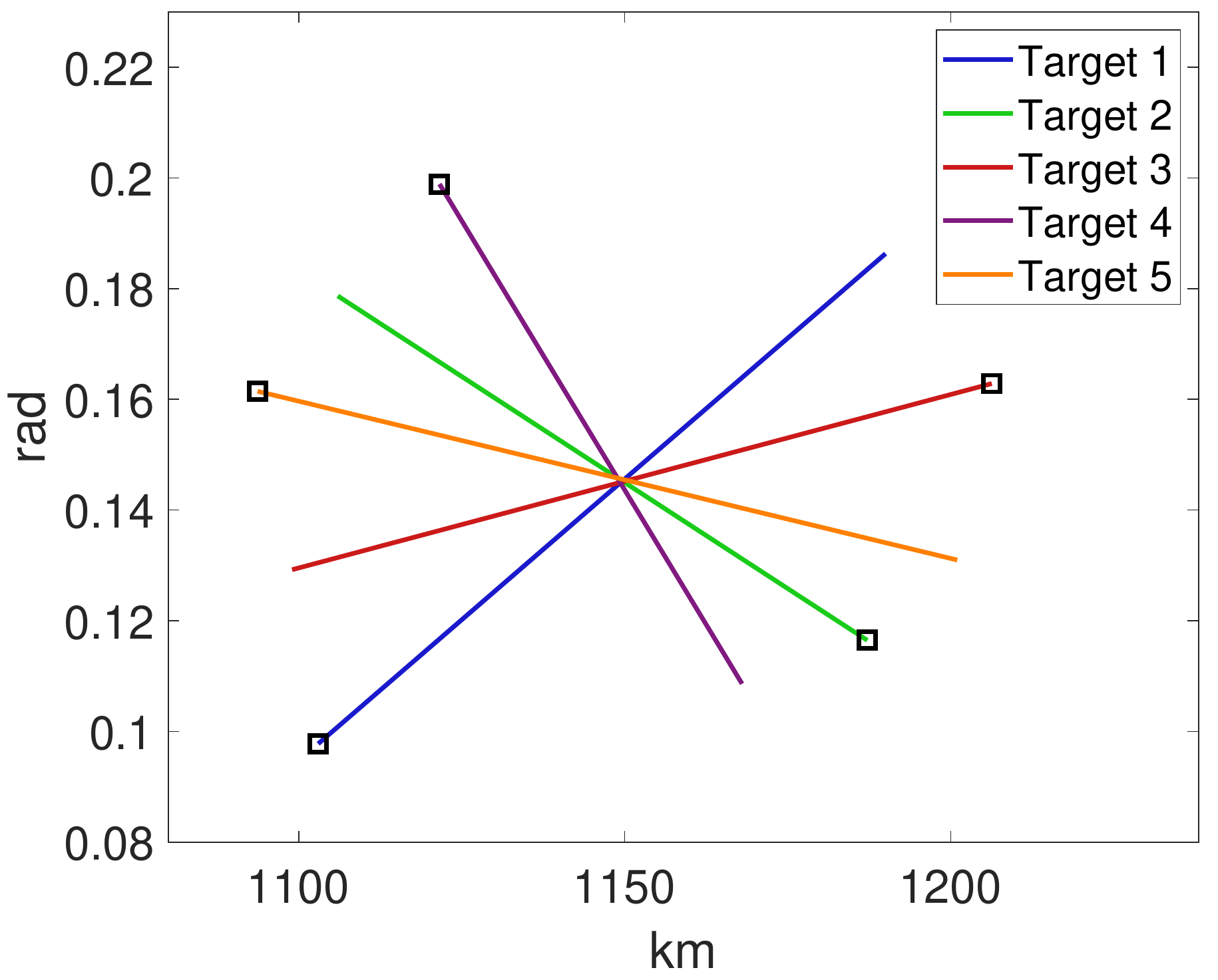}
\caption{True trajectories of all targets in ground coordinate system. $\Box$ represents the start point of a target.}
\label{fig:trueTrackAndusedVIHIndex}
\end{figure}

As shown in Fig.~\ref{fig_OTHR_propagation}, we use the same settings for OTHR as described in~\cite{Pulford1998}, such as surveillance region size, measurement noise level and sampling period.
We assume that there are five targets in the surveillance region of OTHR.
Fig.~\ref{fig:trueTrackAndusedVIHIndex} shows the true trajectories of the five targets in radar ground coordinate system.

\begin{figure}[!htb]
\centering
\includegraphics[width=0.45\textwidth]{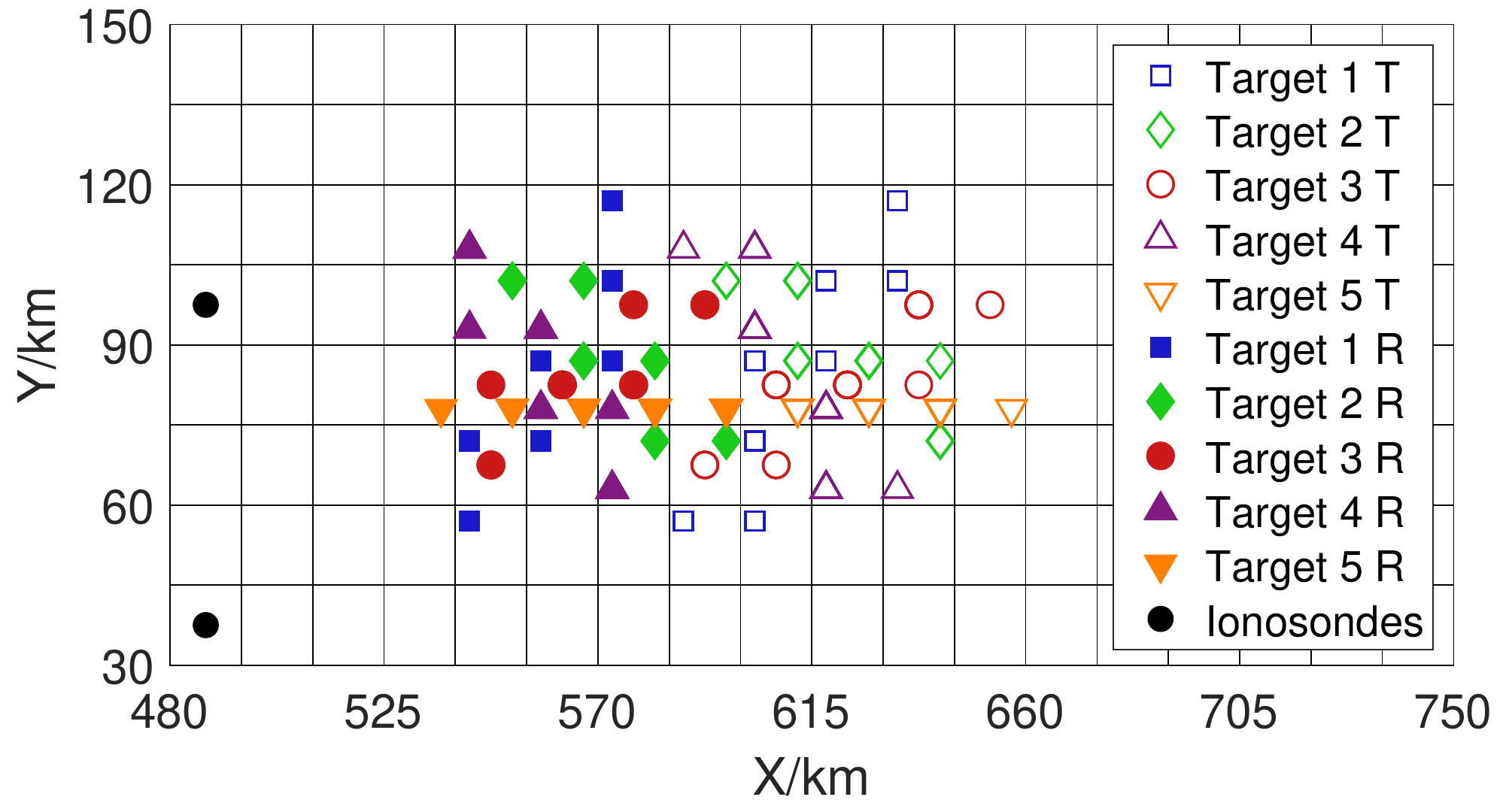}
\caption{Illustration of the partition of the ionosphere and the used VIHs of five targets during 30 scans. For each layer, the ionosphere is divided into 144 subregions. T means the OTHR beam reflects from the transmitter to the target and R means the receiving beam reflects from the target to the receiver.}
\label{fig_region_partition1}
\end{figure}

\begin{figure}[!htb]
\centering
\includegraphics[width=0.40\textwidth]{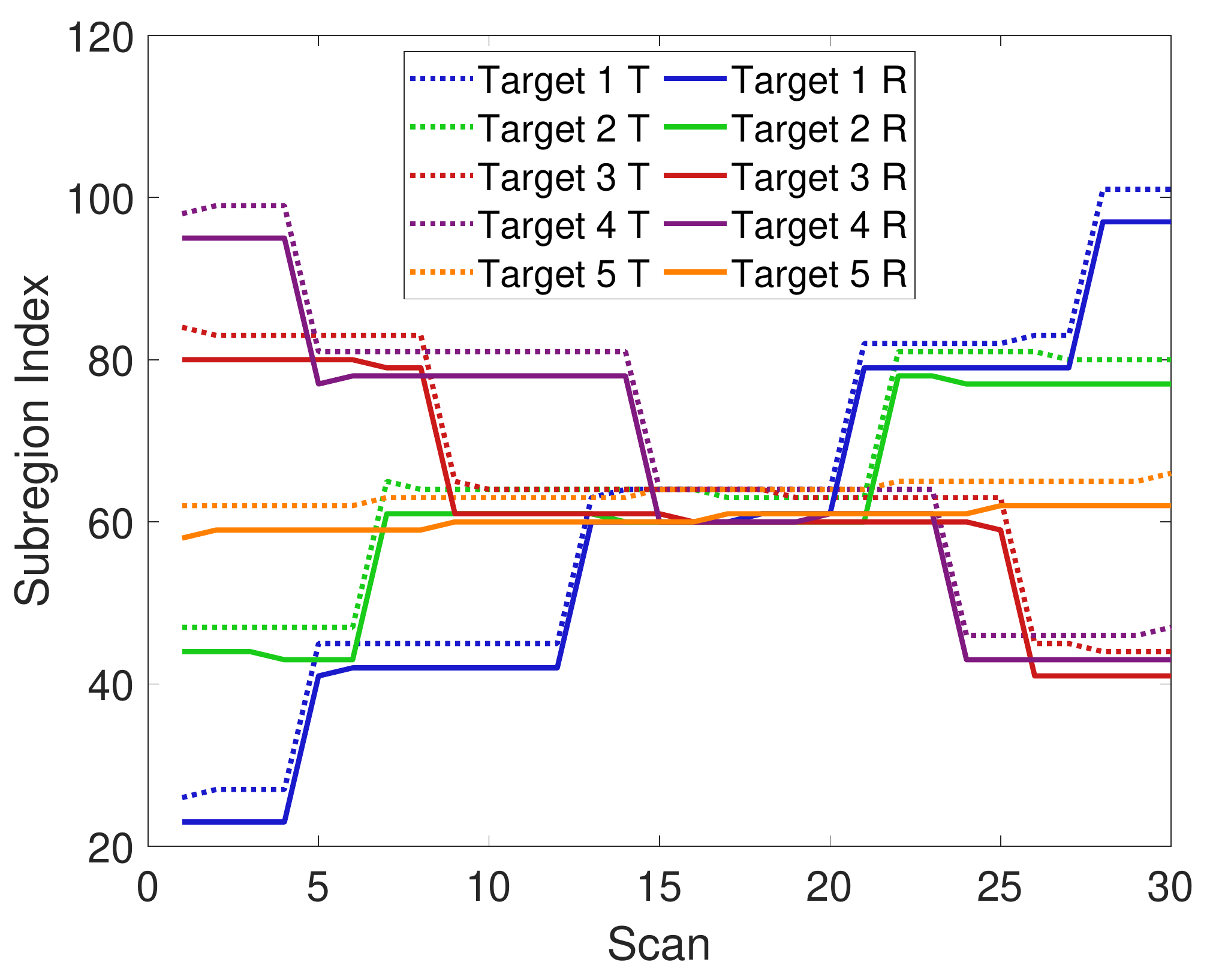}
\caption{Subregion indices (E layer and F layer) of the used VIHs of each target.}
\label{fig_region_partition2}
\end{figure}

To model the ionosphere by a GMRF, without loss of generality, for each layer of the ionosphere,
we divide the corresponding ionosphere region $\mathcal{A}$ into 144 subregions as illustrated in Fig.~\ref{fig_region_partition1}.
The subregions of the used VIHs are determined by the geometry of the CR shown in Fig.~\ref{fig_OTHR_propagation}.
For each target, the locations and the indices (E layer and F layer) of the subregions of the used VIHs are shown in Fig.~\ref{fig_region_partition1} and Fig.~\ref{fig_region_partition2}, respectively.
To directly measure VIHs, we assume that there are two vertical incidence ionospheric sounding ionosondes which are located underneath the subregion $1$ and the subregion $73$, respectively.
Table~\ref{tab:parameters} shows the parameters of the scenario and the initial states of all targets.

In the simulation, the true VIHs at each scan are sampled from two GMRF models with the known mean vector $\bm{\mu}^{\mathrm{E}}$, $\bm{\mu}^{\mathrm{F}}$  and the precision matrix $Q^{\mathrm{E}}$, $Q^{\mathrm{F}}$.
For simplicity, we assume that $\bm{\mu}^{\mathrm{E}} = 110~\text{km}$ and $\bm{\mu}^{\mathrm{F}} = 220~\text{km}$.
We use the approach described in \cite{Norberg2015,Norberg2018} to construct the precision matrix of the first-order approximation GMRF
in the two dimensional latitude-altitude space.
See \cite{Norberg2015,Norberg2018} for more details.
The values of the constructed precision matrix $Q^{\mathrm{E}}$, $Q^{\mathrm{F}}$ can be found in Table~\ref{tab:parameters}.
\begin{table}[!htb]
	\footnotesize
	\caption{\upshape{Parameter settings of the simulation scenario}}
	\label{tab:parameters}
	\centering
	\begin{tabular}{ccccc}
		\hline
		\hline
		\multicolumn{3}{c}{\textbf{Parameters}} & \multicolumn{2}{c}{ \textbf{Value}} \\ \hline
		\multicolumn{3}{l}{Number of scans} & \multicolumn{2}{c}{ 30} \\
		\multicolumn{3}{l}{	Detection probability} & \multicolumn{2}{c}{0.7} \\
		\multicolumn{3}{l}{ Expected number of clutter } & \multicolumn{2}{c}{ 50 per scan} \\
		\multicolumn{3}{l}{	Sampling period} &  \multicolumn{2}{c}{ 20 seconds}    \\
		\multicolumn{3}{l}{Surveillance region size (range) }& \multicolumn{2}{c}{ 1000-1400~km }    \\
		\multicolumn{3}{l}{	Surveillance region size (azimuth) }& \multicolumn{2}{c}{ 4-12~degree }    \\
		\multicolumn{3}{l}{	Ionosphere Region size ($\mathrm{X}$) }& \multicolumn{2}{c}{ 480-750~km}  \\
		\multicolumn{3}{l}{	Ionosphere Region size ($\mathrm{Y}$) }& \multicolumn{2}{c}{30-150~km }   \\
		\multicolumn{3}{l}{	Ionosphere subregion size ($\mathrm{X}\times \mathrm{Y}$) } & \multicolumn{2}{c}{ 15$\times $15~km} \\
		\multicolumn{3}{l}{	Number of subregions per layer} & \multicolumn{2}{c}{144} \\
		\multicolumn{3}{l}{	Standard deviation of the VIHs of E layer } & \multicolumn{2}{c}{11~km       }  \\
		\multicolumn{3}{l}{	Standard deviation of the VIHs of F layer   } & \multicolumn{2}{c}{13~km   } \\
		\hline
		\multirow{2}*{$Q^{\mathrm{E}}$} & \multicolumn{2}{l}{Diagonal element } & \multicolumn{2}{c}{0.082} \\
		& \multicolumn{2}{l}{Off diagonal element } & \multicolumn{2}{c}{-0.0205} \\
		\multirow{2}*{$Q^{\mathrm{F}}$} & \multicolumn{2}{l}{Diagonal element } & \multicolumn{2}{c}{0.0587} \\
		& \multicolumn{2}{l}{Off diagonal element } & \multicolumn{2}{c}{-0.0147} \\
		\hline
		\multicolumn{3}{l}{\textbf{Measurement noise}} & \multicolumn{2}{c}{ \textbf{Standard deviation}} \\
		\multicolumn{3}{l}{  Slant range of OTHR}        & \multicolumn{2}{c}{5~km} \\
		\multicolumn{3}{l}{  Slant range rate of OTHR}        & \multicolumn{2}{c}{  0.001~km/s } \\
		\multicolumn{3}{l}{  Azimuth of OTHR}        & \multicolumn{2}{c}{  0.003~rad } \\
		\multicolumn{3}{l}{Vertical incidence ionosondes } & \multicolumn{2}{c}{10~km} \\
		\hline
		\textbf{ Initial state}	  &\textbf{km}   &\textbf{ km/s} & \textbf{ rad}  & \textbf{ rad/s} 		\\ 
		Target 1	      &1100          &0.15           & 0.09472          & 1.52665$\times10^{-4}$ 	\\ 
		Target 2	      &1190          &-0.14           & 0.11432          & 1.07266 $\times10^{-4}$     \\ 	
		Target 3	      &1210          &-0.185           & 0.16401          & -5.79865$\times10^{-5}$     \\ 	
		Target 4	      &1120          &0.08               & 0.20201          & -1.55665 $\times10^{-4}$     \\ 	
		Target 5	      &1090          &0.185           & 0.16251          & -5.25665$\times10^{-5}$     \\ 	
		\hline
		\hline
	\end{tabular}
\end{table}

Next, we first run ECM-GMRF only to analyze its performance considering a single target~(e.g., Target 1) tracking in Section~\ref{subsection:singTargetResult}.
Then we explore the improvements of the used VIH estimation and the target state estimation when multiple targets are tracked, and
compare ECM-GMRF with MD-JPDAF in Section~\ref{subsec:multipleTargetResults}.

\subsection{Single target tracking results}\label{subsection:singTargetResult}
Firstly, we let $\kappa$ (slide window) of ECM-GMRF be one and
consider the following three cases with different information sources on VIHs.
\begin{itemize}
\item Case 1: The VIHs of layer E and layer F are fixed at $110$~km and $220$~km, respectively.
\item Case 2: Only the measurements from ionosondes are used to estimate the used VIHs.
\item Case 3: Both the measurements from ionosondes and the measurements from OTHR are used to estimate the used VIHs.
\end{itemize}

\begin{figure}[!htb]
	\centering
	\subfigure{
		\begin{minipage}[t]{0.499\linewidth}
			\centering
			\includegraphics[width=0.99\textwidth]{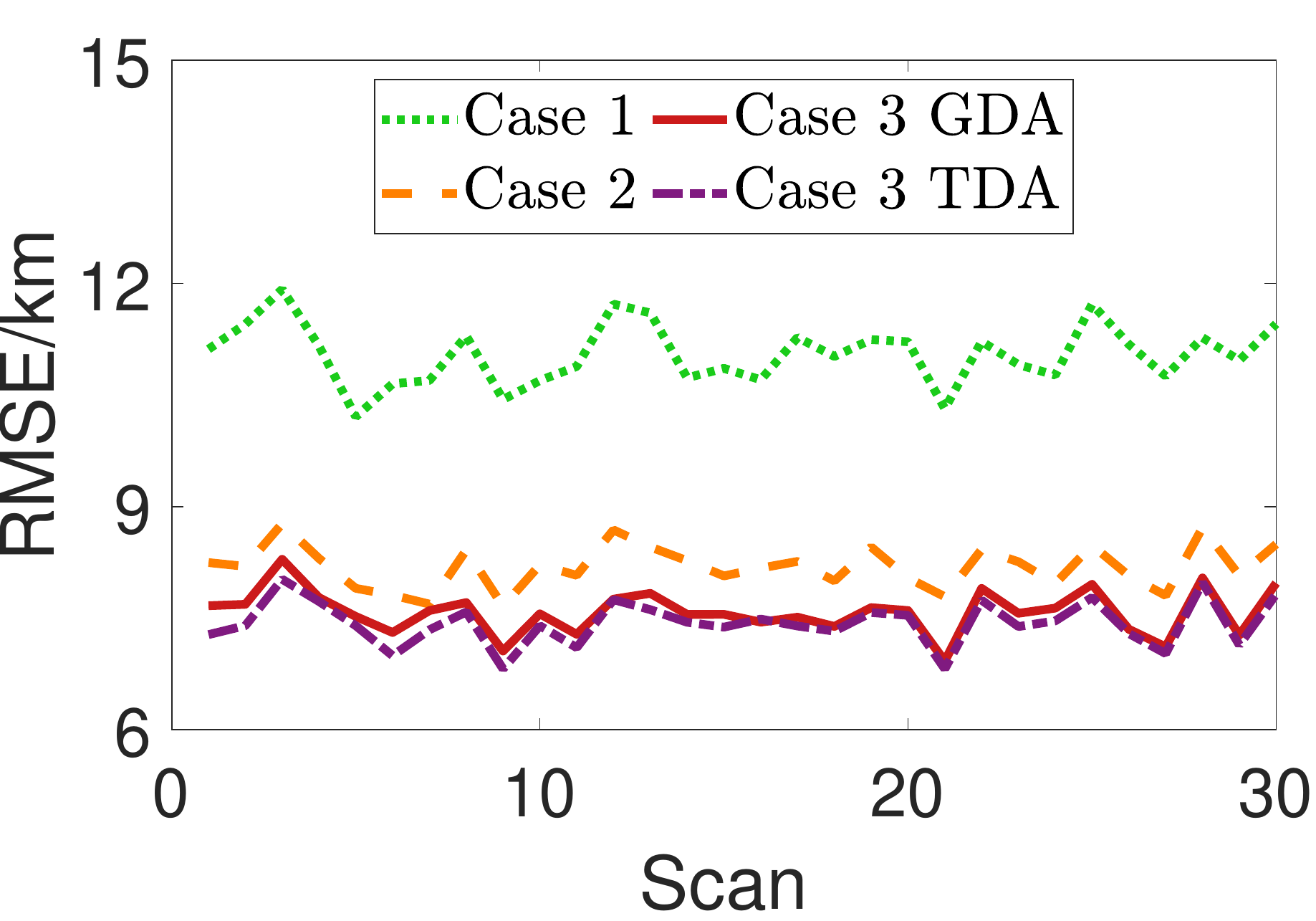}
			\includegraphics[width=0.99\textwidth]{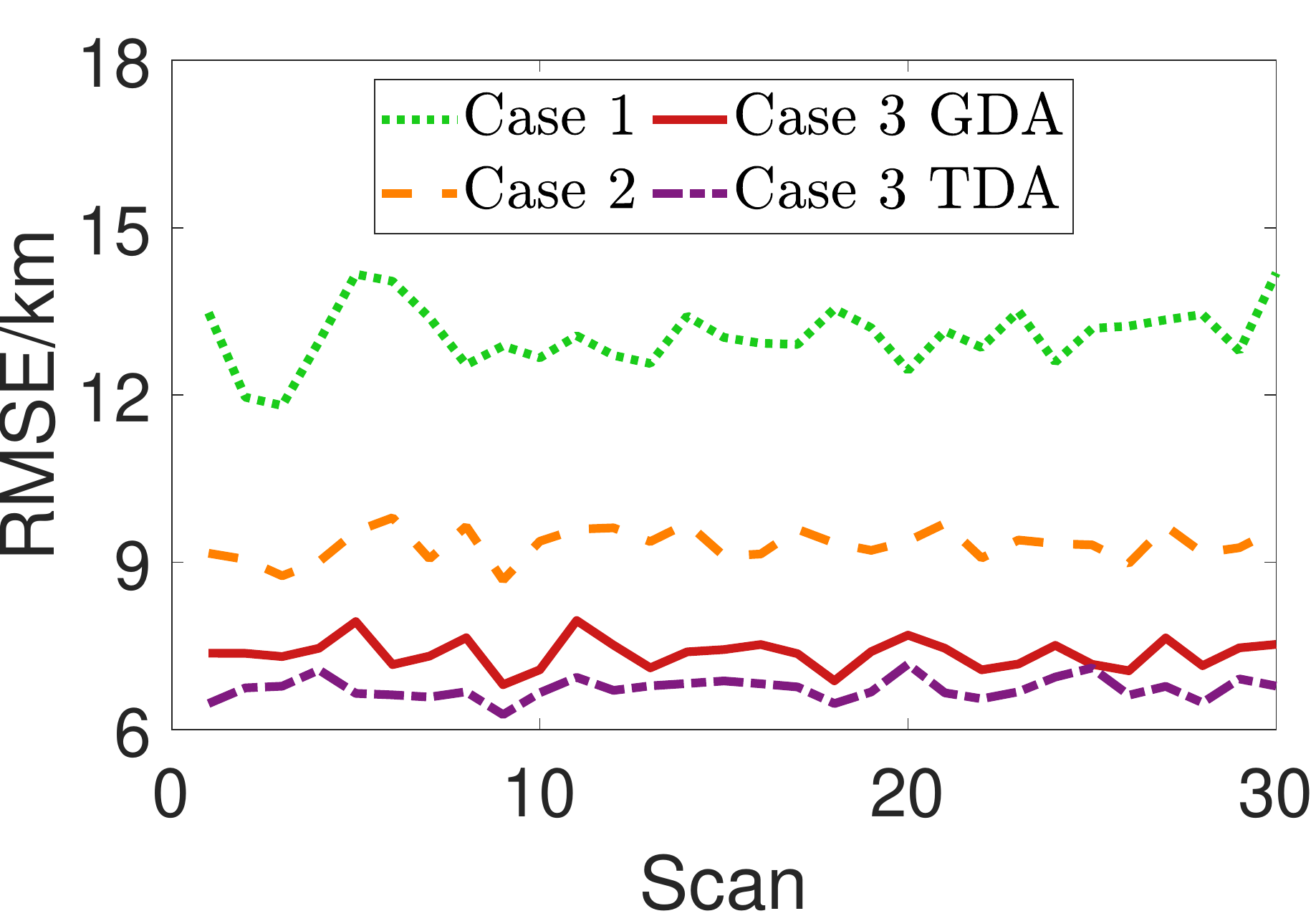}
		\end{minipage}%
	}%
	\subfigure{
		\begin{minipage}[t]{0.499\linewidth}
			\centering
			\includegraphics[width=0.99\textwidth]{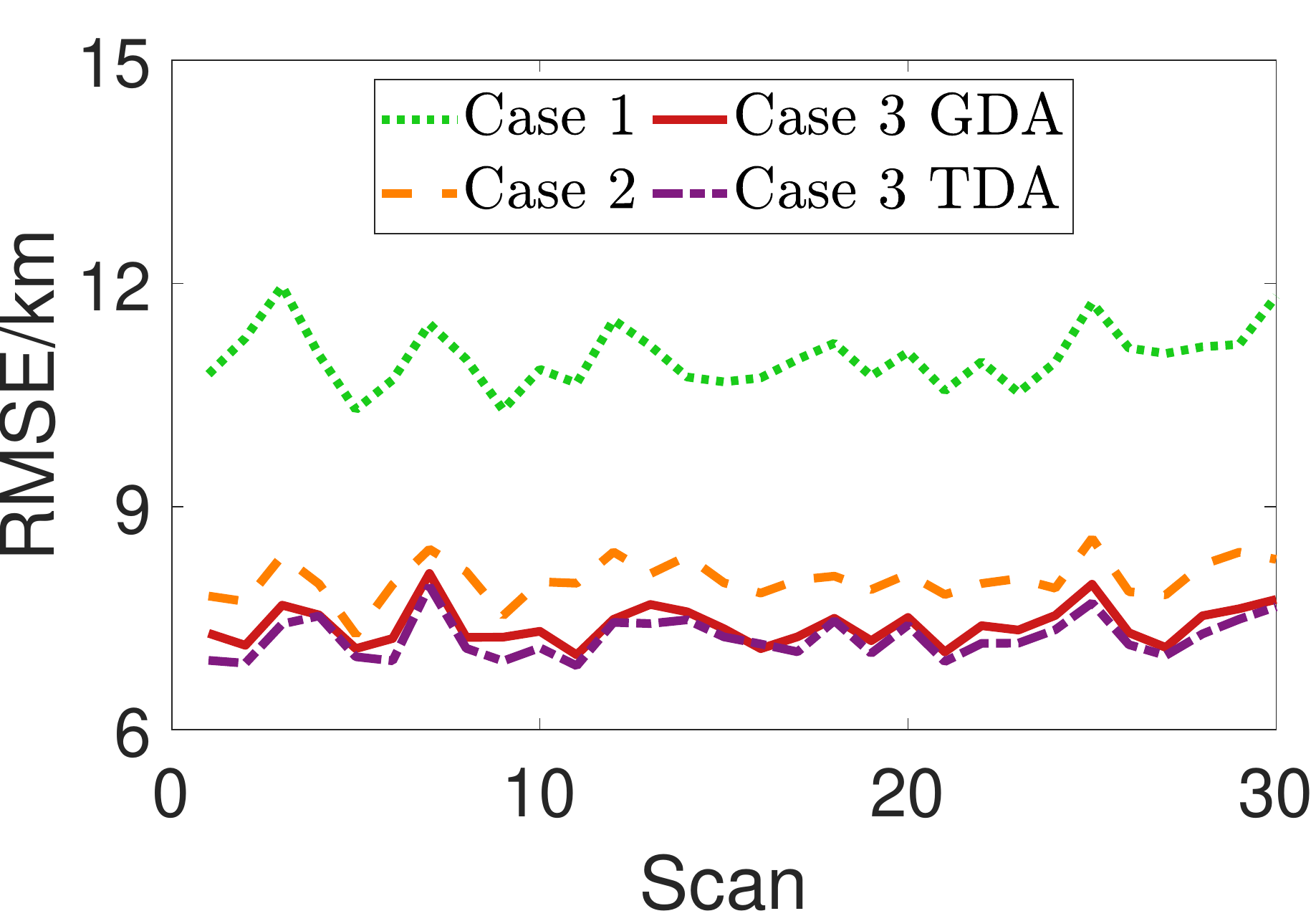}
			\includegraphics[width=0.99\textwidth]{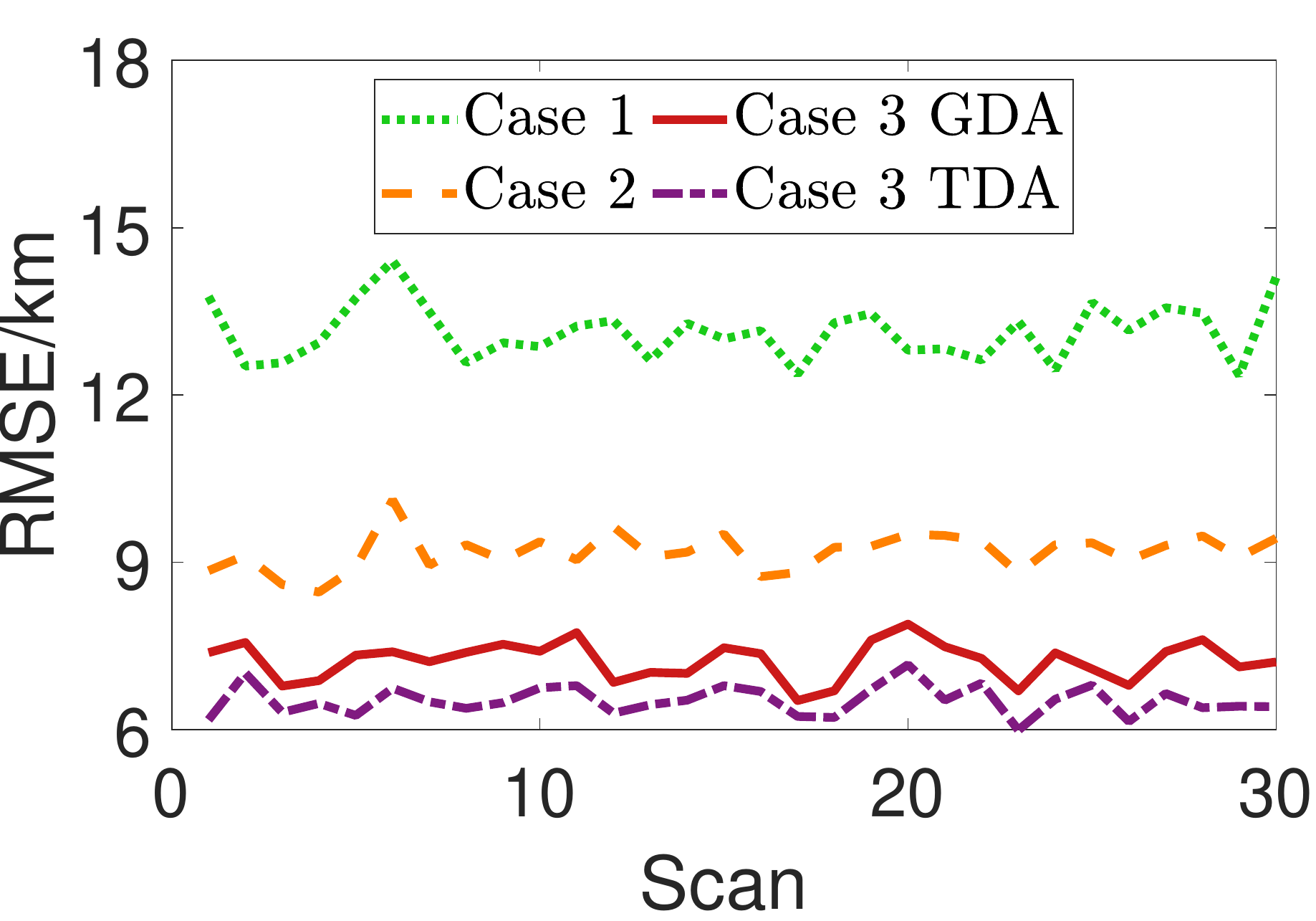}
		\end{minipage}%
	}%
	\centering
	\caption{Statistical results of the used VIHs estimation under different cases: from left to right, from top to bottom
		are the results of $ { \bm{h}^{\mathrm{E}}(i_t) }$, ${ \bm{h}^{\mathrm{E}}(i_r) }$, ${ \bm{h}^{\mathrm{F}}(i_t) } $ and ${ \bm{h}^{\mathrm{F}}(i_r) }$, respectively.
		The abbreviations GDA and TDA represent that validation gate data association and true data association are used in ECM-GMRF, respectively.
	}
	\label{fig:oneTargetFirstHRMSE}
\end{figure}

\begin{figure}[!htb]
	\centering
	\subfigure{
		\begin{minipage}[t]{0.49\linewidth}
			\centering
			\includegraphics[width=0.99\textwidth]{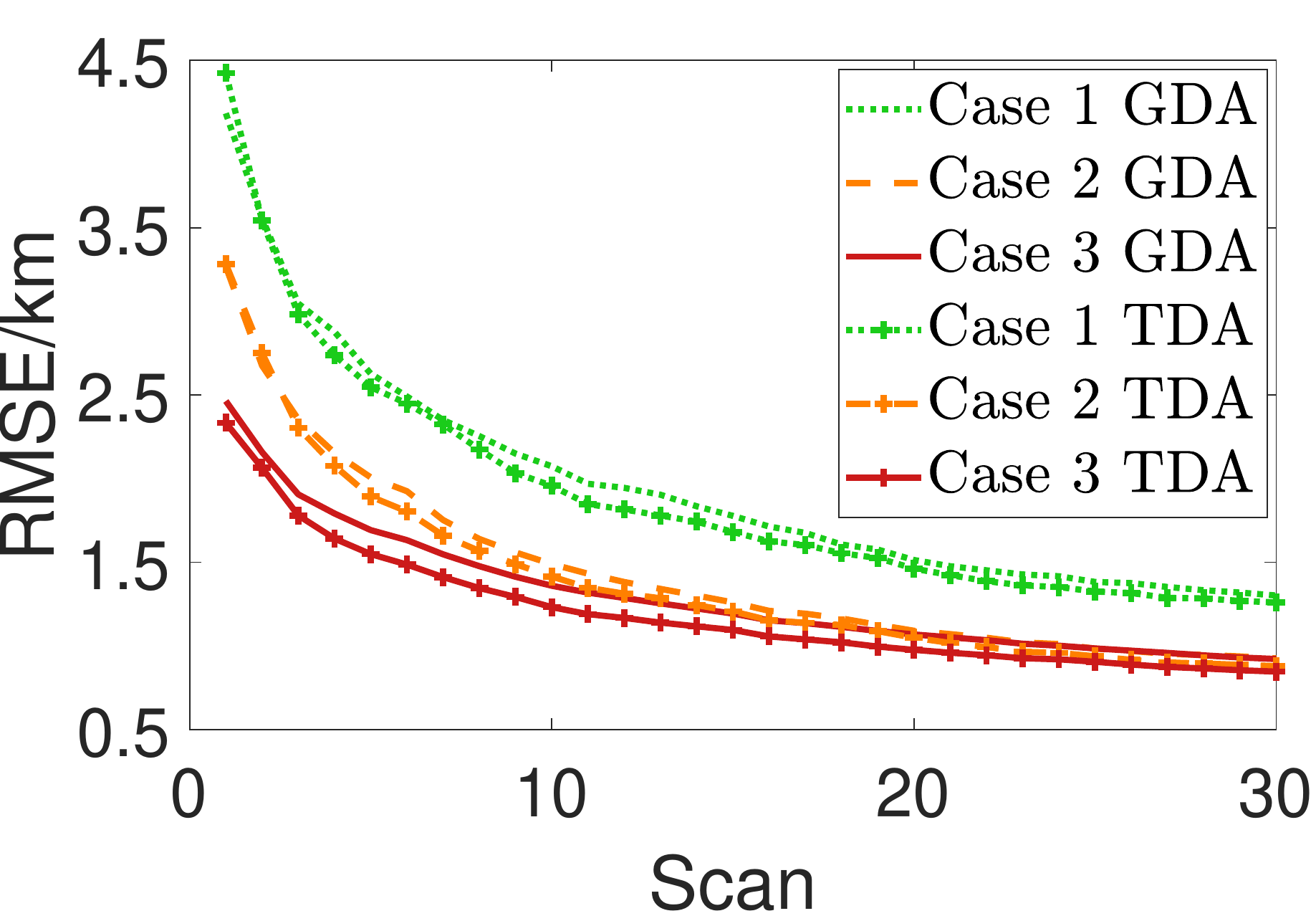}
			\label{Subfig: oneTargetFirstXRMSEDim1}
		\end{minipage}%
	}%
	\subfigure{
		\begin{minipage}[t]{0.49\linewidth}
			\centering
			\includegraphics[width=0.99\textwidth]{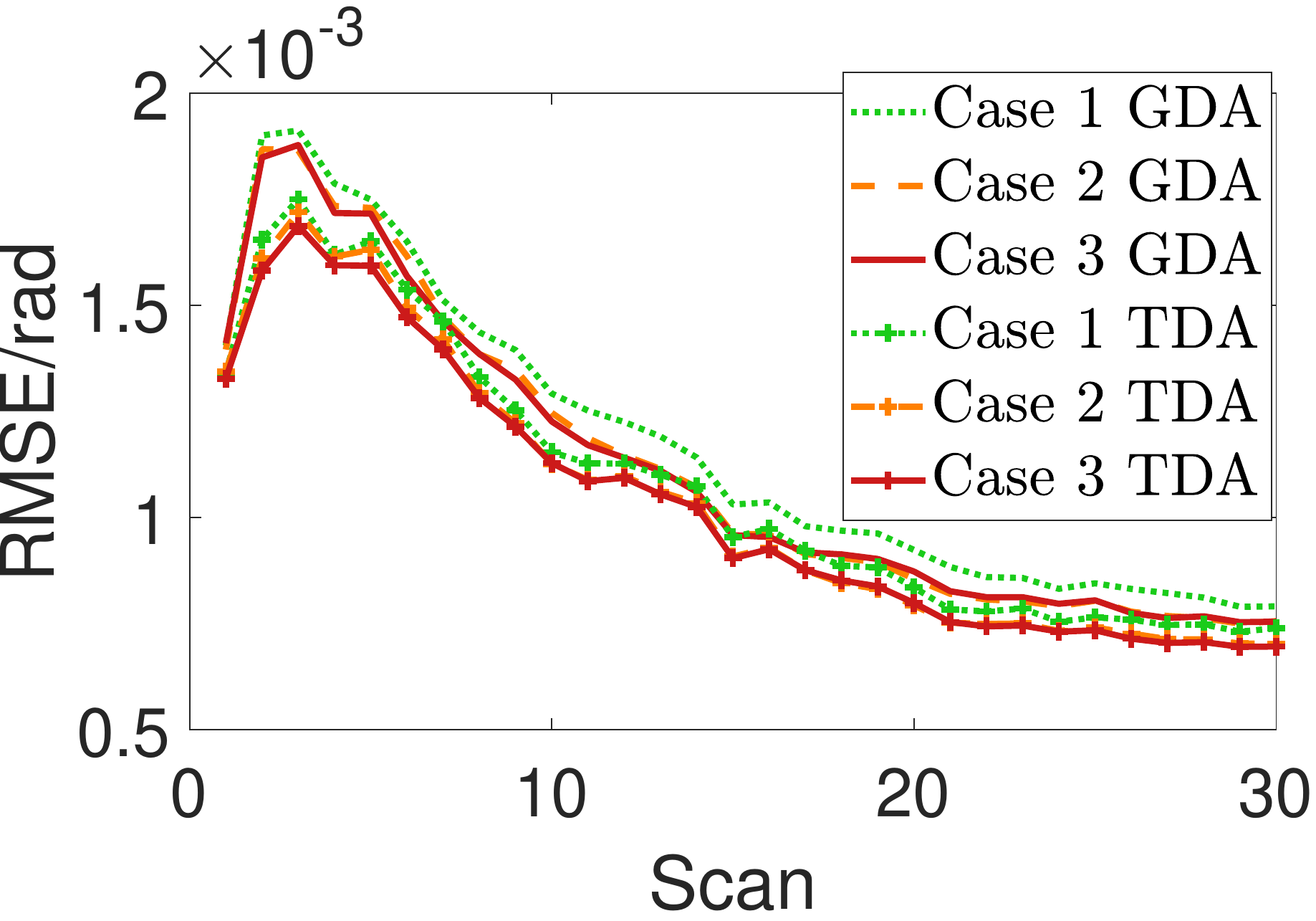}
			\label{Subfig: oneTargetFirstXRMSEDim3}
		\end{minipage}%
	}%
	\centering
	\caption{Statistical results of target state estimation with different information source on the VIHs:~{ground range~(left panel)~and~bearing~(right panel).}
	}
	\label{fig:oneTargetFirstXRMSE}
\end{figure}

The statistical results of VIHs estimation and target state estimation obtained by ECM-GMRF are shown in Fig.~\ref{fig:oneTargetFirstHRMSE} and Fig.~\ref{fig:oneTargetFirstXRMSE}, respectively.
For comparison, the results of VIHs estimation and target state estimation with true data association are also
presented in Fig.~\ref{fig:oneTargetFirstHRMSE} and Fig.~\ref{fig:oneTargetFirstXRMSE}.

From Fig.~\ref{fig:oneTargetFirstHRMSE}, it is seen that
using ionosonde measurements~(Case 2 and Case 3) can significantly reduce the estimation error of VIHs comparing with using constant VIHs~(Case 1).
By observing the RMSE curves of VIHs of Case 2 and Case 3, it is concluded that, as we expected,
using the additional measurements of the target from OTHR can improve the estimation of VIHs as well,
especially for  ${ \bm{h}^{\mathrm{F}}(i_t) }$ and  ${ \bm{h}^{\mathrm{F}}(i_r) }$.
The difference in the increase of the used VIHs of the E layer and the F layer is due to the difference in the values and parameters of the two layers. For example, the linearization errors in Eq.~(\ref{equ_tarlorExpansionOTHRMeasureFunc}) for mode EE and mode FF are different due to the different values of the two layers.
The results in Fig.~\ref{fig:oneTargetFirstHRMSE} verify our standpoints by the following facts.
The comparison between Case 1 and Case 2 shows that the correlation among the VIHs contributes to the estimation of the used VIH. Fig.~\ref{fig_region_partition1} shows that the used VIHs are not directly measured by ionosondes.
The comparison between Case 2 and Case 3 shows that OTHR measurements  are helpful to the estimation of the used VIHs.

The RMSE curves of VIHs of Case 3 with true data association and gate data association indicate that erroneous data association
indeed occurred to ECM-GMRF using gate data association and it can deteriorate the estimation performance of ECM-GMRF on VIHs.
Note that since the temporal correlation of VIHs is not considered in this paper,
the estimation accuracy of VIHs has not been improved with the accumulation of~(both OTHR and ionosonde)~measurements over time.

By observing the RMSE curves of ground range and bearing for all the three cases in Fig.~\ref{fig:oneTargetFirstXRMSE}
and the above analysis of VIHs estimation based on Fig.~\ref{fig:oneTargetFirstHRMSE},
we conclude that better estimation of VIHs is beneficial to target state estimation.
The mean RMSE of the ground range of the three cases using gate data association are 1.97~km, 1.44~km  and 1.3~km, respectively;
the improvement ratio of Case 2 and Case 3 over Case 1 are 26.7\% and 33.7\%, respectively.
Overall, joint estimation of target state and VIHs using both OTHR measurements and ionosonde measurements
can greatly improve the estimation accuracy of the used VIHs as well as the target state.

\begin{figure}[!htb]
	\centering
	\subfigure{
		\begin{minipage}[t]{0.499\linewidth}
			\centering
			\includegraphics[width=0.99\textwidth]{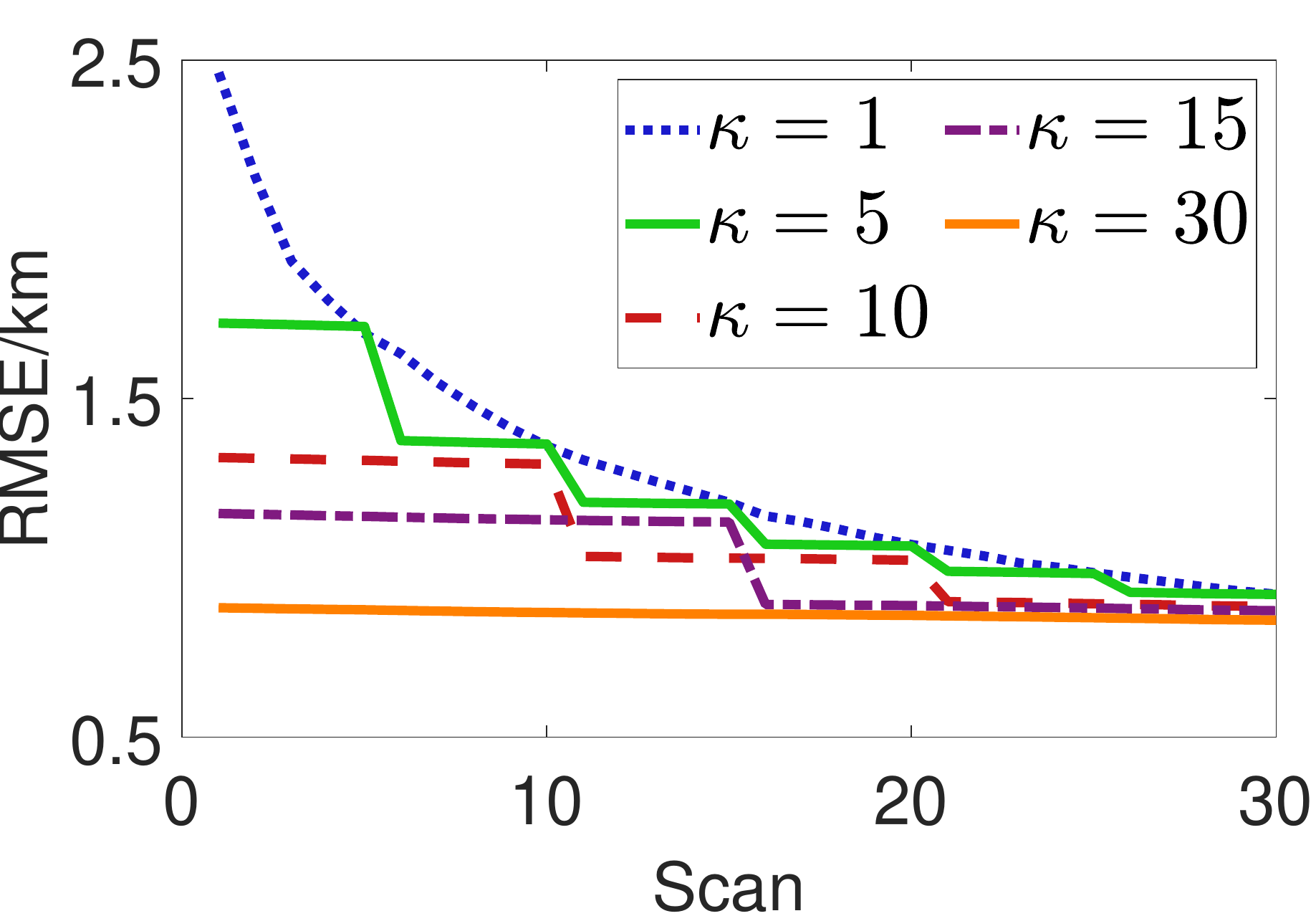}
		\end{minipage}%
	}%
	\subfigure{
		\begin{minipage}[t]{0.499\linewidth}
			\centering
			\includegraphics[width=0.99\textwidth]{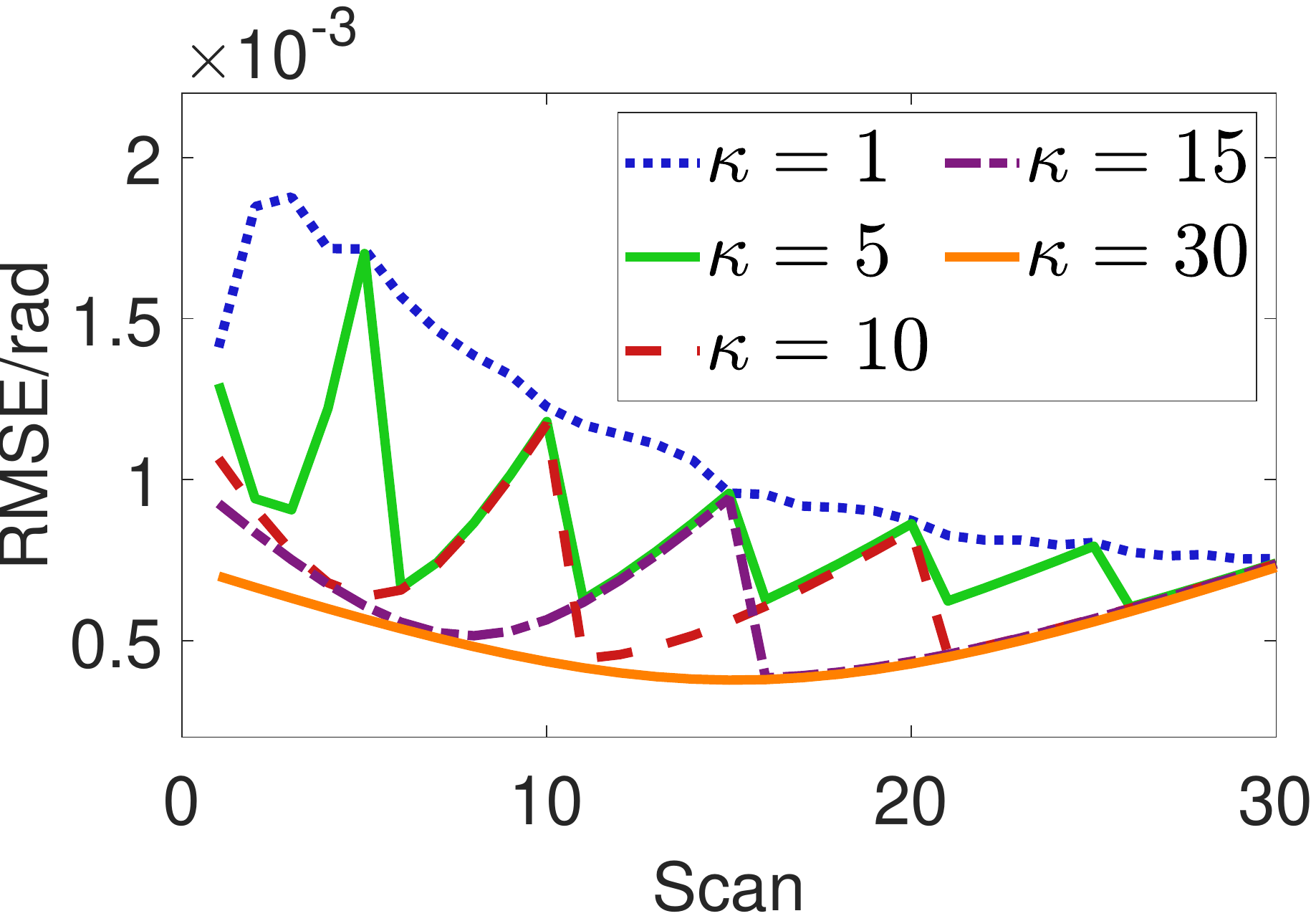}
		\end{minipage}%
	}%
	\centering
	\caption{Statistical results of target state estimation with different sequence length $\kappa$:~ground range~(left panel)~and~bearing~(right panel). }
	\label{fig:oneTargetSecondXRMSE}
\end{figure}

\begin{figure}[!htb]
	\centering
	\subfigure{
		\begin{minipage}[t]{0.499\linewidth}
			\centering
			\includegraphics[width=0.99\textwidth]{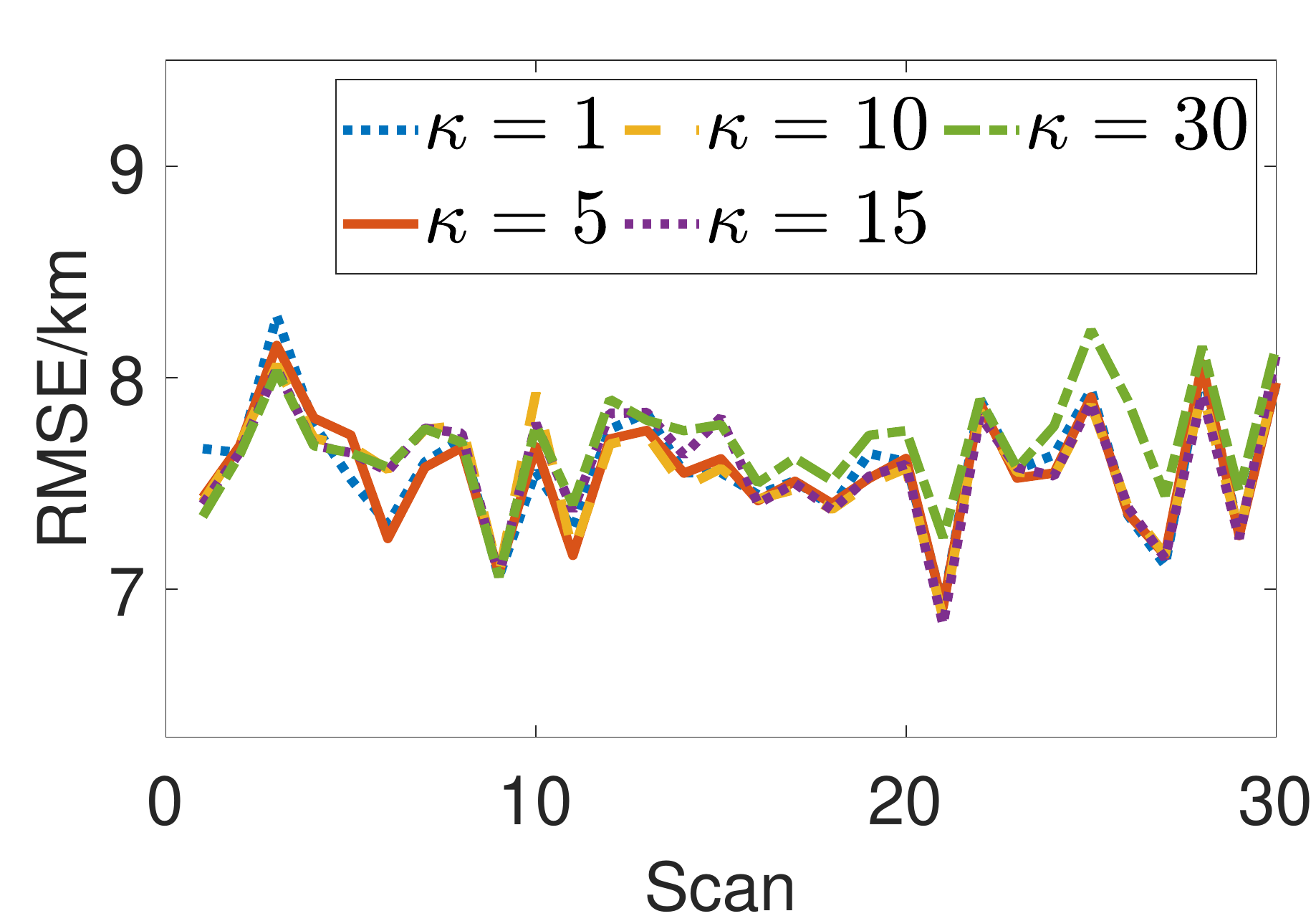}
			\includegraphics[width=0.99\textwidth]{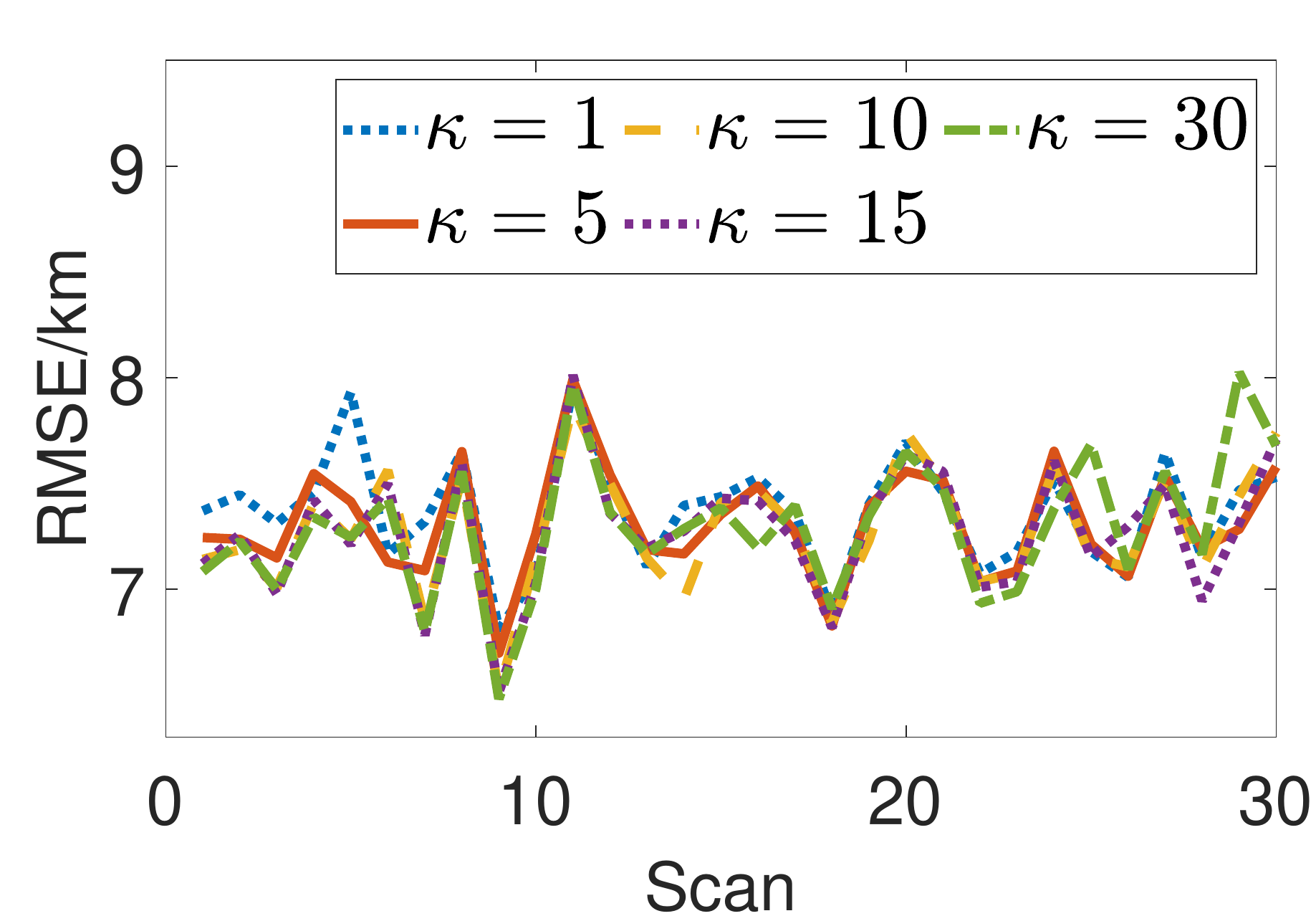}
			\label{Subfig: VIHrmseDim1}
		\end{minipage}%
	}%
	\subfigure{
		\begin{minipage}[t]{0.499\linewidth}
			\centering
			\includegraphics[width=0.99\textwidth]{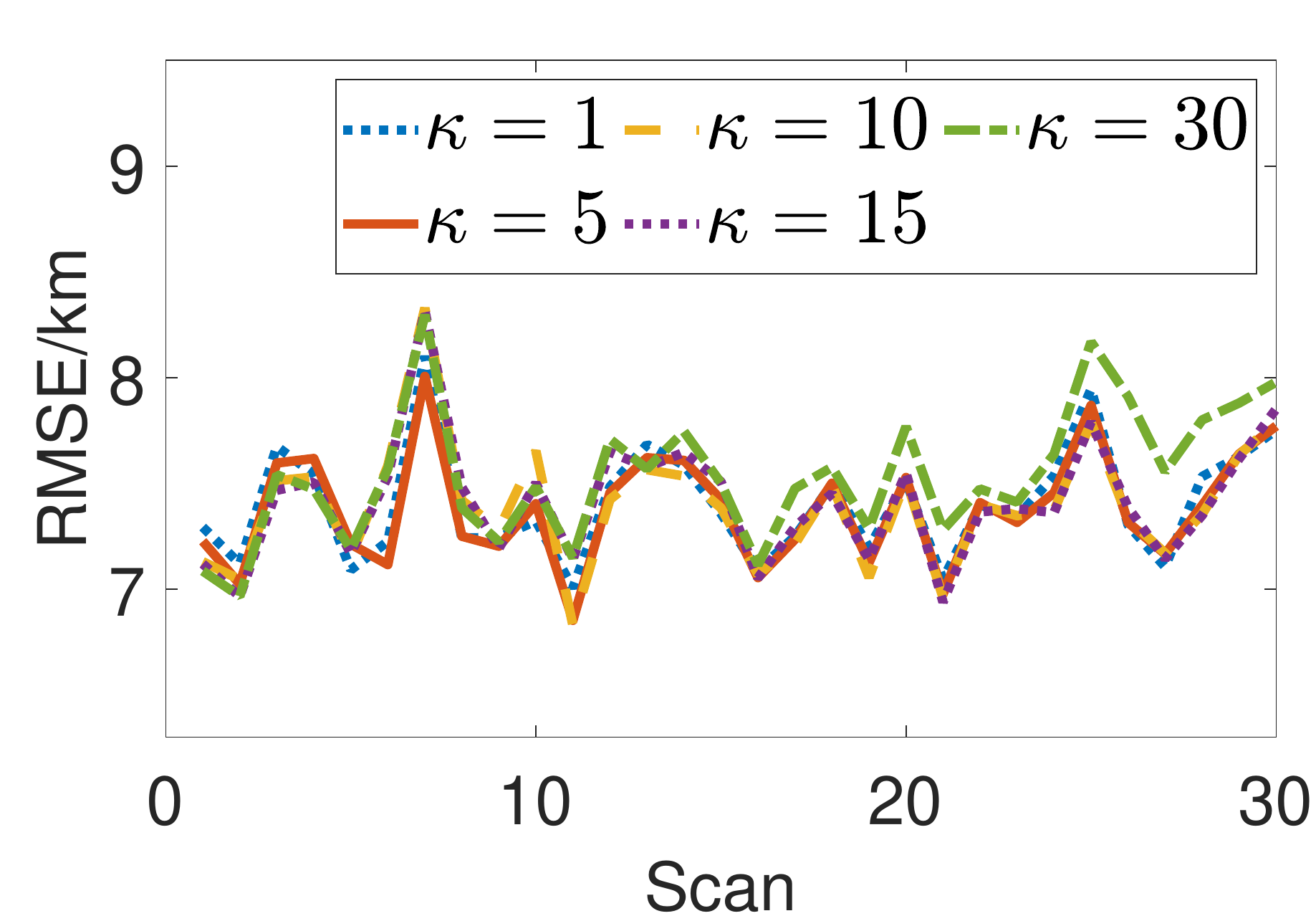}
			\includegraphics[width=0.99\textwidth]{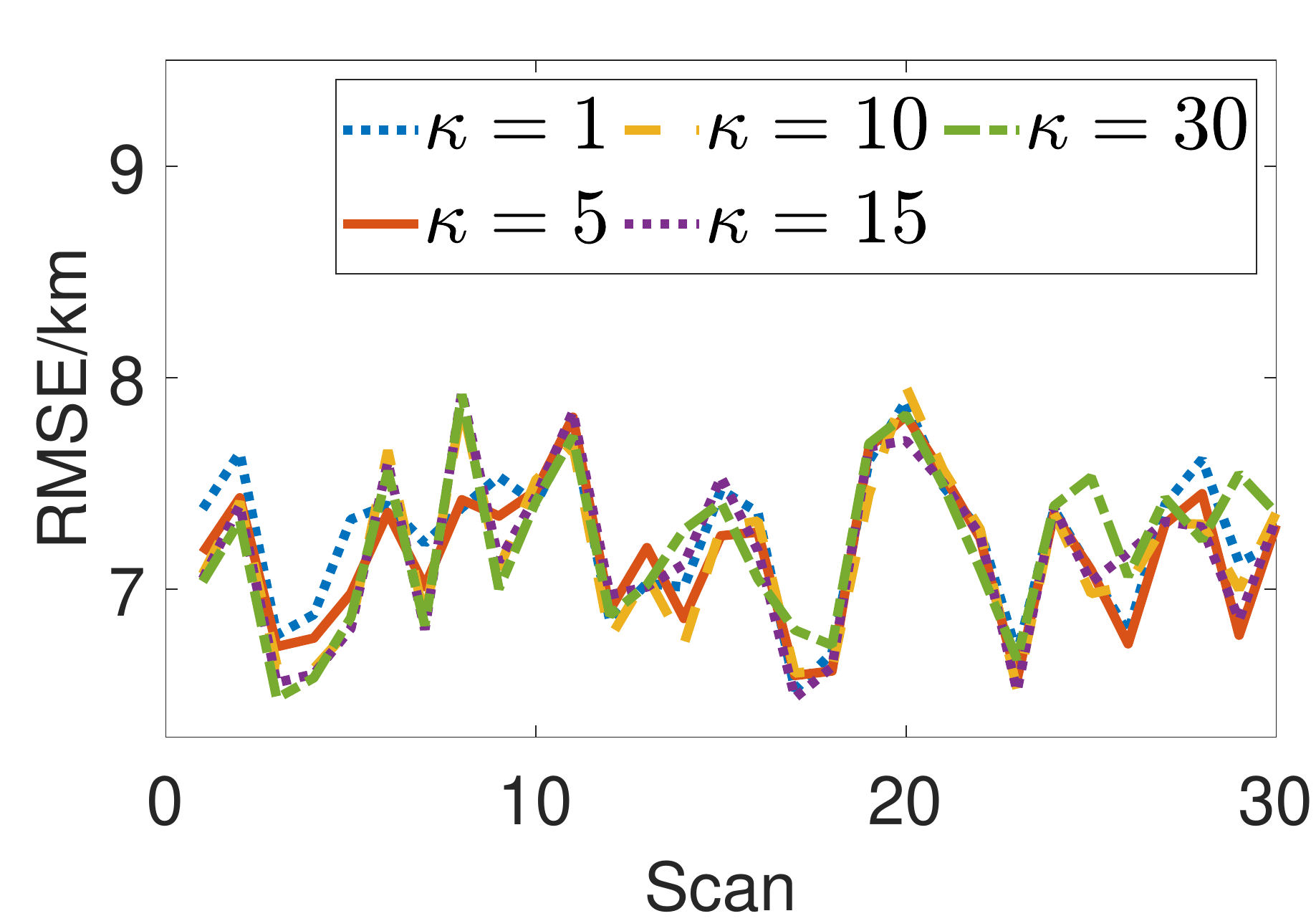}
			\label{Subfig: VIHrmseDim3}
		\end{minipage}%
	}%
	\centering
	\caption{Statistical results of the used VIHs estimation with different sequence length $\kappa$:
		from left to right, from top to bottom are the results of ${ \bm{h}^{\mathrm{E}}(i_t) }$, ${ \bm{h}^{\mathrm{E}}(i_r) }$, ${ \bm{h}^{\mathrm{F}}(i_t) } $ and ${ \bm{h}^{\mathrm{F}}(i_r) }$, respectively.
	}
	\label{fig:oneTargetSecondHRMSE}
\end{figure}

Next, the performance of ECM-GMRF with different sequence length $\kappa$ is explored.
We run ECM-GMRF on Case 3 with gate data association.
The statistical results of target state estimation and the used VIHs estimation are shown in Fig.~\ref{fig:oneTargetSecondXRMSE}
and Fig.~\ref{fig:oneTargetSecondHRMSE}, from which it can be seen that as $\kappa$ increases,
the accuracy of target state estimation increases since more measurements have been used.
When $\kappa = 30$, the RMSE of the ground range of the target approaches $1$~km.
Note that a greater $\kappa$ means a greater output delay of a tracker.

\subsection{Multitarget tracking results}\label{subsec:multipleTargetResults}
In order to compare ECM-GMRF with MD-JPDAF and verify the performance improvement on a target brought by using
the OTHR measurements of other targets which is achieved indirectly by the improved VIHs used by the target,
we consider the following three cases.
\begin{itemize}
	\item Case 4: MD-JPDAF is performed for target tracking. The VIHs of layer E and layer F are fixed at $110$~km and $220$~km, respectively.
	\item Case 5: Each target is tracked individually using ECM-GMRF.
	\item Case 6: All the targets are tracked simultaneously using ECM-GMRF.
\end{itemize}
The parameter settings remain unchanged in Table~\ref{tab:parameters} and $\kappa = 30$ for ECM-GMRF.

\begin{figure*}[!htb]
	\centering
	\subfigure[RMSE of  $h^{E}_{i_t}$  for each target.]{
		\begin{minipage}[t]{0.24\linewidth}
			\centering
			\includegraphics[width=0.99\textwidth]{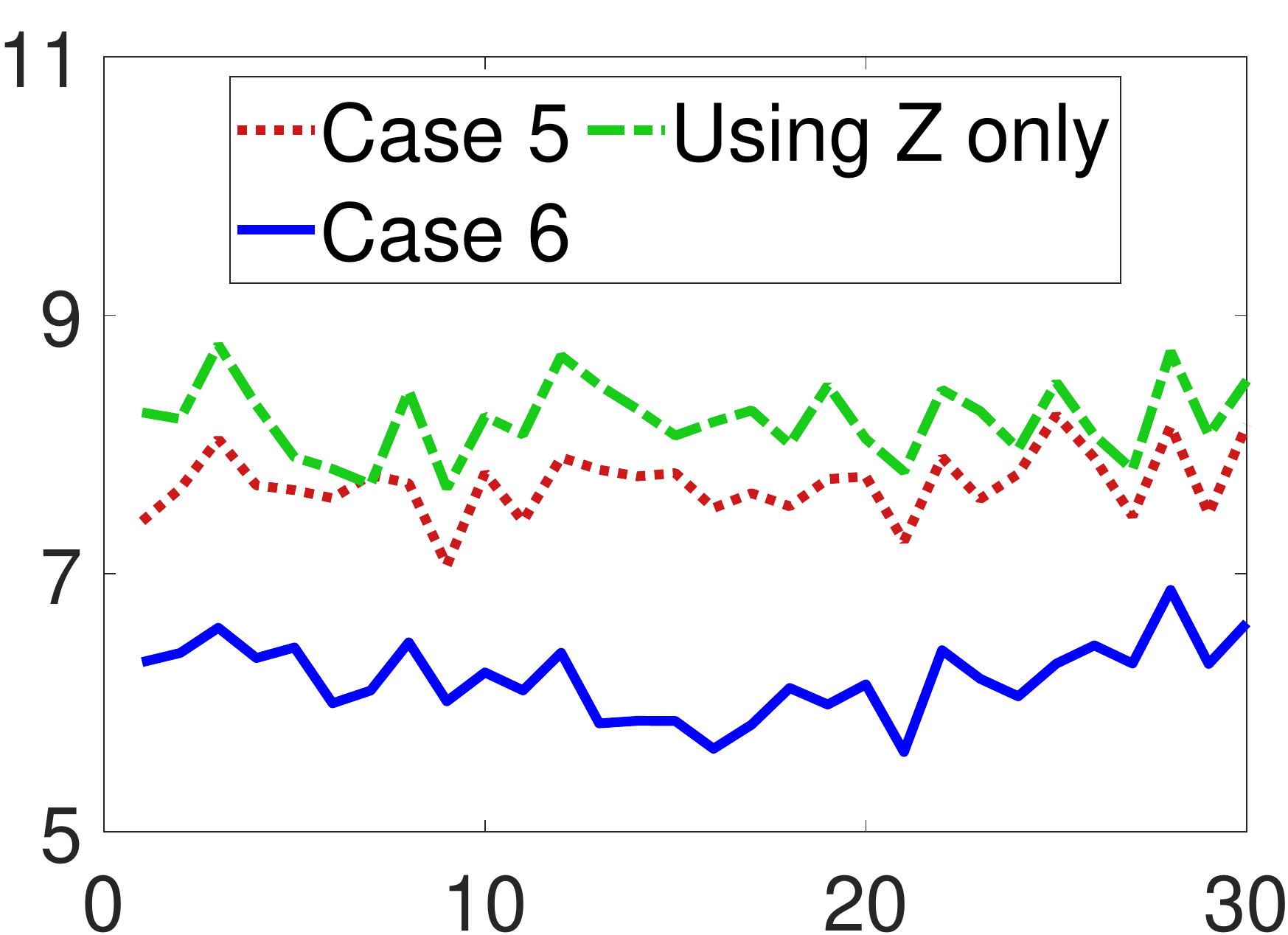}
			\includegraphics[width=0.99\textwidth]{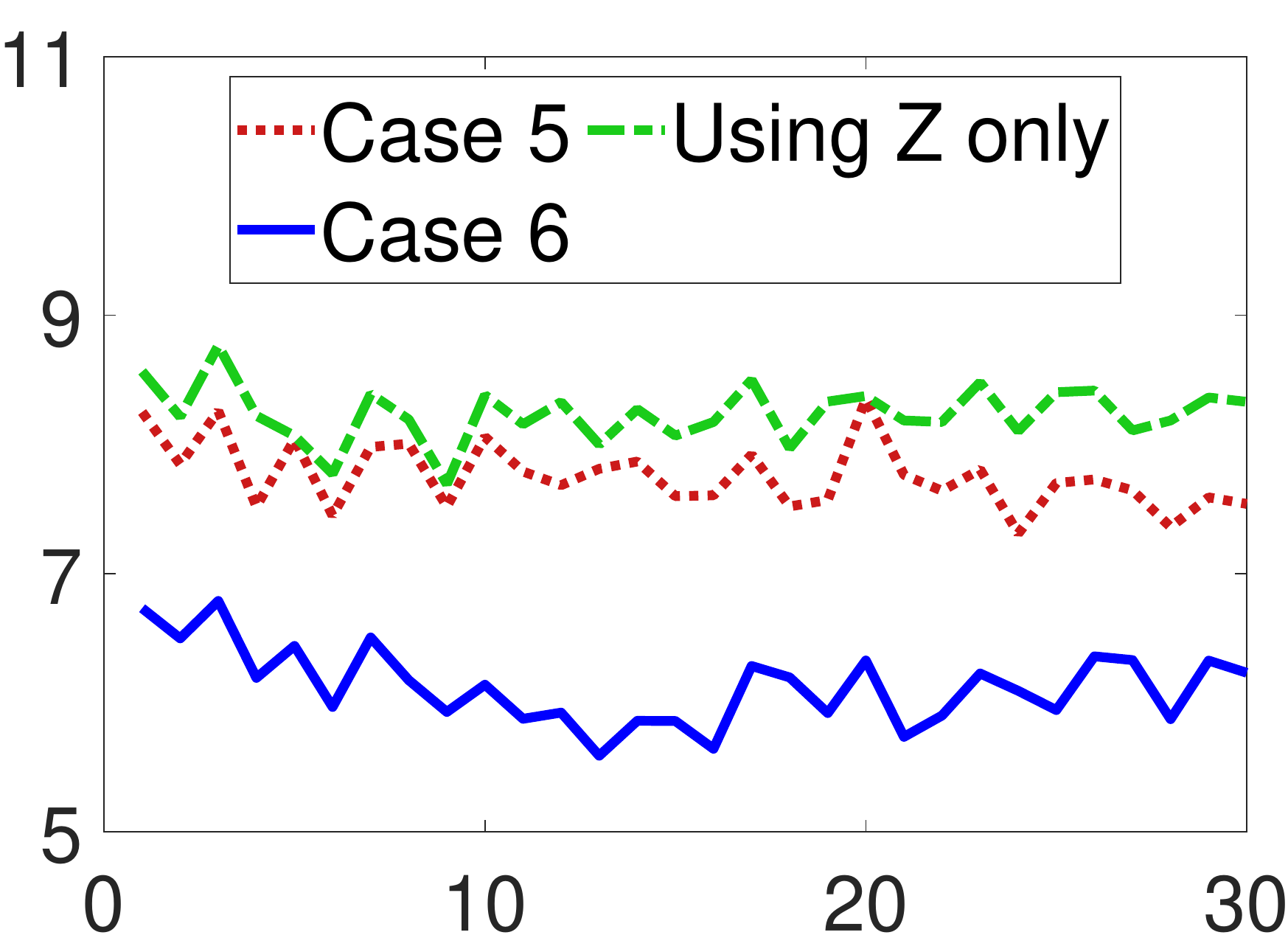}
			\includegraphics[width=0.99\textwidth]{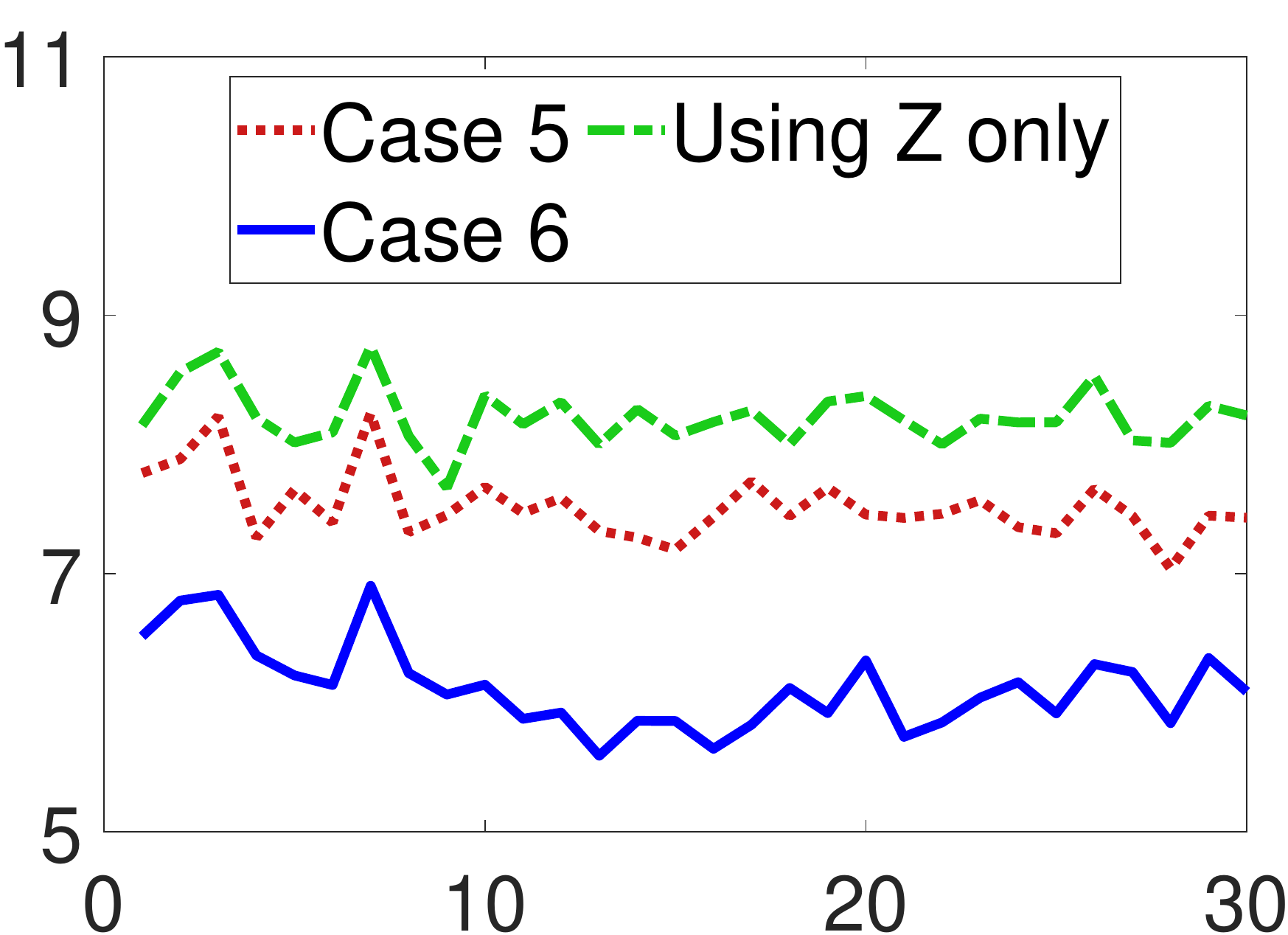}
			\includegraphics[width=0.99\textwidth]{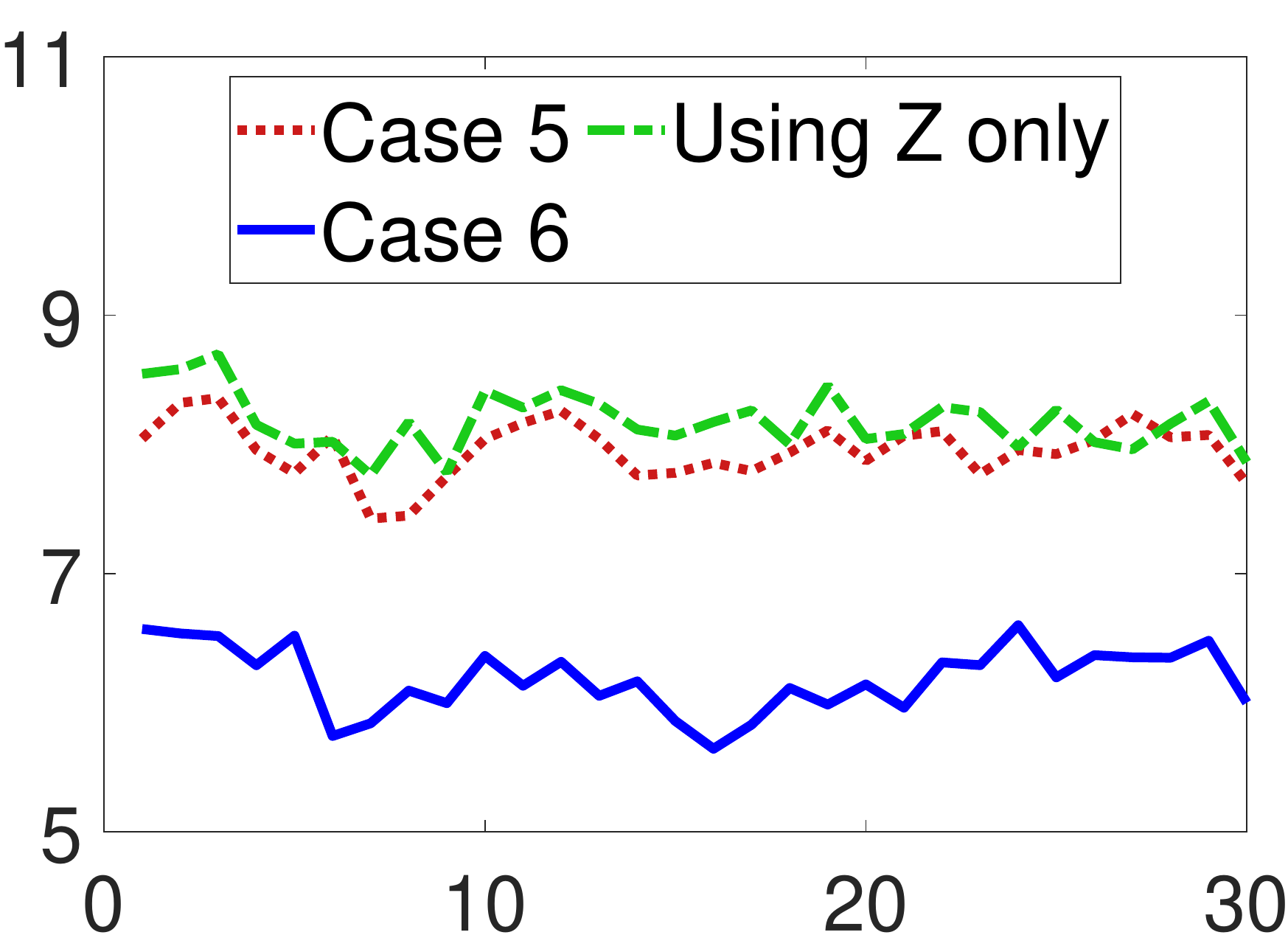}
			\includegraphics[width=0.99\textwidth]{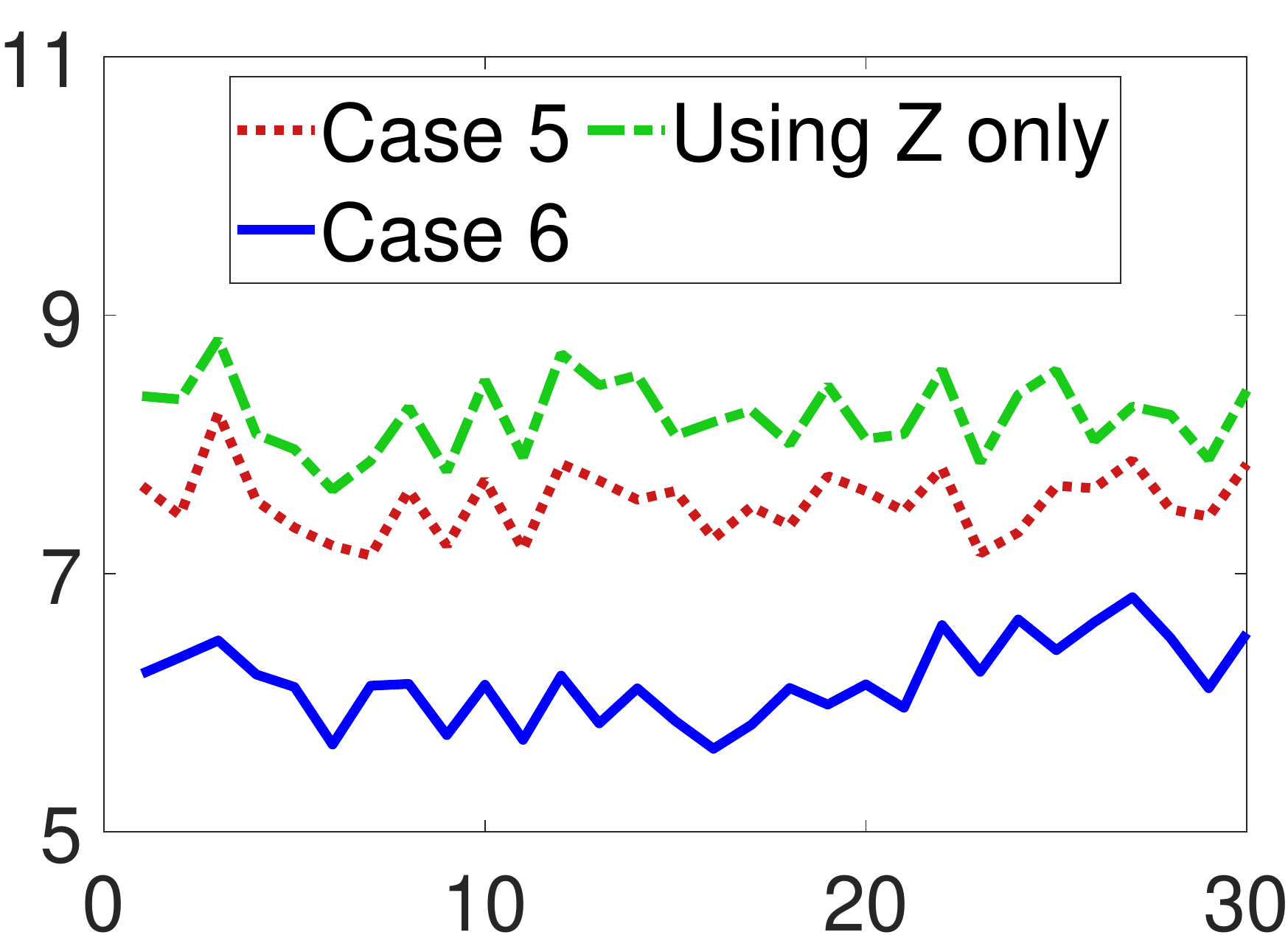}
		\end{minipage}%
	}%
	\subfigure[RMSE of  $h^{E}_{i_r}$  for each target.]{
		\begin{minipage}[t]{0.24\linewidth}
			\centering
			\includegraphics[width=0.99\textwidth]{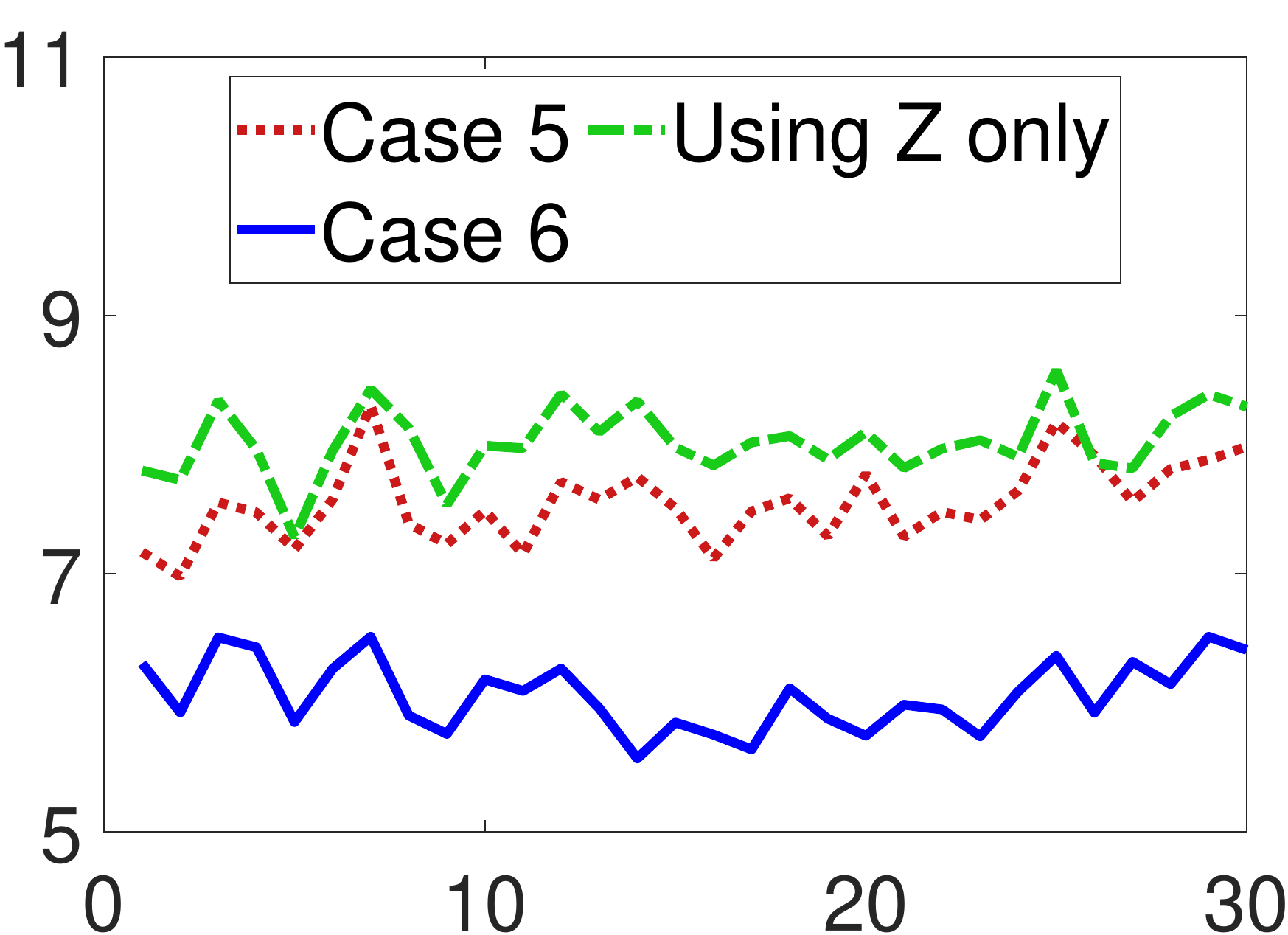}
			\includegraphics[width=0.99\textwidth]{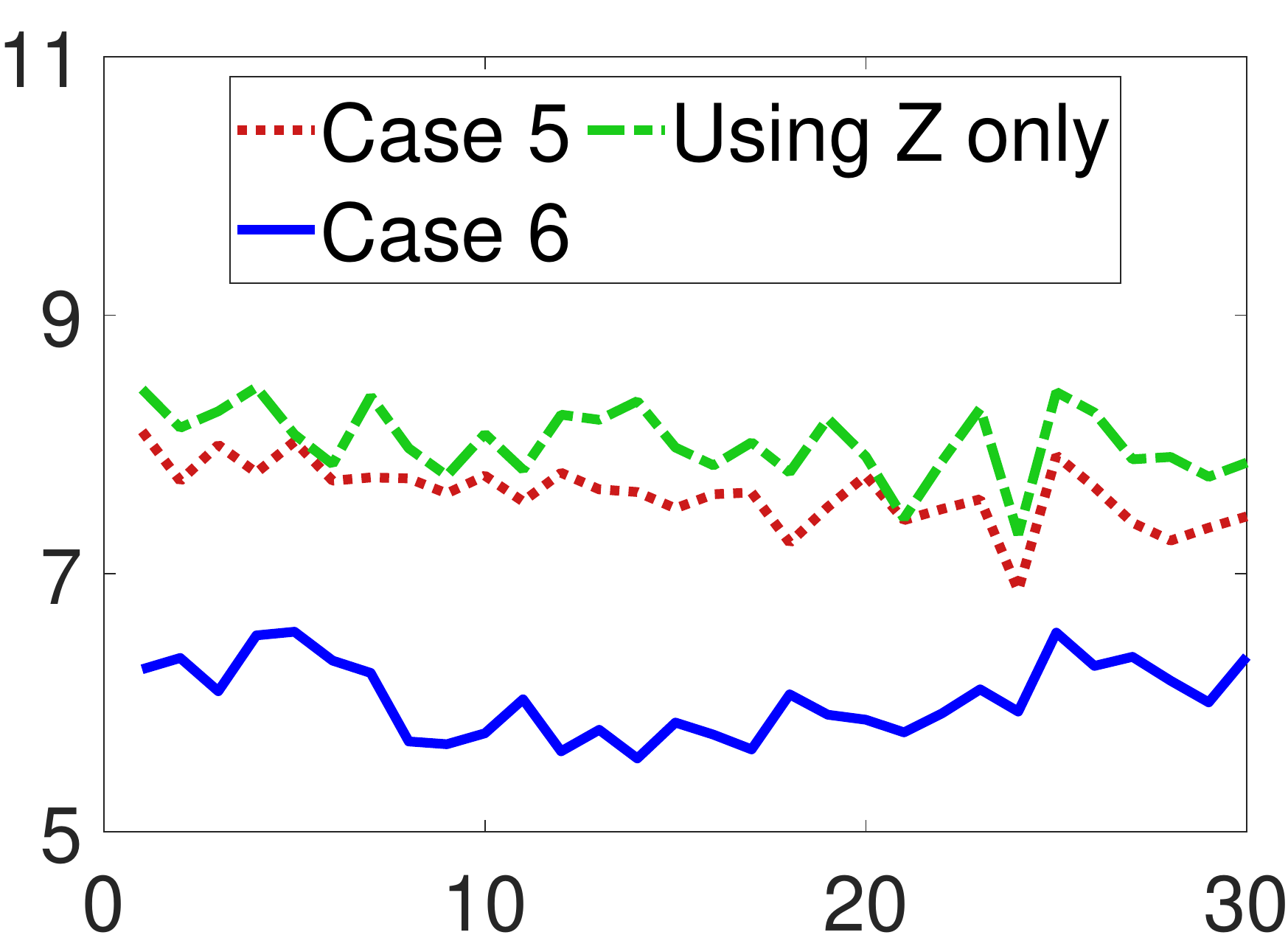}
			\includegraphics[width=0.99\textwidth]{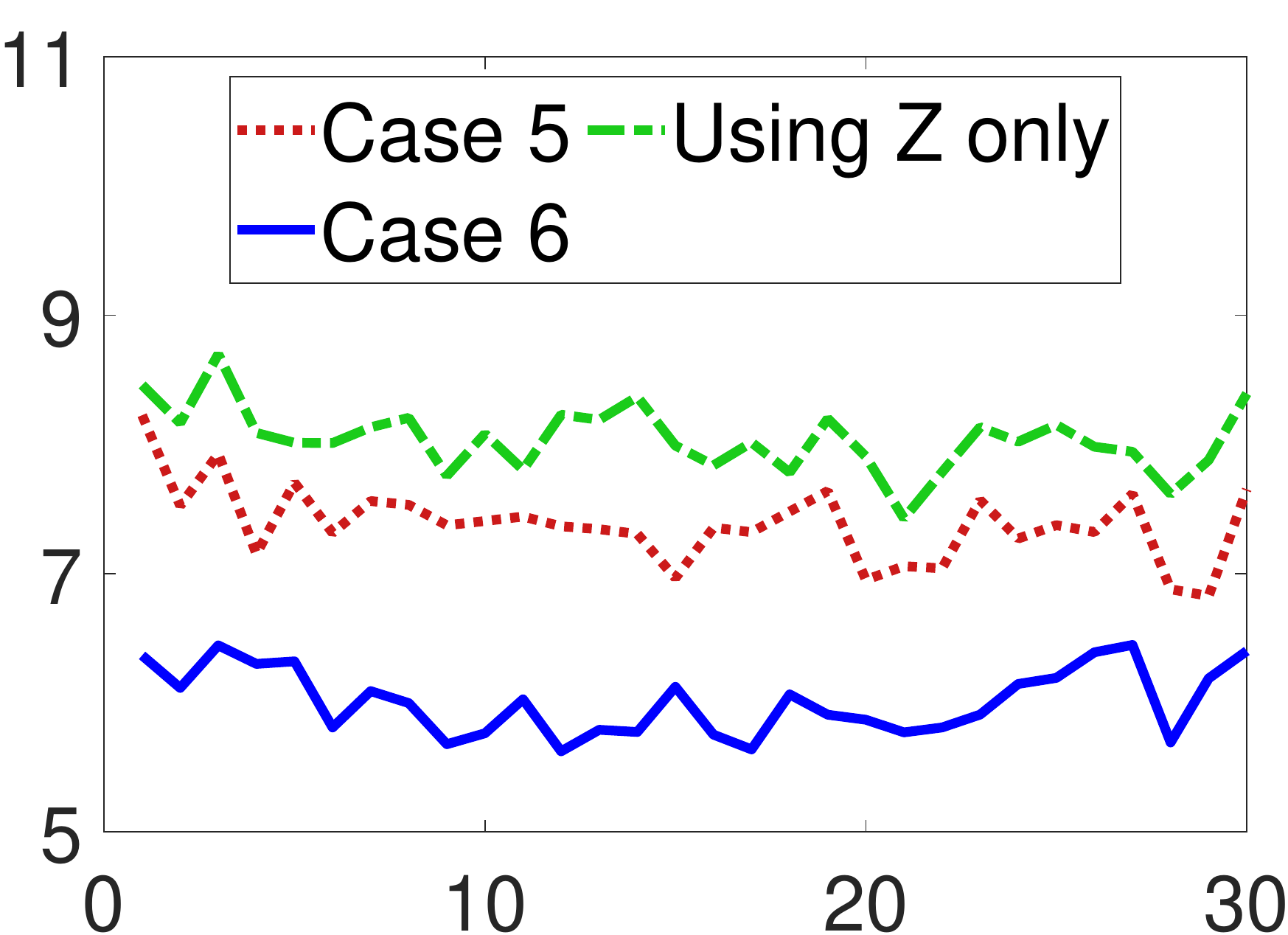}
			\includegraphics[width=0.99\textwidth]{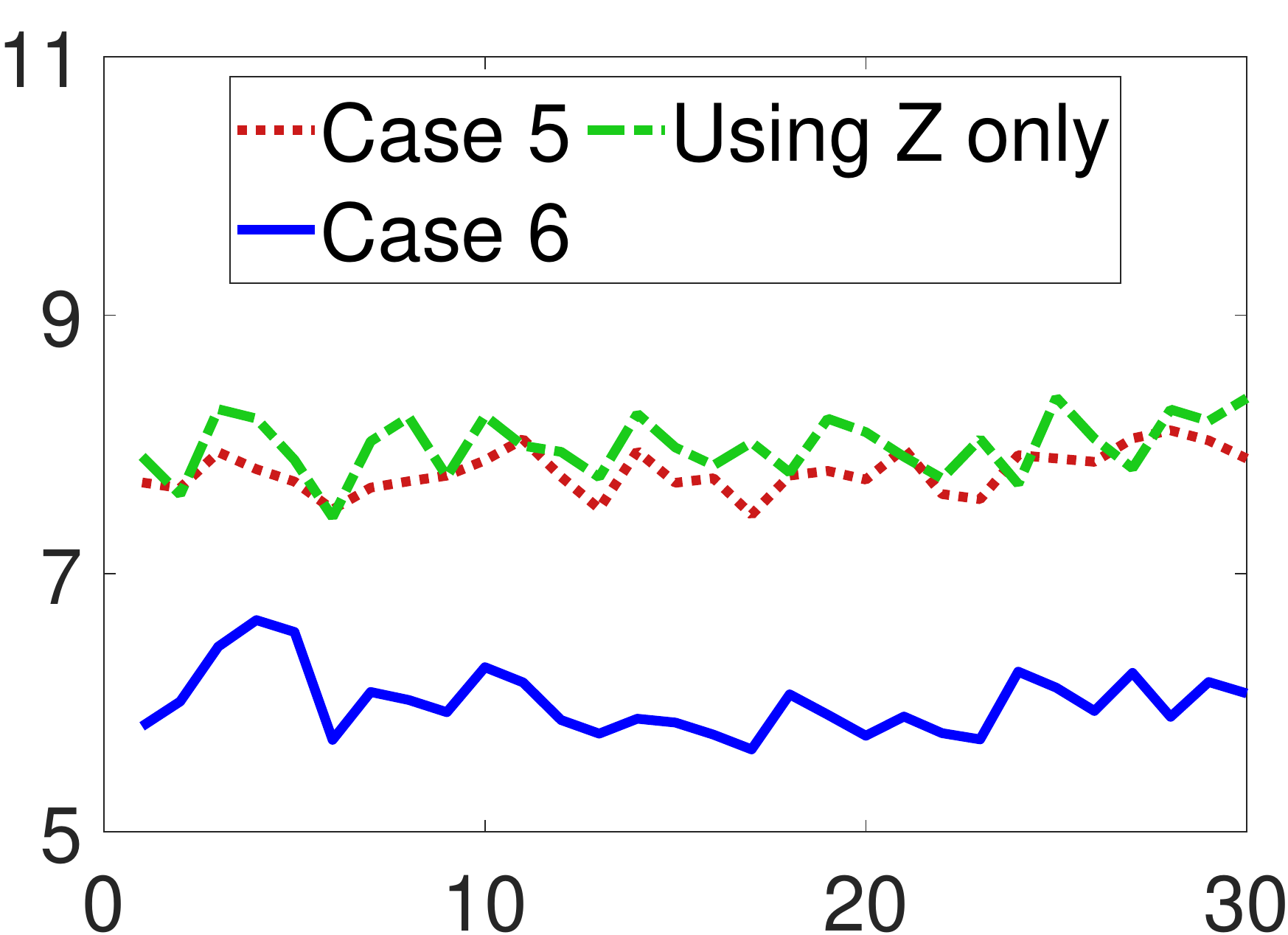}
			\includegraphics[width=0.99\textwidth]{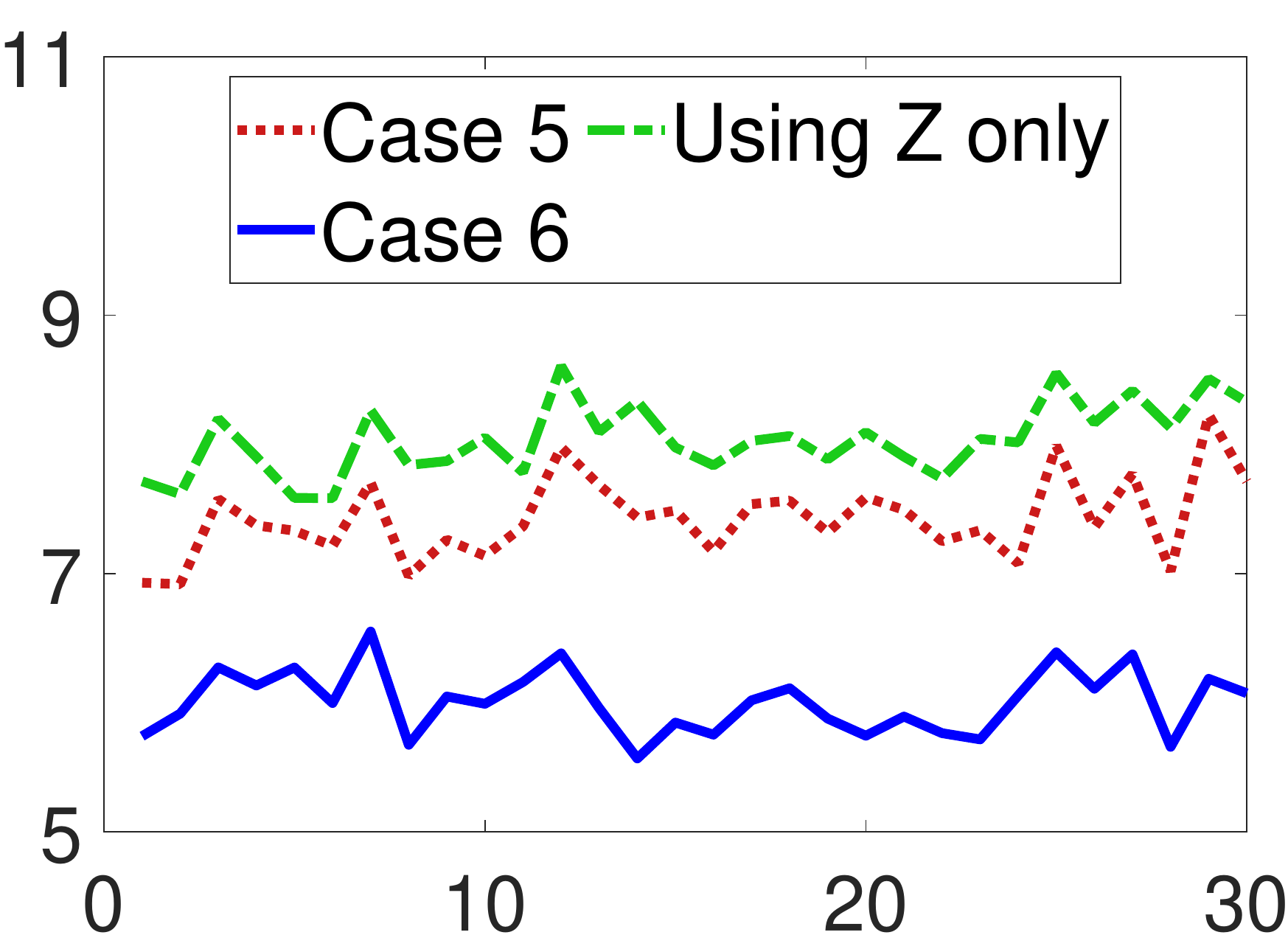}
		\end{minipage}%
	}%
	\subfigure[RMSE of  $h^{F}_{i_t}$  for each target.]{
		\begin{minipage}[t]{0.24\linewidth}
			\centering
			\includegraphics[width=0.99\textwidth]{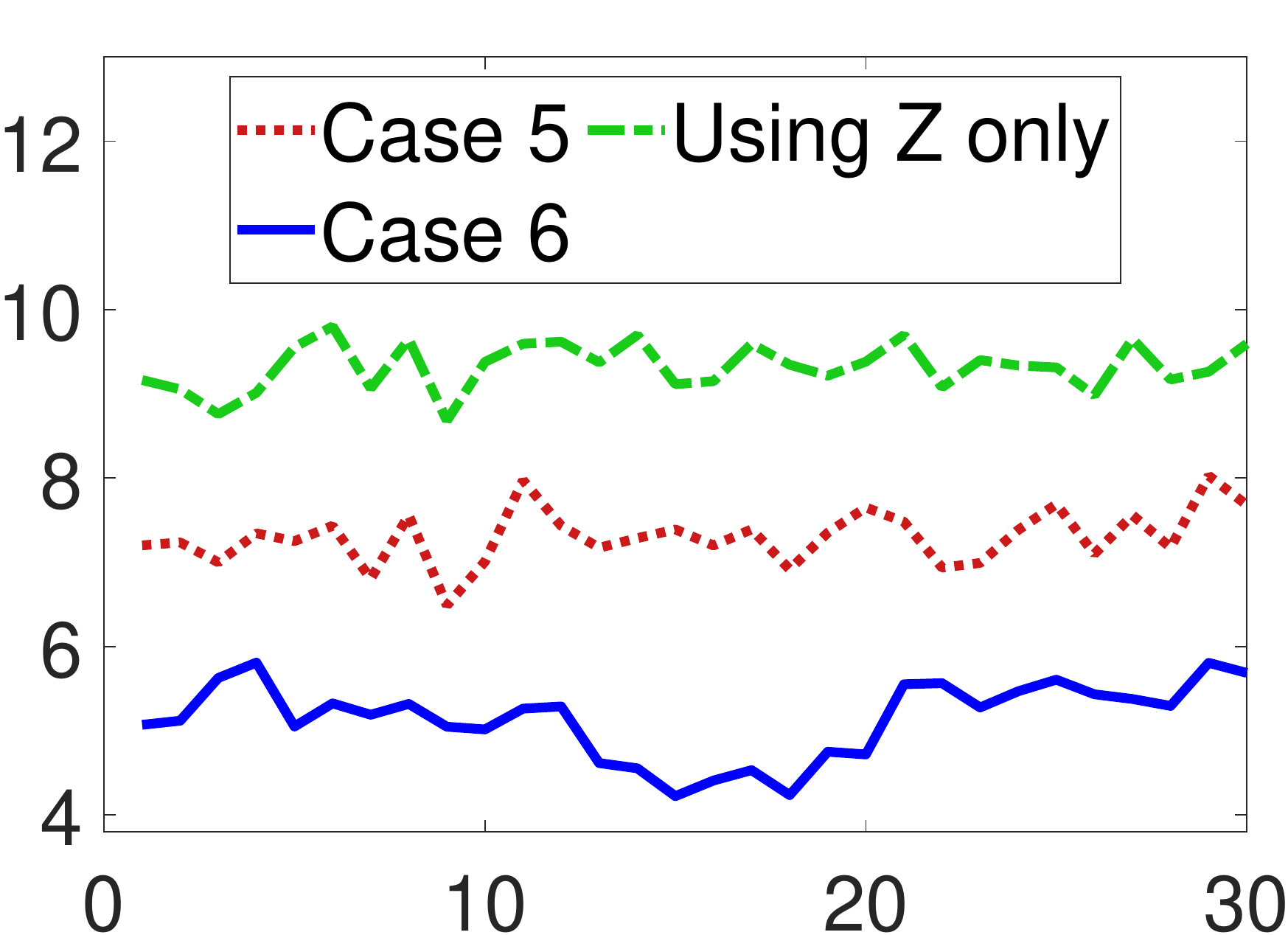}
			\includegraphics[width=0.99\textwidth]{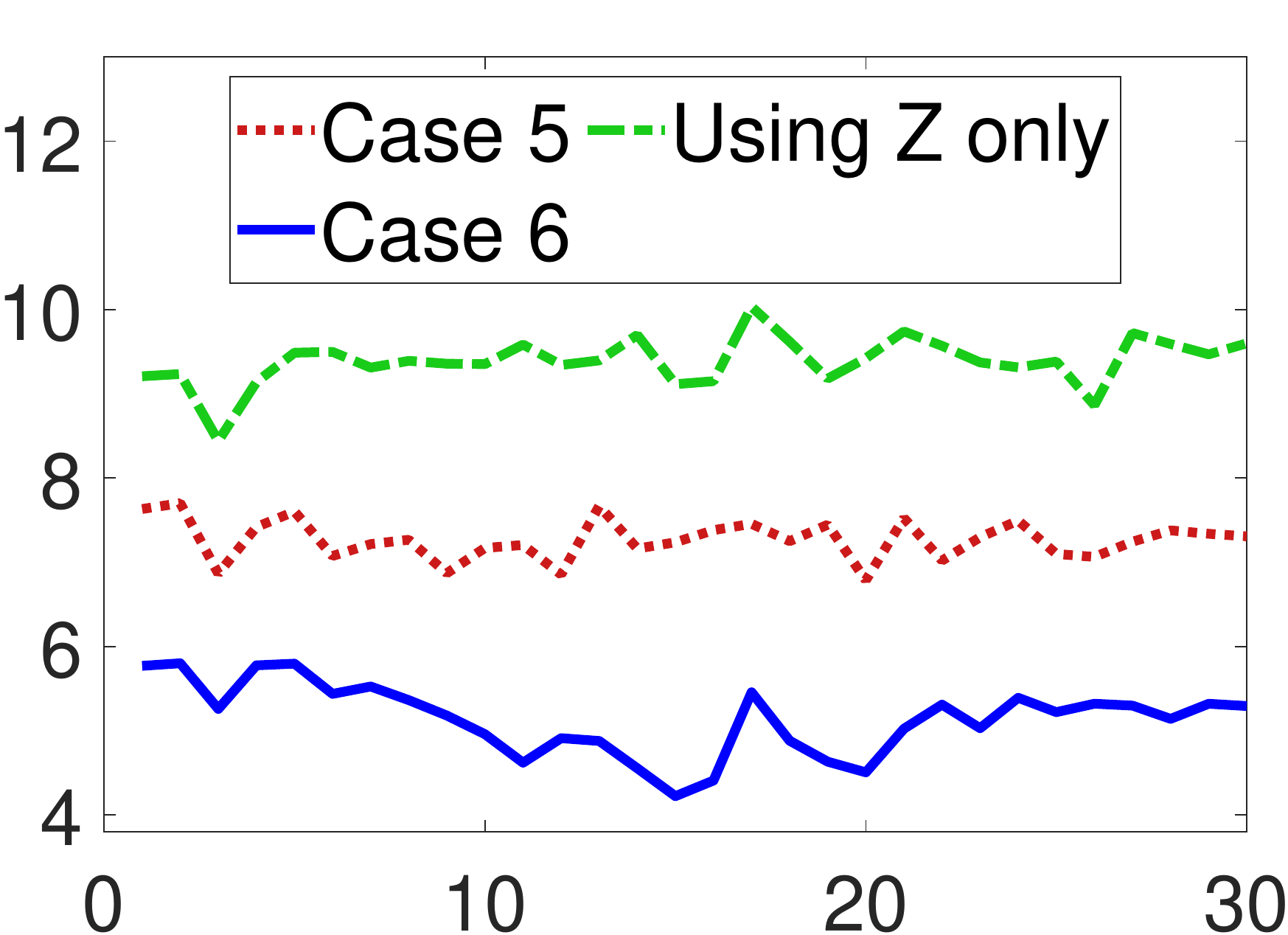}
			\includegraphics[width=0.99\textwidth]{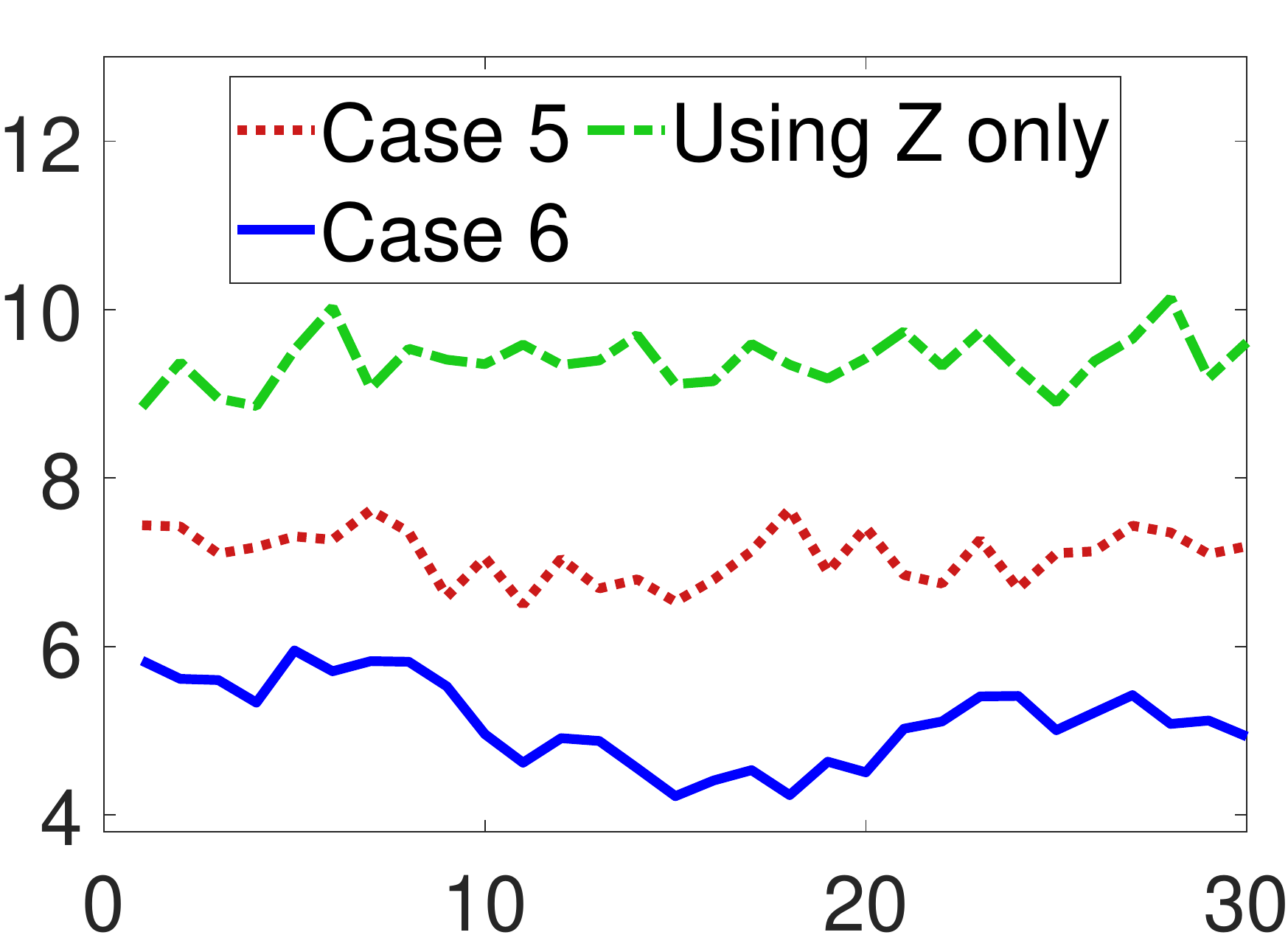}
			\includegraphics[width=0.99\textwidth]{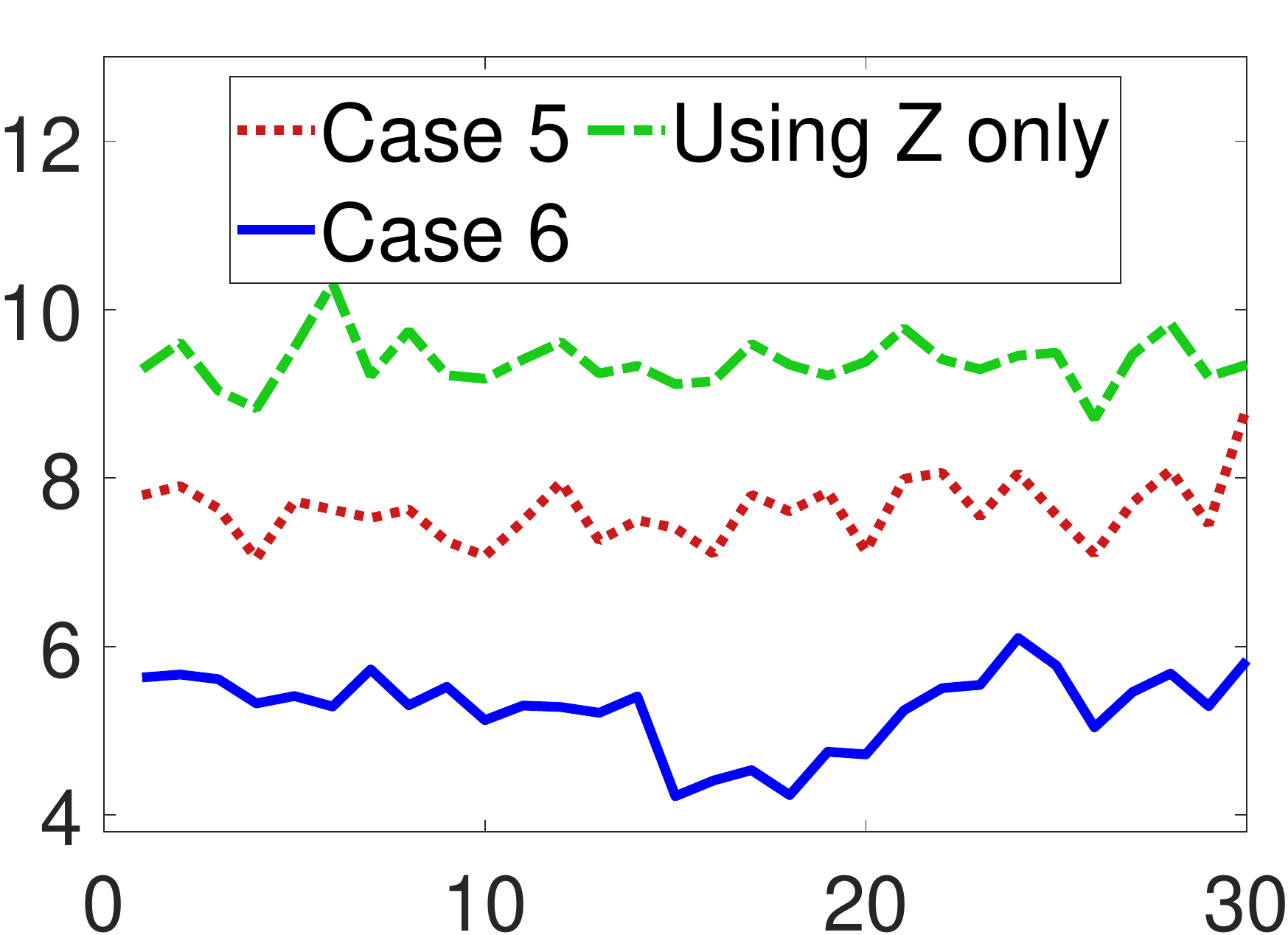}
			\includegraphics[width=0.99\textwidth]{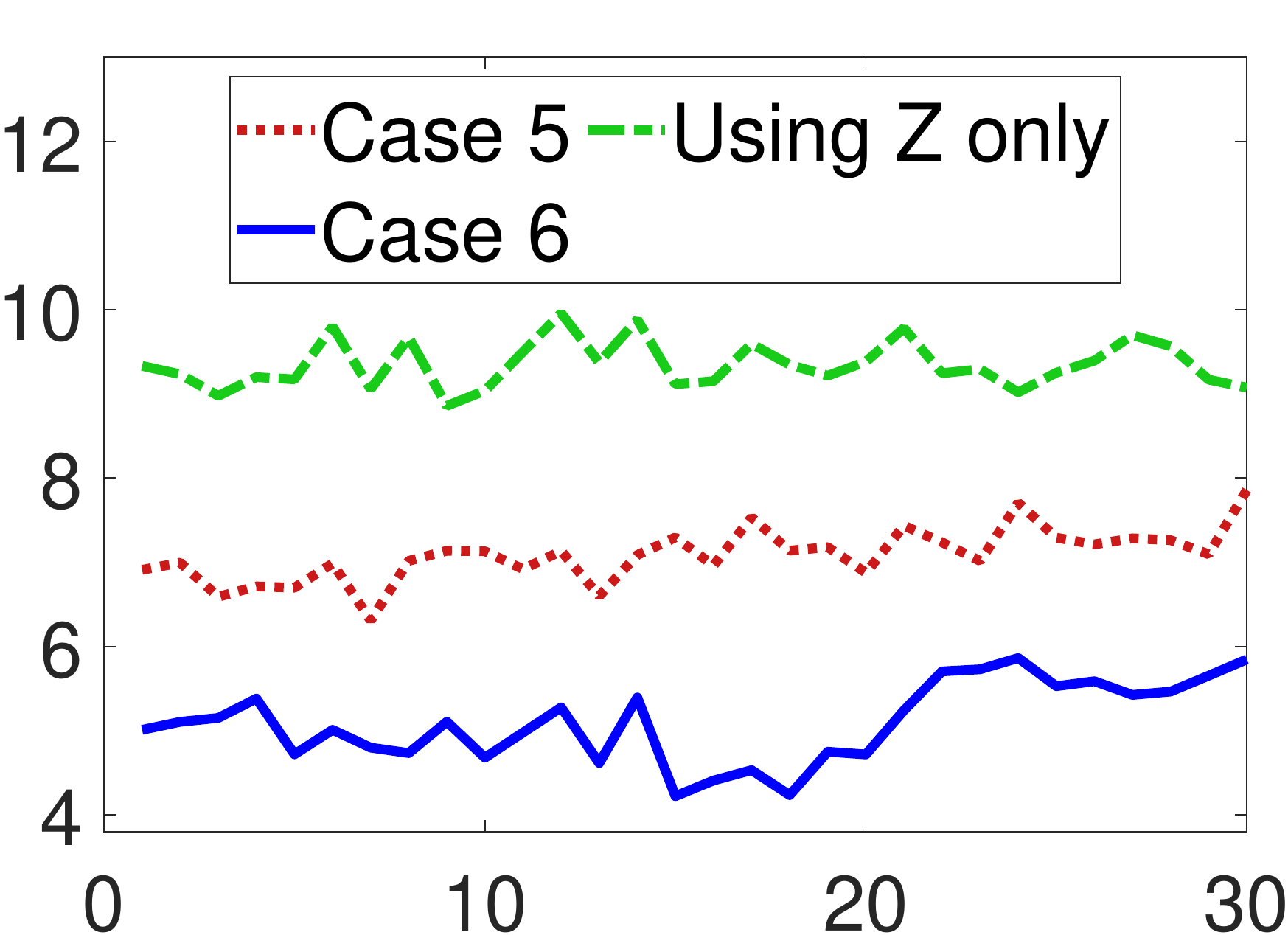}
		\end{minipage}%
	}%
	\subfigure[RMSE of  $h^{F}_{i_r}$  for each target.]{
		\begin{minipage}[t]{0.24\linewidth}
			\centering
			\includegraphics[width=0.99\textwidth]{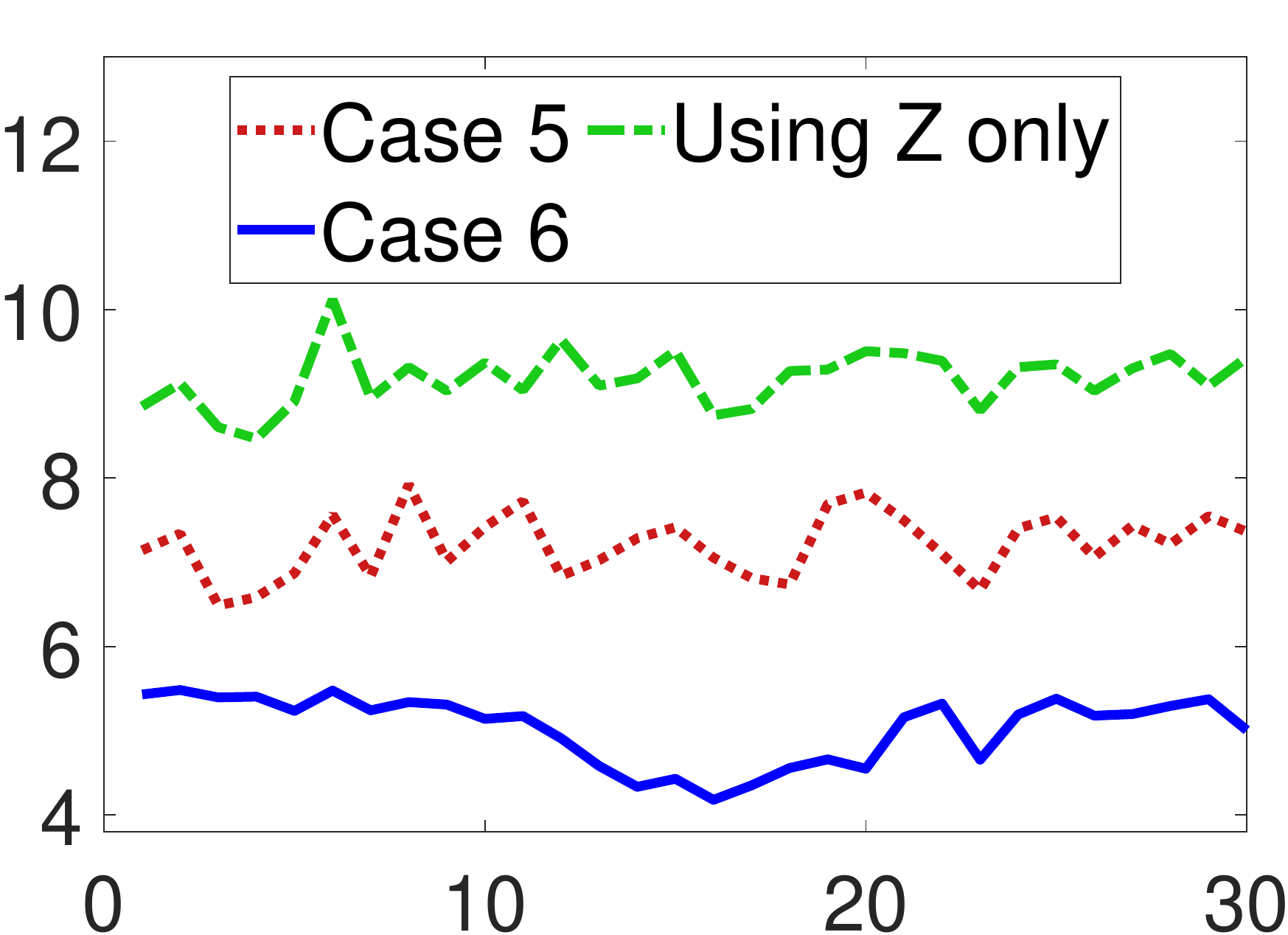}
			\includegraphics[width=0.99\textwidth]{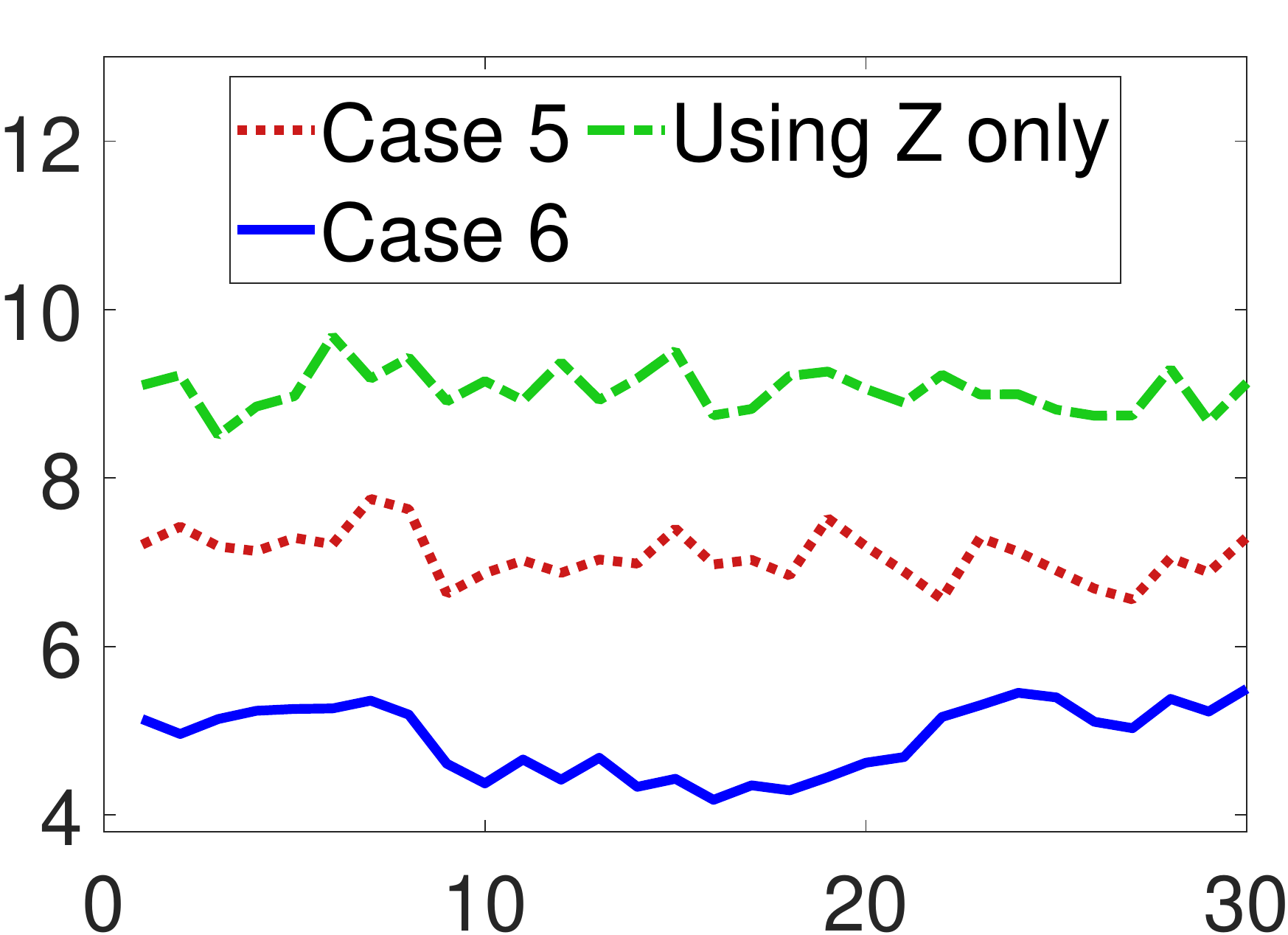}
			\includegraphics[width=0.99\textwidth]{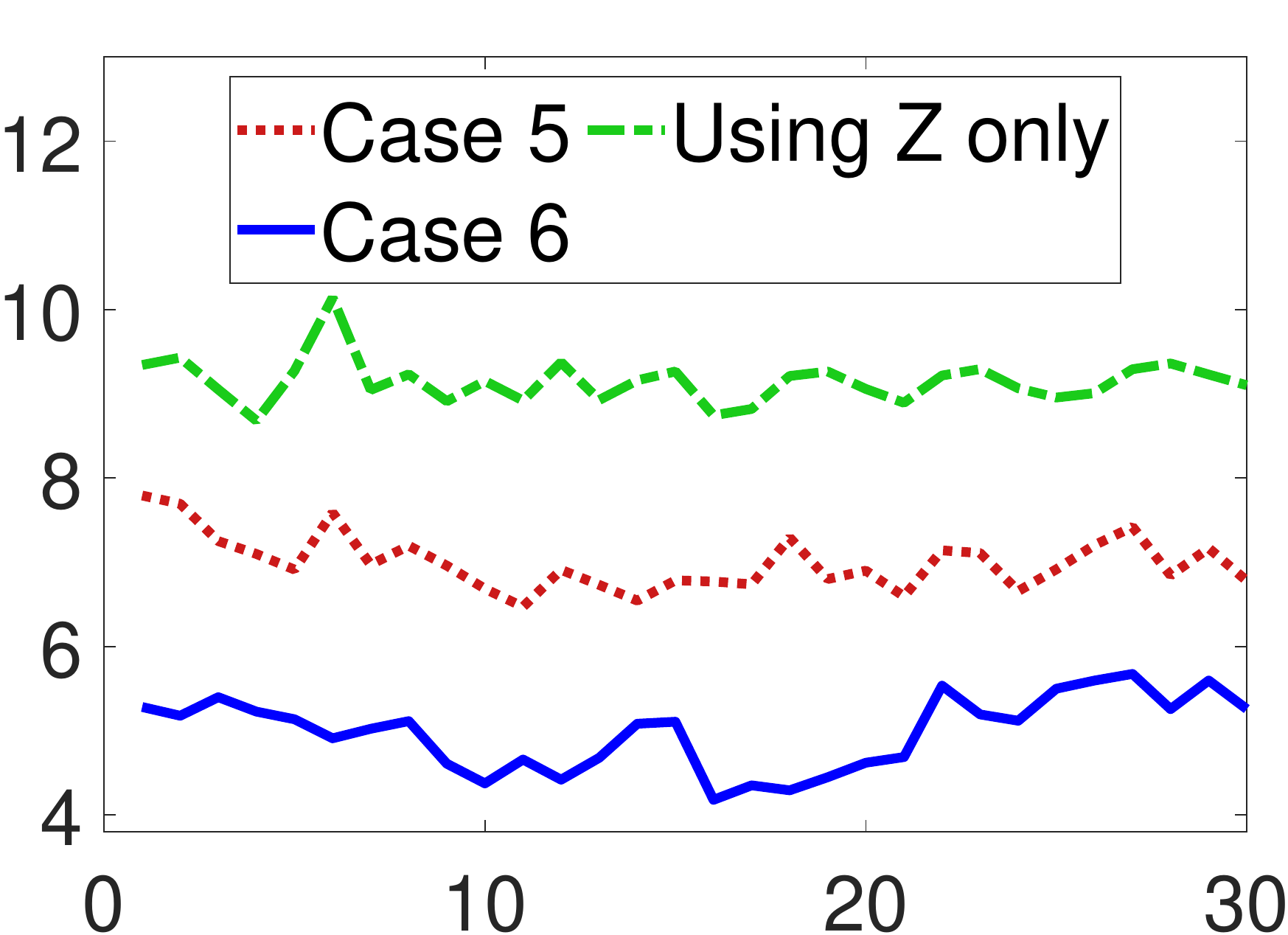}
			\includegraphics[width=0.99\textwidth]{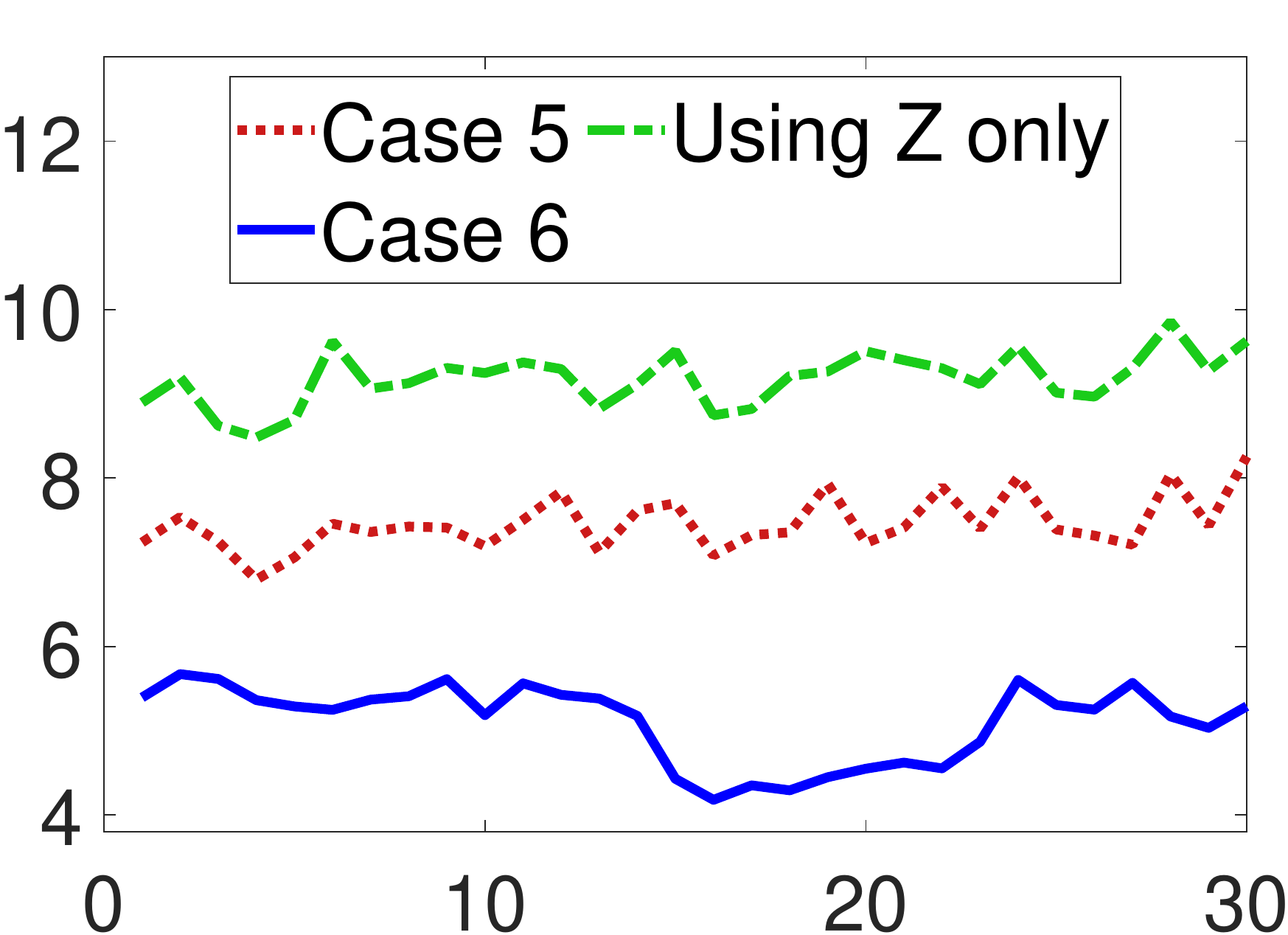}
			\includegraphics[width=0.99\textwidth]{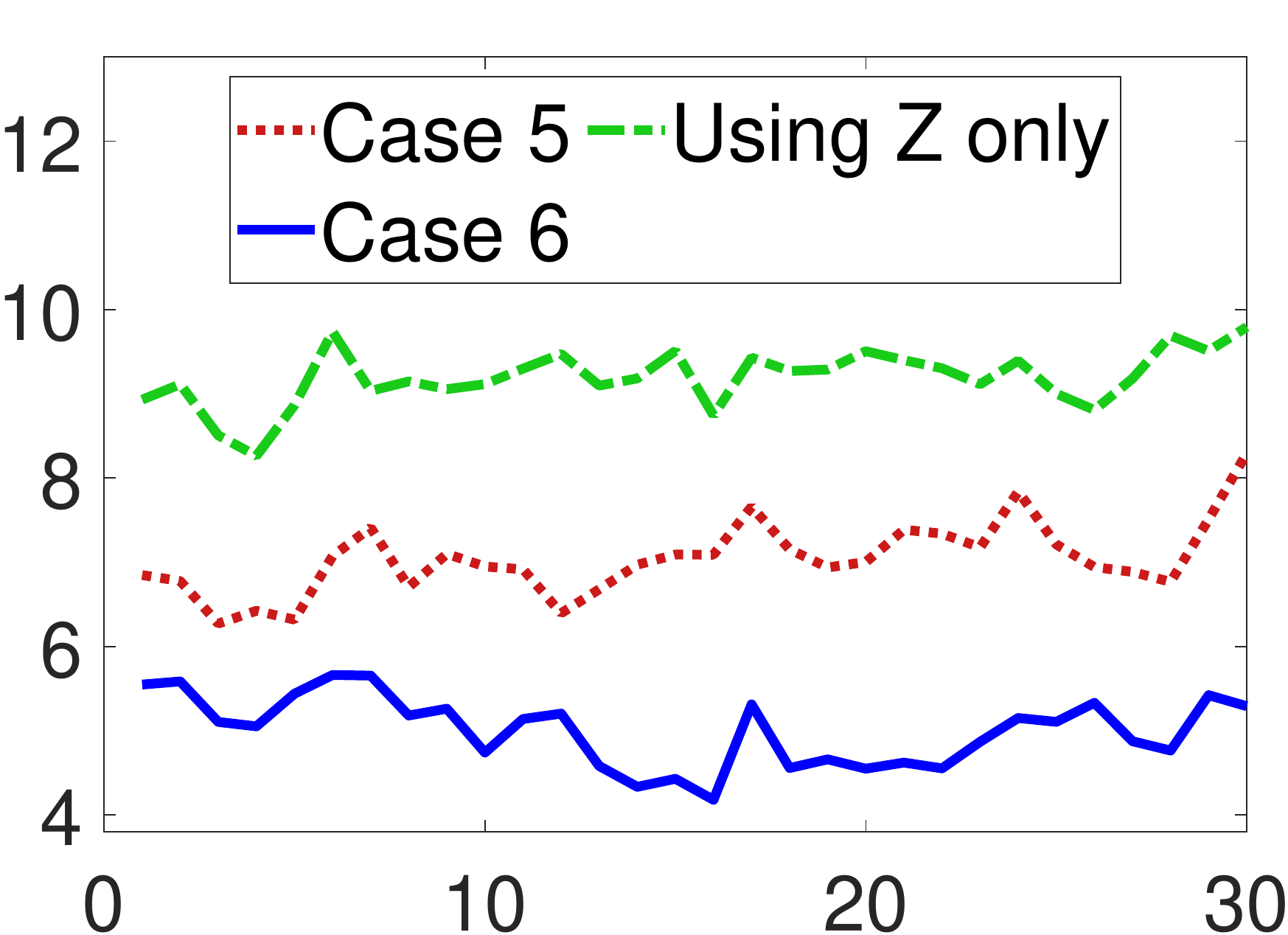}
		\end{minipage}%
	}%
	
	\centering
	\caption{RMSE of the VIHs for each target.
		The figures in each a single row are the results belong to the same target. From top to bottom are the results of Target 1-Target 5.
        From left to right are the results of ${ \bm{h}^{\mathrm{E}}(i_t) }$, ${ \bm{h}^{\mathrm{E}}(i_r) }$, ${ \bm{h}^{\mathrm{F}}(i_t) } $ and ${ \bm{h}^{\mathrm{F}}(i_r) }$ for each target.
		The legend 'Using Z only' represents the case that only ionosonde measurements are used to estimate the VIHs.
		The unit of the ordinate axis is km, and the abscissa axis represents the scan. }
	\label{Fig:usedVIHofEachDimensionofEachTarget}
\end{figure*}

Fig.~\ref{Fig:usedVIHofEachDimensionofEachTarget} shows the comparison of the used VIHs estimation obtained by ECM-GMRF for each target under Case 5, Case 6 and the case that only ionosonde measurements are used to estimate the used VIHs.
In Fig.~\ref{Fig:usedVIHofEachDimensionofEachTarget}, the green curves represent the RMSE of the estimated VIHs by only using ionosonde measurements $Z$, the red curves represent the RMSE of the estimated VIHs by using ionosonde measurements $Z$ and OTHR measurements $Y$ when targets are tracked
individually (i.e., Case 5), and the blue curves represent  the RMSE of the estimated VIHs by using ionosonde measurements $Z$ and OTHR measurements $Y$ when targets are tracked simultaneously.
From Fig.~\ref{Fig:usedVIHofEachDimensionofEachTarget}, it is seen that comparing with only using ionosonde measurements,
using both ionosonde measurements and OTHR measurements can improve the estimation of VIHs,
especially when all the targets are tracked simultaneously since OTHR measurements of all targets are used to
infer the used VIHs through the GMRF model.
In the GMRF model we constructed, the closer the subregions are, the stronger the correlation is.
As we can see from Fig.~\ref{fig:trueTrackAndusedVIHIndex},
the targets move toward the same point and they are very close around scan 15 as well as the VIHs.
Correspondingly, the lowest value of the estimated RMSE of VIHs when tracking simultaneously~(i.e., the green curve) in Fig.~\ref{Fig:usedVIHofEachDimensionofEachTarget} is at around scan 15.
Table~\ref{tab:improveRatioMultiTarget} shows the RMSE mean values~(km) of each curves of 30 scans in Fig.~\ref{Fig:usedVIHofEachDimensionofEachTarget}, and the improvement ratios compared with the standard deviations of the VIHs.

\begin{figure}[!htb]
	\centering
	\subfigure[RMSE of range for each target.]{
		\begin{minipage}[t]{0.499\linewidth}
			\centering
			\includegraphics[width=0.99\textwidth]{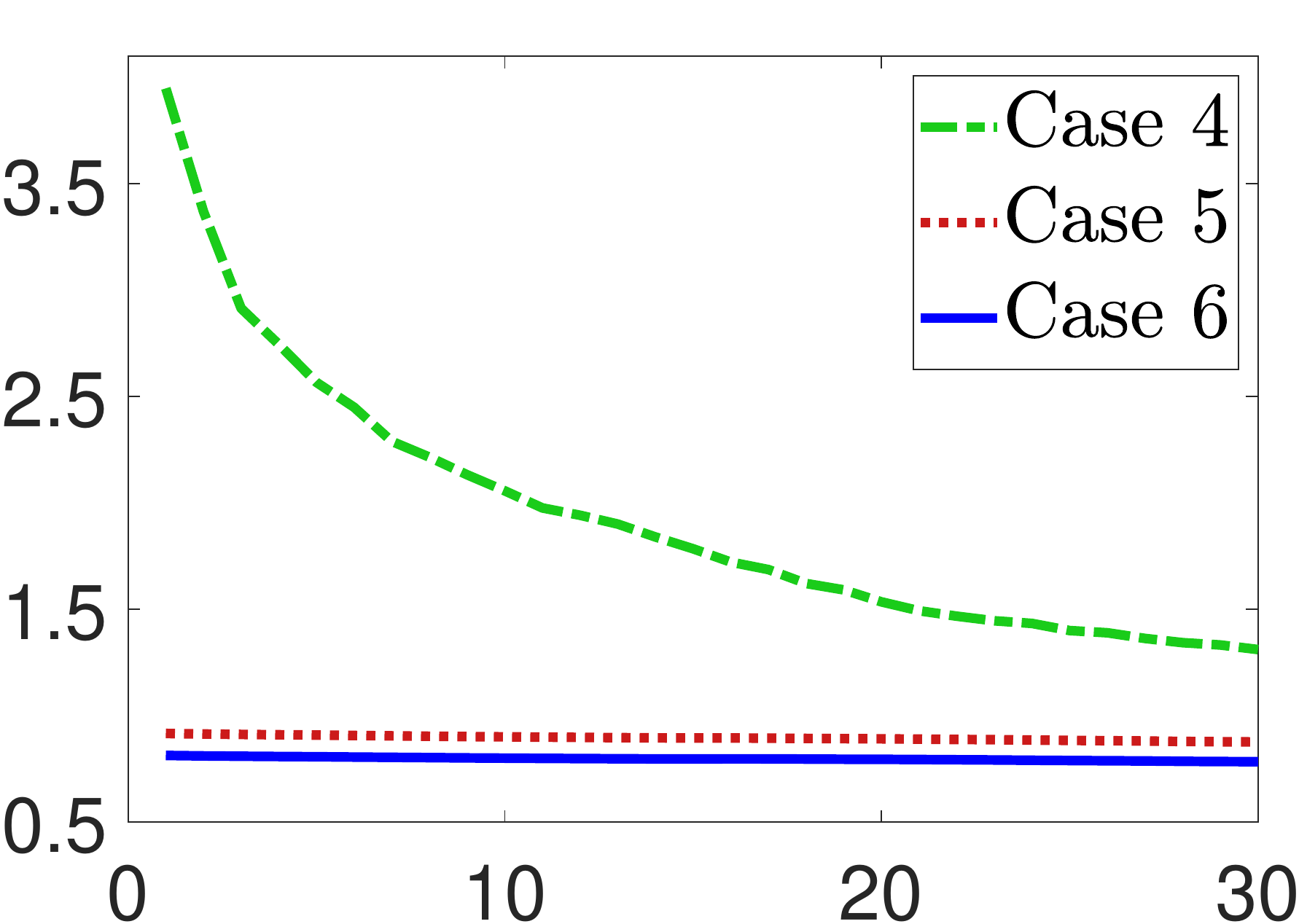}
			\includegraphics[width=0.99\textwidth]{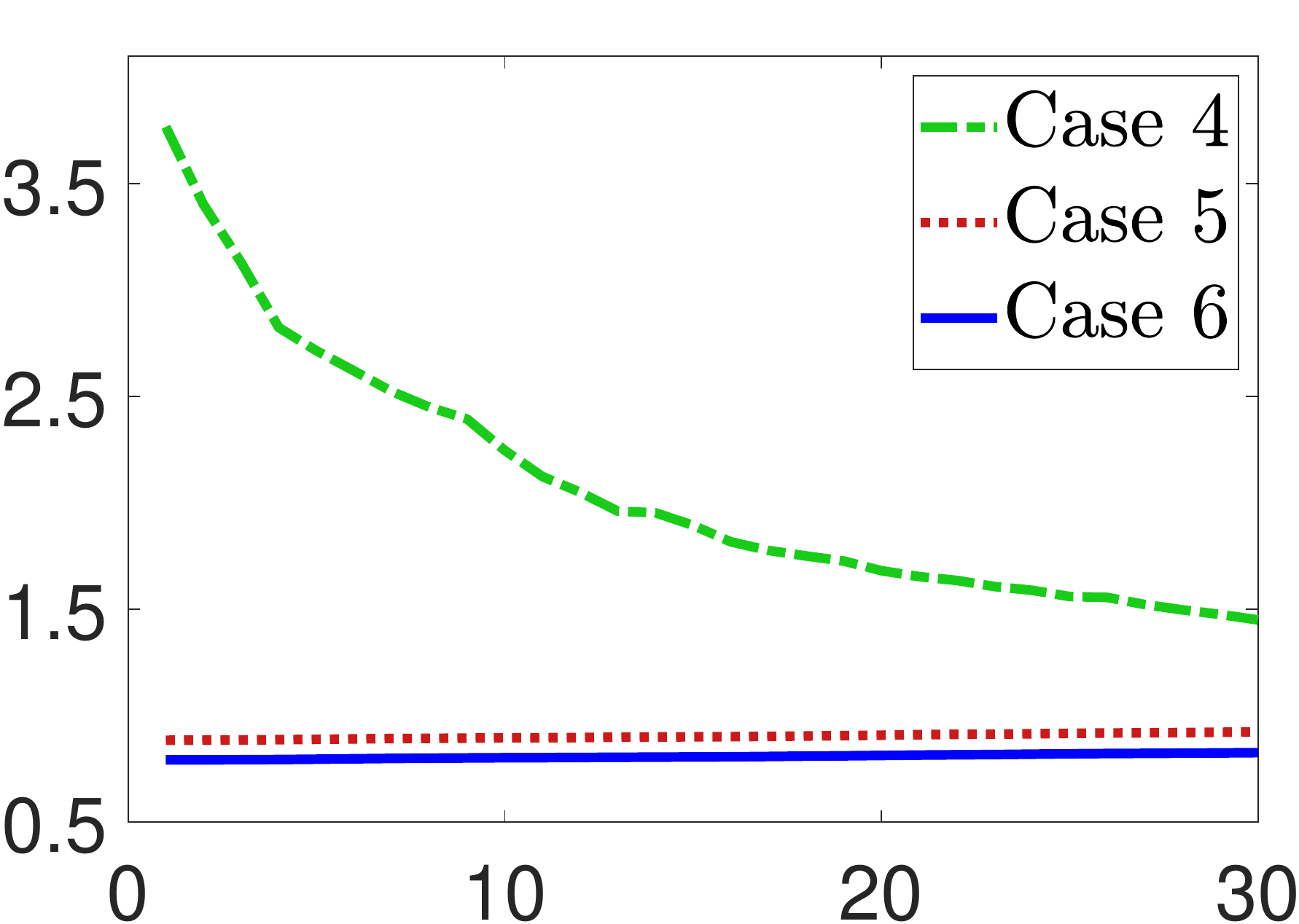}
			\includegraphics[width=0.99\textwidth]{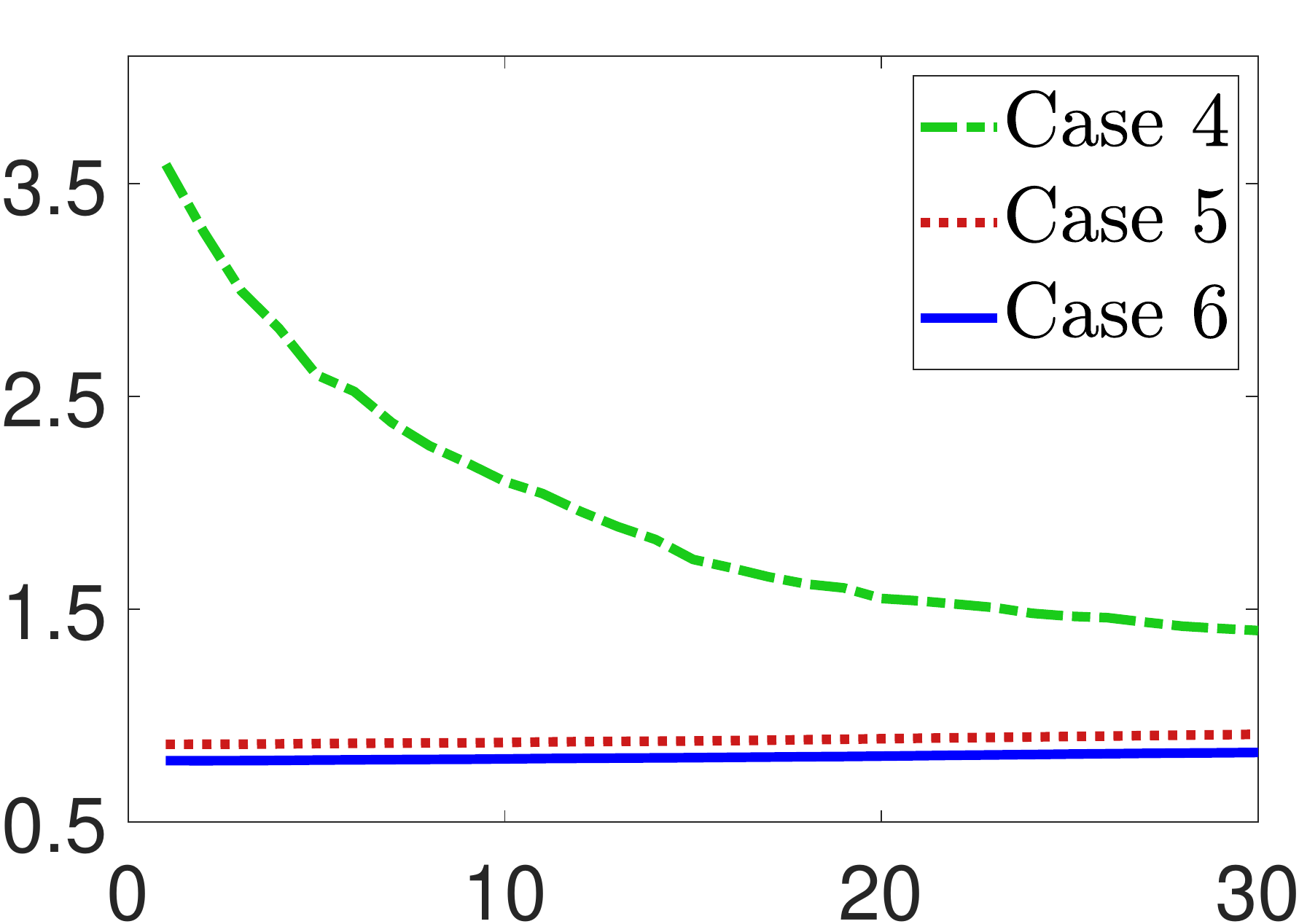}
			\includegraphics[width=0.99\textwidth]{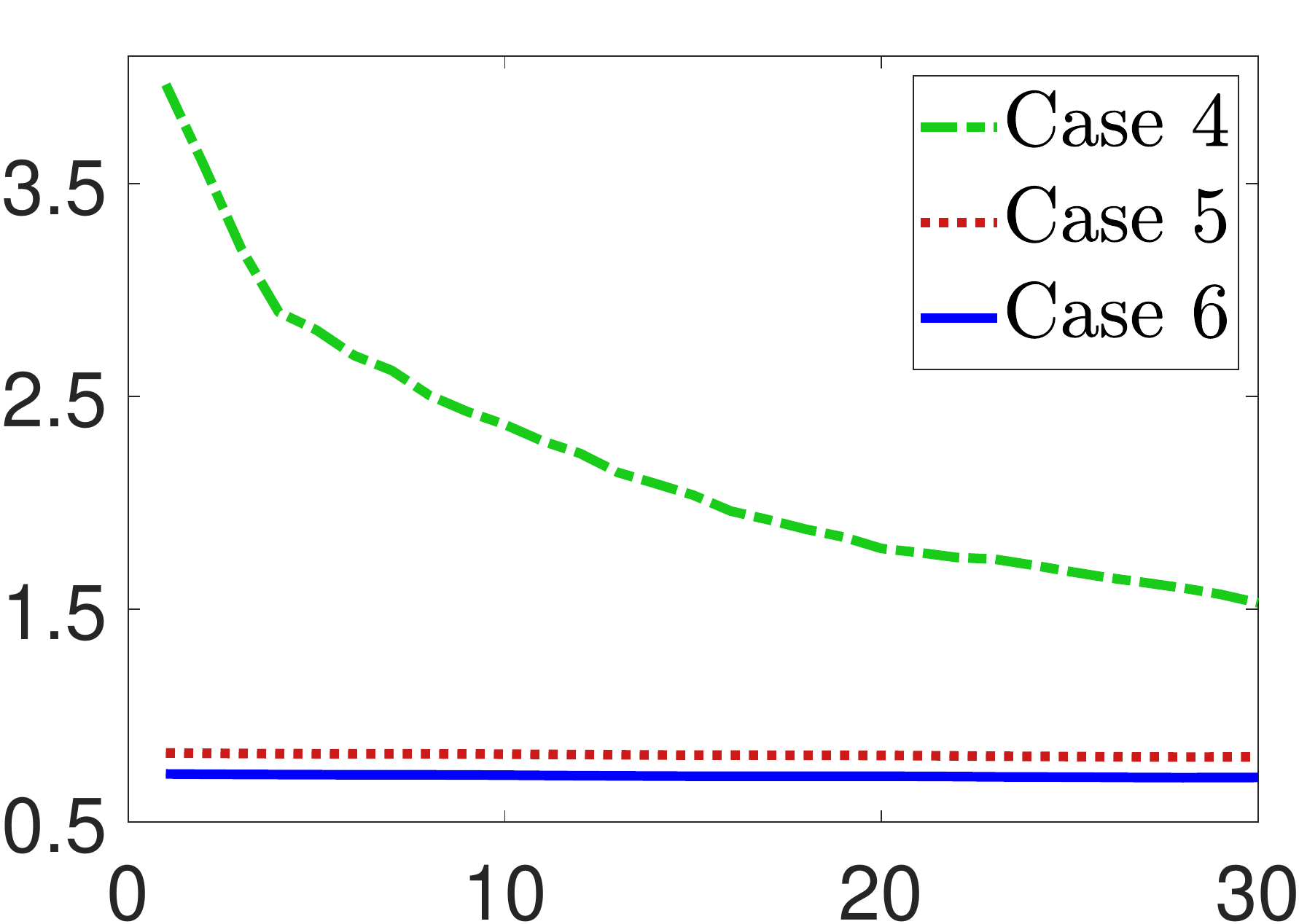}
			\includegraphics[width=0.99\textwidth]{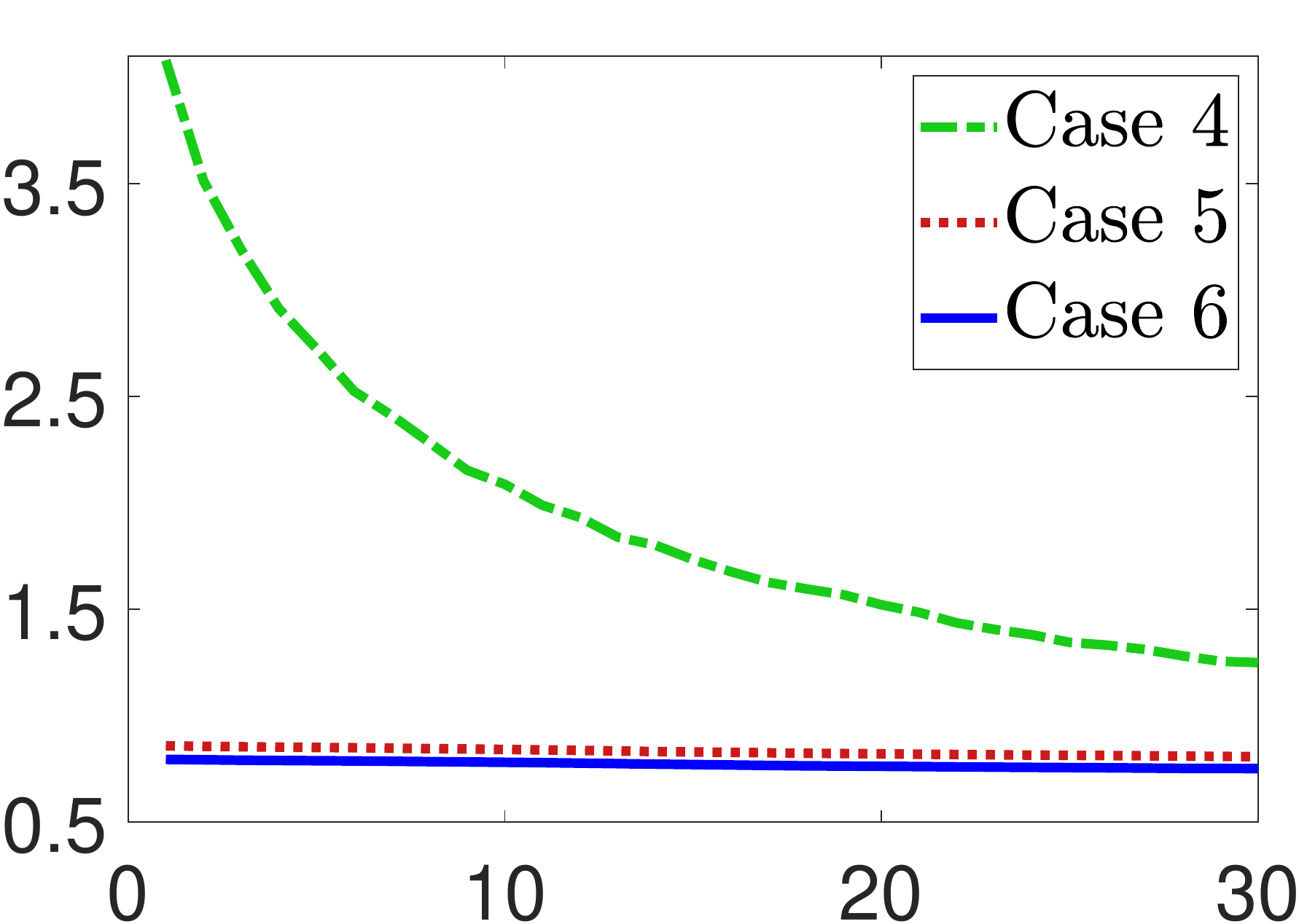}
			\label{Subfig: XRMSEDim1}
		\end{minipage}%
	}%
	\subfigure[RMSE of bearing for each target.]{
		\begin{minipage}[t]{0.499\linewidth}
			\centering
			\includegraphics[width=0.99\textwidth]{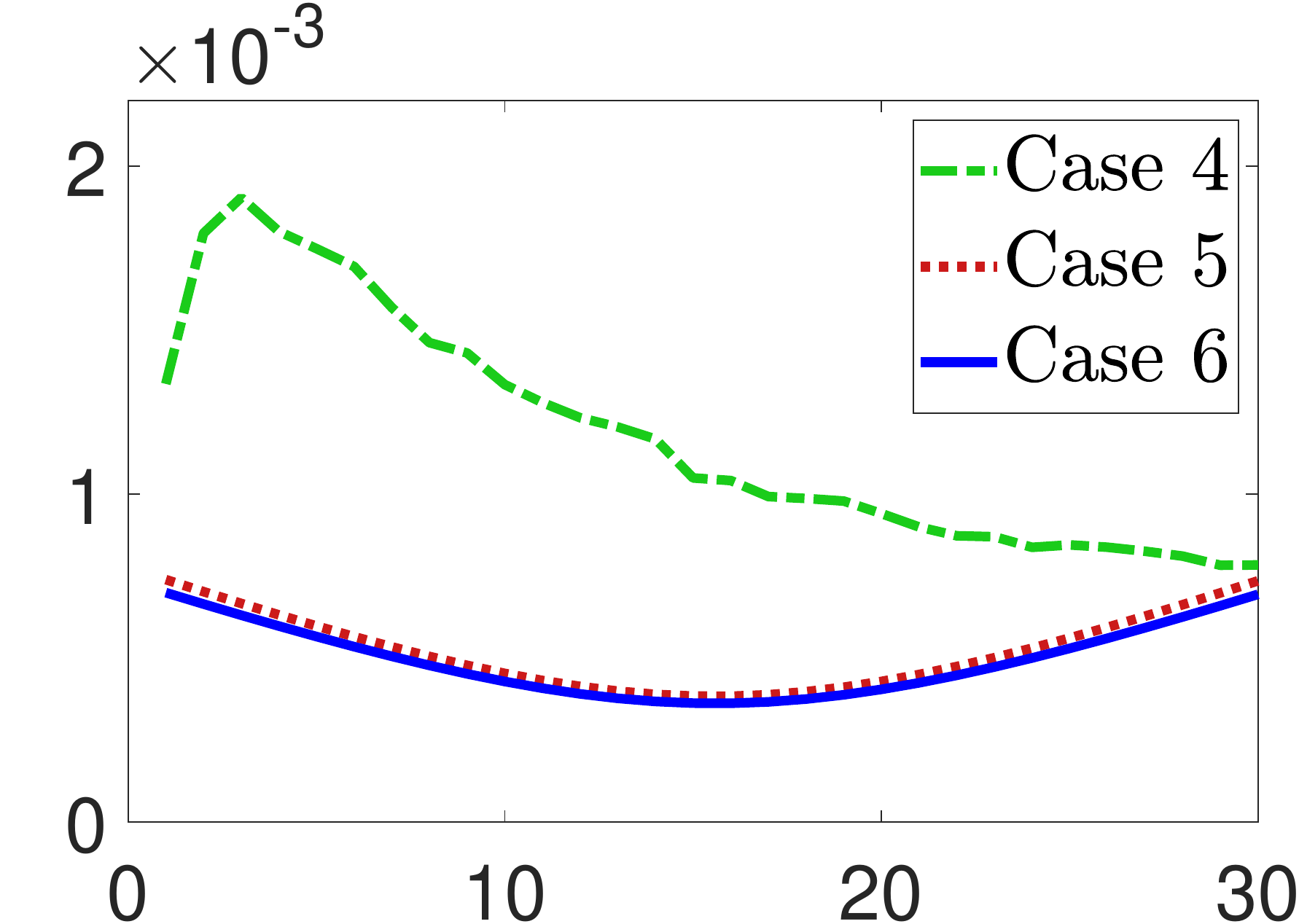}
			\includegraphics[width=0.99\textwidth]{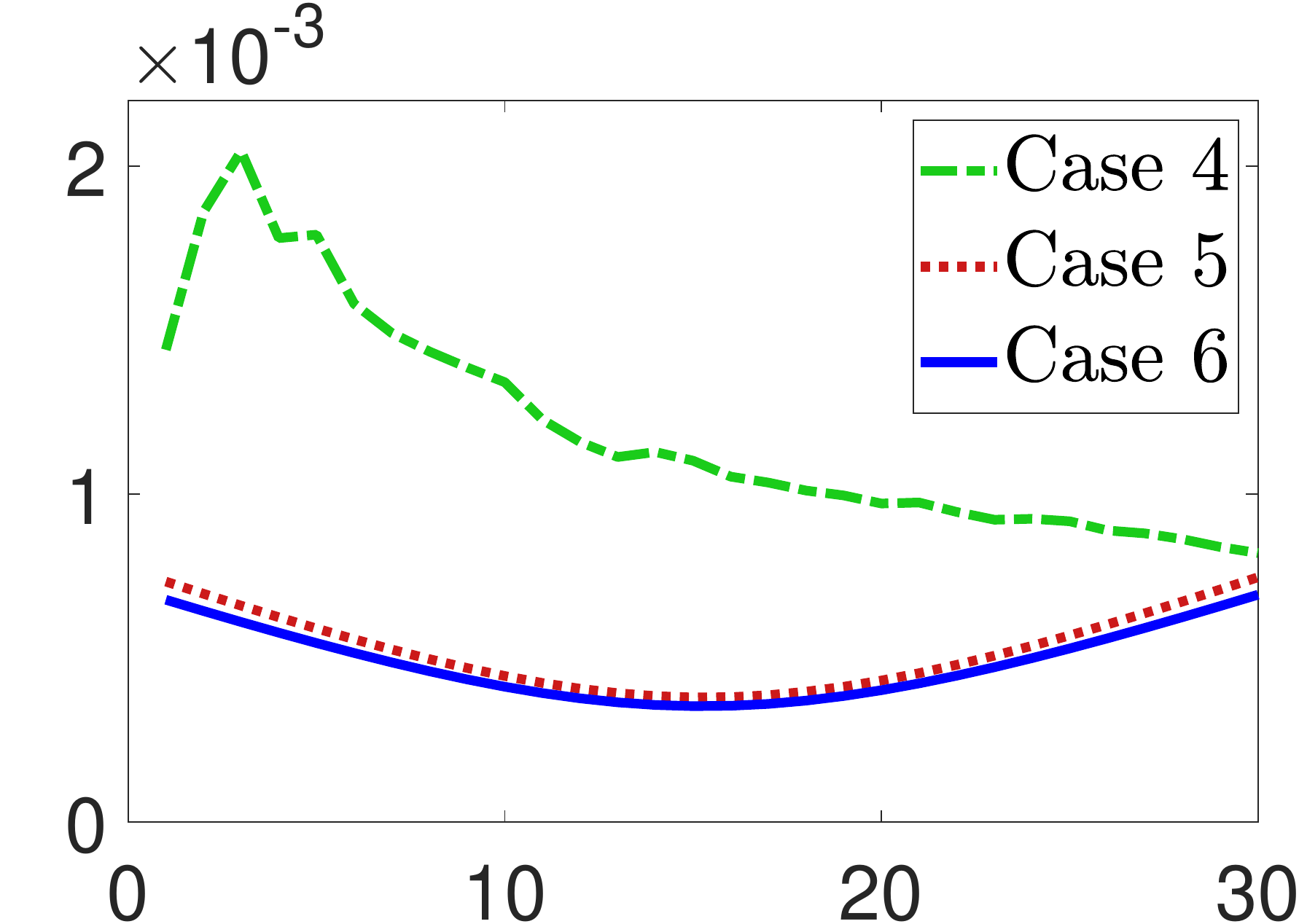}
			\includegraphics[width=0.99\textwidth]{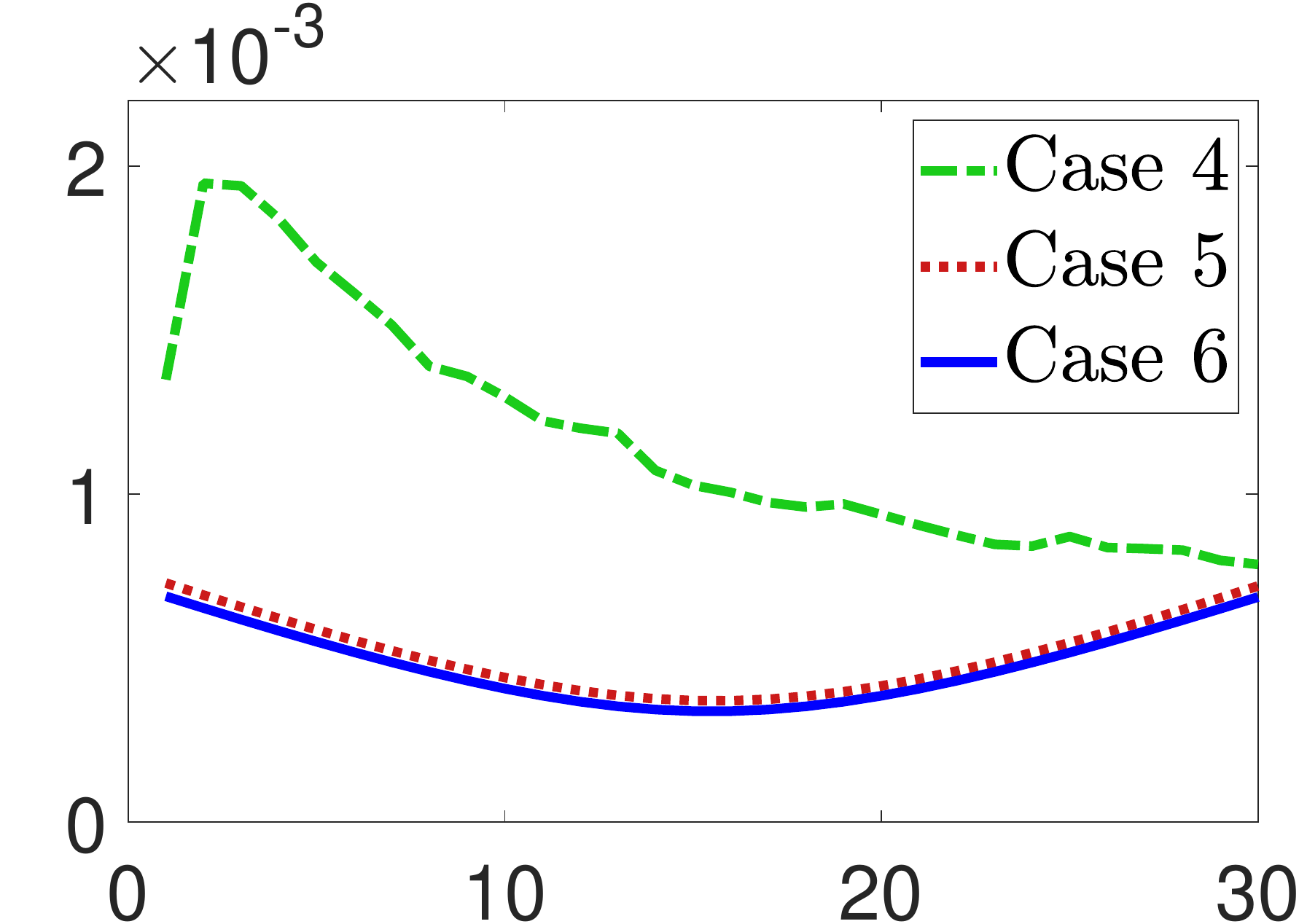}
			\includegraphics[width=0.99\textwidth]{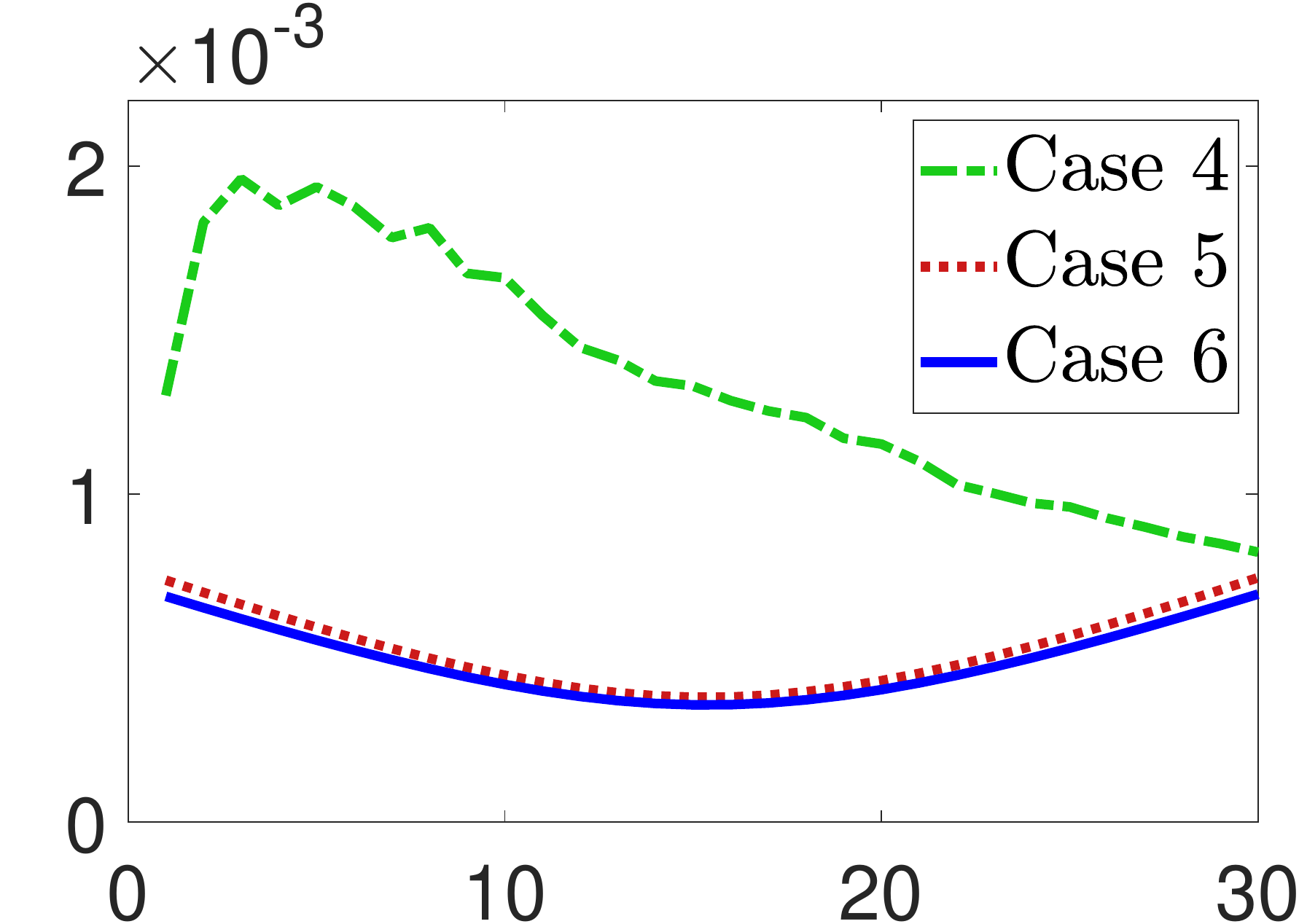}
			\includegraphics[width=0.99\textwidth]{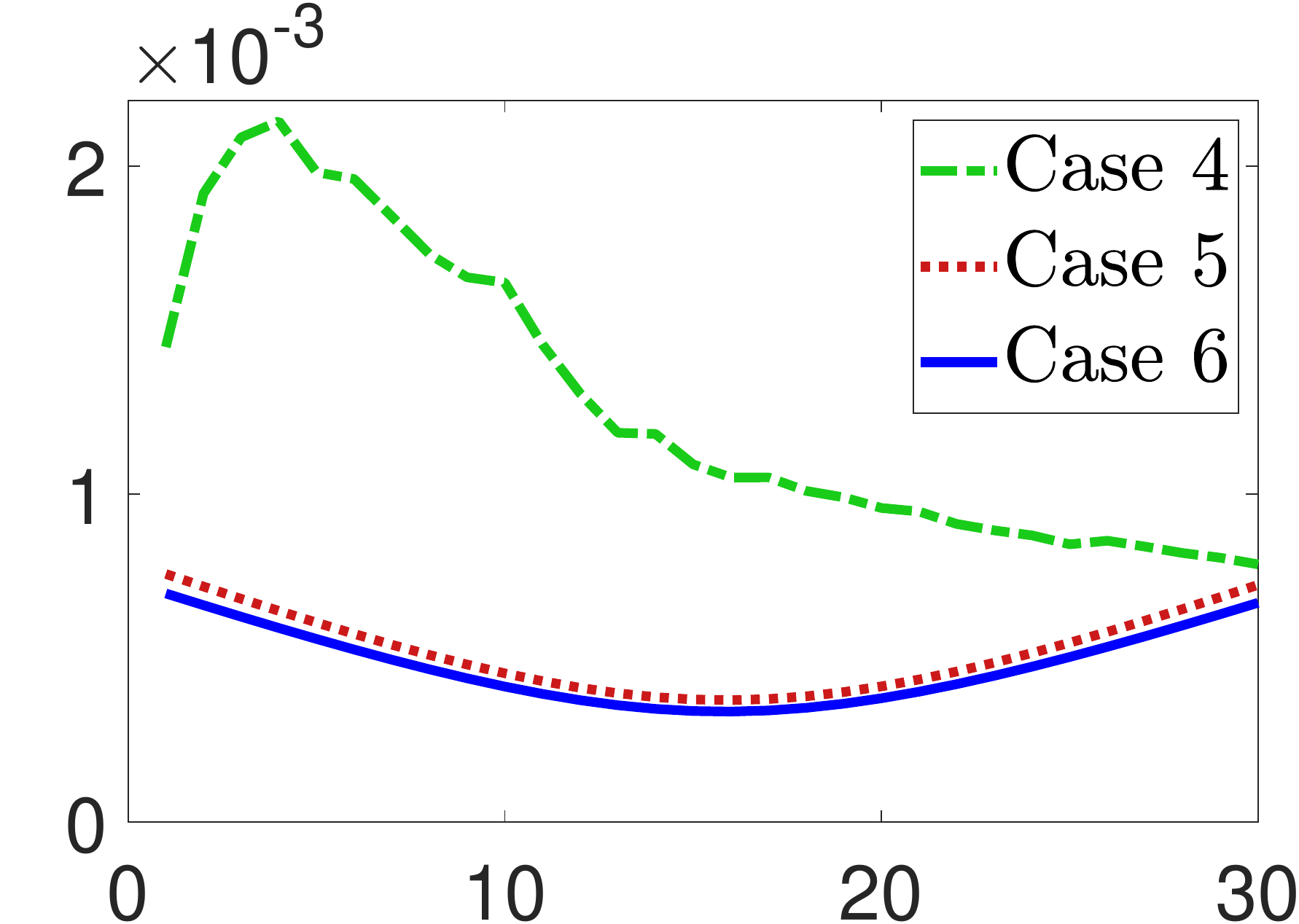}
			\label{Subfig: XRMSEDim3}
		\end{minipage}%
	}%
	
	\centering
	\caption{The comparison of the target state RMSE under different cases. From top to bottom are the results of Target 1-Target 5.
		The units of the ordinate axis in (a) and (b) are km and rad, respectively. The abscissa axis in both (a) and (b) represents the scan. }
	\label{Fig:XRMSE}
\end{figure}

For the evaluation of target tracking accuracy, Fig.~\ref{Fig:XRMSE} depicts the RMSEs for the range estimation and bearing estimation of the targets under different cases.
As we can see in Fig.~\ref{Fig:XRMSE}, for all targets, the estimation result of Case 6 is the best.
The tracking accuracy of Case 5 is slightly worse than that of Case 6.
Using the OTHR measurements of targets simultaneously can improve the tracking accuracy up to $9.95\%$ for all targets averagely.
ECM-GMRF performs better than MD-JPDAF (Case 4) under both Case 5 and Case 6.
The smoother in the ECM-GMRF improves the tracking accuracy.
Intuitively, since the target state and the VIHs are coupled in the radar measurement function,
accurate estimation of the VIHs leads to improvement of the target state estimation.
\begin{table}[!htb]
	\caption{\upshape{The statistics of the estimated VIHs in Fig.~\ref{Fig:usedVIHofEachDimensionofEachTarget}}}
	\footnotesize
	\renewcommand\arraystretch{1.1}
	\label{tab:improveRatioMultiTarget}
	\centering
	\begin{tabular}{l|l|ccc}
		\toprule
		\multicolumn{2}{c|}{}& Using Z only  & Case 5 & Case 6  \\
		\hline
		\multirow{2}*{layer E} &
		Mean value~(km)        &8.16       & 7.62 & 6.09 \\
		&Improvement ratio~(\%)&26.22      &30.69 & 44.63   \\
		\hline
		\multirow{2}*{layer F} &
		Mean value~(km)        &9.26      & 7.22 & 5.08 \\
		&Improvement ratio~(\%)&28.79     & 44.5 & 60.93  \\
		\bottomrule
	\end{tabular}
\end{table}

\section{Conclusion}\label{sec:conclusion}
We have addressed the problem of OTHR target state estimation and VIHs identification taking account of
the variation of the VIHs with location and the spatial correlation of the VIHs.
We have used GMRF to model the VIHs.
The problem has been formulated as a maximum a posteriori estimation problem based on both OTHR measurements and ionosonde measurements.
By applying ECM, we have proposed a joint joint optimization algorithm to perform target state estimation, multipath data association and VIHs estimation simultaneously.
In our proposed algorithm, both ionosonde measurements and OTHR measurements are exploited to estimate the VIHs,
leading to the reduction of the VIHs error and the improvement of target state estimation in return.
For future work, we plan to develop learning algorithms for the GMRF model and investigate the GMRF model considering both the horizontal and the vertical correlation of the VIHs.

\section*{Acknowledgments}
This work was in part supported by the National Natural Science Foundation of China~(grant no. 61503305, 61873211, 61790552)

\bibliographystyle{elsarticle-num}
\bibliography{MyCollection}

\end{document}